\documentclass[aps, prd, onecolumn, tightenlines, notitlepage, superscriptaddress, nofootinbib, preprintnumbers, floatfix,showkeys,11pt]{revtex4-2}

\usepackage[normalem]{ulem}
\usepackage{amstext}
\usepackage{amssymb}
\usepackage{amsmath}
\usepackage{graphicx}
\graphicspath{{plots/}}
\usepackage{url}
\usepackage{color}
\usepackage{ulem}
\usepackage[utf8]{inputenc}
\pdfoutput=1
\usepackage{textcomp}
\usepackage{booktabs}
\usepackage{booktabs,siunitx,array,threeparttable}
\usepackage[T1]{fontenc}
\usepackage{ae,aecompl}
\usepackage{appendix}

\usepackage{epsfig,amsfonts,mathrsfs,amsmath,amssymb,graphicx,color,slashed,multirow}
\usepackage{amsmath,latexsym,amssymb,graphicx,slashed,color,enumerate,url,cancel,gensymb}
\usepackage{textcomp}

\usepackage[x11names]{xcolor}
\usepackage[colorlinks]{hyperref}

\usepackage[paperwidth=210mm,paperheight=297mm,centering,hmargin=2.5cm,vmargin=2.5cm]{geometry}
\usepackage{tikz-feynman}
\tikzfeynmanset{compat=1.1.0}
\usetikzlibrary{patterns,decorations.pathreplacing}
\usetikzlibrary{decorations.pathmorphing}
\usetikzlibrary{decorations.markings}
\usetikzlibrary{arrows,shapes}
\usetikzlibrary{shapes.misc}
\tikzset{
  gluon/.style={decorate, draw=black,
    decoration={coil,amplitude=4pt, segment length=4pt,aspect=0.7}} 
}
\tikzset{
  photon/.style={decorate, decoration={snake}},
}

\usepackage{textgreek} 
\usepackage{booktabs} 

\hypersetup{colorlinks,citecolor= nicered,linkcolor= blue}
\definecolor{nicered}{rgb}{0.7,0.1,0.1}
\definecolor{nicegreen}{rgb}{0.1,0.5,0.1}

\def\cevns{CE$\nu$NS}
\def\d{\mathrm{d}}
\newcommand{\qtransfer}{\left|\mathbf{q}\right|}

\definecolor{vdrgreen}{rgb}{0.0, 0.6, 0.0}

\definecolor{myblue}{cmyk}{0.65, 0.37, 0.0, 0.19}

\definecolor{blue(ncs)}{rgb}{0.0, 0.53, 0.74}

\AtBeginDocument{\hypersetup{citecolor=myblue,linkcolor=myblue,urlcolor=myblue}}

\begin{document}
\begin{flushright}
  IFIC/23-27
\end{flushright}

\title{{\LARGE 
A neutrino window to scalar leptoquarks:\\
from low energy to colliders}}
\author{Valentina De Romeri}\email{deromeri@ific.uv.es}
\affiliation{Instituto de F\'{i}sica Corpuscular (CSIC-Universitat de Val\`{e}ncia), Parc Cient\'ific UV C/ Catedr\'atico Jos\'e Beltr\'an, 2 E-46980 Paterna (Valencia) - Spain}
\author{Víctor Martín Lozano }\email{victor.lozano@ific.uv.es}
\affiliation{Instituto de F\'{i}sica Corpuscular (CSIC-Universitat de Val\`{e}ncia), Parc Cient\'ific UV C/ Catedr\'atico Jos\'e Beltr\'an, 2 E-46980 Paterna (Valencia) - Spain}
\affiliation{Departament de F\'isica Teòrica, Universitat de Val\`{e}ncia, 46100 Burjassot, Spain}
\author{G. Sanchez Garcia}\email{gsanchez@ific.uv.es}
\affiliation{Instituto de F\'{i}sica Corpuscular (CSIC-Universitat de Val\`{e}ncia), Parc Cient\'ific UV C/ Catedr\'atico Jos\'e Beltr\'an, 2 E-46980 Paterna (Valencia) - Spain}


\begin{abstract}
Leptoquarks are theorized particles of either scalar or vector nature that couple simultaneously to quarks and leptons. Motivated by recent measurements of coherent elastic neutrino-nucleus scattering, we consider the impact of scalar leptoquarks coupling to neutrinos on a few complementary processes, from low energy to colliders. In particular, we set competitive constraints on the typical mass and coupling of scalar leptoquarks by analyzing recent COHERENT data. We compare these constraints with bounds from atomic parity violation experiments, deep inelastic neutrino-nucleon scattering and collider data. Our results highlight a strong complementarity between different facilities and demonstrate the  power of coherent elastic neutrino-nucleus scattering experiments to probe leptoquark masses in the sub-TeV range. Finally, we also present prospects for improving current bounds with future upgrades of the COHERENT detectors and the planned European Spallation Source. 
\end{abstract}

\maketitle

\section{Introduction}
Leptoquarks (LQs) are hypothetical particles that carry both lepton
and baryon numbers, and can arise in many extensions of the Standard Model (SM) unifying matter~\cite{Pati:1973uk,Georgi:1974sy,Georgi:1974my,Fritzsch:1974nn} with the unique property of connecting leptons and quarks. This peculiar property could induce rapid proton decay. However, as they arise in many Grand Unified Theories (GUT), their mass is expected to be close to the GUT scale preventing this process from happening~\cite{Nath:2006ut}. On the other hand, there are models where the operator that gives rise to proton decay, the diquark operator ($qqlq$), is suppressed or even forbidden by a symmetry, allowing LQs to have low masses~\cite{Buchmuller:1986zs,Belyaev:2005ew,Dorsner:2005fq,Arnold:2013cva,FileviezPerez:2013zmv}.
LQ properties and signatures have been extensively studied in the literature. We refer the reader to~\cite{Dorsner:2016wpm} for a comprehensive review on the LQ phenomenology at precision experiments and colliders.
Moreover, LQs coupling to third-generation fermions have received much attention
lately as likely candidates to account for flavor anomalies, see for instance~\cite{Tanaka:2012nw,Sakaki:2013bfa,Dorsner:2013tla,Gripaios:2014tna,Bauer:2015knc,Buttazzo:2017ixm,Dorsner:2018ynv,Hiller:2014yaa,ColuccioLeskow:2016dox,Crivellin:2017zlb,Hiller:2018wbv,Crivellin:2020tsz,Angelescu:2021lln,Nomura:2021oeu,Marzocca:2021azj,FileviezPerez:2021lkq,Murgui:2021bdy,Singirala:2021gok,Crivellin:2022mff}.

At present, LQ models have been studied through a variety of processes involving different ranges of energy. Depending on certain assumptions, different observables can test particular regions of masses and coupling strengths in the parameter space. On the one hand, Atomic Parity Violation (APV) in cesium  nuclei has allowed to test the effects of LQs at low energy. Particle colliders, on the other hand, like LEP and LHC, have allowed to probe LQs in relatively large energy ranges through processes like Drell Yan, as well as through single and double pair production. Another important test of LQ interactions are electroweak precision observables. The presence of LQs could induce effects on the self-energy of the $Z$ and $W$ bosons constrained by the oblique parameters, $S$, $T$ and $U$. However, such corrections  mostly depend on the difference of masses between states \cite{Dorsner:2016wpm}. Since we are interested in LQ multiplets with degenerate states,  we will not take these constraints into account in our study. A more complete investigation of the LQ parameter space, focusing on the intermediate energy region, is now possible thanks to the recent observation of Coherent Elastic Neutrino-Nucleus Scattering (\cevns). In this work, we provide a general overview of the constraints that can be obtained for LQ properties from different processes, at different energy ranges, and under certain specific assumptions. Our main motivation is to investigate LQ signatures at low-energy facilities devoted to the study of \cevns~\cite{Abdullah:2022zue}.  Unless explicitly stated, in the rest of this manuscript we will mainly focus on LQs coupling to the first and second generation of leptons. 

The first observation of \cevns~was achieved by the COHERENT collaboration in 2017 using a cesium iodide (CsI) detector~\cite{COHERENT:2017ipa},
about forty years after its theoretical prediction~\cite{Freedman:1973yd}.
More recently, the same collaboration observed \cevns~using a liquid argon (LAr) detector~\cite{COHERENT:2020iec} and, in 2021, they released a more extended CsI data set~\cite{COHERENT:2021xmm} with improved statistical analysis. Complementary to COHERENT, which uses neutrinos from a $\pi$-DAR (pion decay-at-rest) source, further \cevns~measurements are soon expected by other facilities, either exploiting neutrinos from nuclear reactors~\cite{CONNIE:2021ggh,CONUS:2021dwh,nuGeN:2022bmg,MINER:2016igy,Billard:2016giu,Strauss:2017cuu,Wong:2015kgl,Fernandez-Moroni:2020yyl,Akimov:2022xvr,NEON:2022hbk,NEWS-G:2021mhf,SBC:2021yal,Colaresi:2022obx}, from $\pi$-DARs~\cite{CCM:2021leg,Baxter:2019mcx} or decay-in-flight neutrinos produced at the Long Baseline Neutrino Facility~\cite{AristizabalSierra:2021uob,AristizabalSierra:2022jgg}. 
As a consequence, current and upcoming \cevns~results provide plenty of opportunities for both testing the SM parameters~\cite{Papoulias:2019lfi,Coloma:2020nhf,Cadeddu:2021ijh,Majumdar:2022nby,AristizabalSierra:2022axl,Fernandez-Moroni:2020yyl,DeRomeri:2022twg,Sierra:2023pnf,AtzoriCorona:2023ktl} and searching for new physics (see for instance~\cite{Barranco:2005yy,Ohlsson:2012kf,Miranda:2015dra,Dent:2016wcr,Farzan:2017xzy,Liao:2017uzy,AristizabalSierra:2018eqm,Abdullah:2018ykz,Billard:2018jnl,Denton:2018xmq,Han:2019zkz,Abdullah:2018ykz,Giunti:2019xpr,AristizabalSierra:2019ykk,AristizabalSierra:2019ufd,Denton:2020hop,Flores:2020lji,Cadeddu:2020nbr,Amaral:2021rzw,Flores:2021kzl,AtzoriCorona:2022moj,DeRomeri:2022twg,Schechter:1981hw,Pal:1981rm,Kayser:1982br,Nieves:1981zt,Shrock:1982sc,Kosmas:2015sqa,Canas:2015yoa,Miranda:2019wdy,Miranda:2020kwy,Kosmas:2017zbh,Brdar:2018qqj,Blanco:2019vyp,Miranda:2020syh,Miranda:2021kre,Bolton:2021pey,Chang:2020jwl,Chao:2021bvq,Chen:2021uuw,AristizabalSierra:2021fuc,DeRomeri:2022twg,Calabrese:2022mnp,Candela:2023rvt,Felkl:2023nan, Breso-Pla:2023tnz}).
Among a variety of applications, \cevns~data can be used to probe LQs connecting neutrinos to quarks. Assuming effective interactions, model-independent constraints on neutrino nonstandard interactions (NSI) can be recast into bounds on masses and couplings of heavy LQs as shown in~\cite{Barranco:2007tz,Billard:2018jnl}. \cevns~limits on first-generation LQs have also been obtained in~\cite{Crivellin:2021bkd} under the assumption of a SM effective field theory. Finally, bounds on scalar LQs have been obtained in~\cite{Calabrese:2022mnp} using COHERENT CsI (2017) and LAr (2020) data. Here we investigate the impact of scalar LQs on current \cevns~data from the COHERENT experiment. In particular, we perform a combined analysis of the most recent CsI (2021) and LAr (2020) data sets, and we include timing information in the data analysis. We improve upon previous results in the literature~\cite{Barranco:2007tz,Billard:2018jnl,Crivellin:2021bkd} by considering an explicit dependence on the LQ mass (and not assuming an effective interaction), and upon~\cite{Calabrese:2022mnp} by improving the statistical analysis in several ways: a newer CsI data set, and the inclusion of timing information together with all relevant systematic effects. In addition, we further provide sensitivities for planned upgrades of the CsI and LAr COHERENT detectors~\cite{Asaadi:2022ojm,Akimov:2022oyb} and for a future proposal at the European Spallation Source (ESS)~\cite{Baxter:2019mcx}.

Complementary to the \cevns~analysis, but still focusing on the low-energy sector, additional information on first-generation LQs can be extracted from APV~\cite{Bouchiat:1974kt,Safronova:2017xyt} experiments using cesium atoms~\cite{Wood:1997zq,Guena:2004sq}. APV can be originated as a manifestation of the weak interaction either by the exchange of a $Z$-boson between electrons and nucleus or by P-violating inter-nuclear forces. APV limits on LQs have been obtained previously in the literature~\cite{Langacker:1990jf,Davidson:1993qk,Barger:2000gv,Dorsner:2016wpm,Crivellin:2021bkd}  parameterizing the LQ effects by means of an effective Lagrangian.
We reevaluate APV constraints on first-generation LQs by
estimating their effect on the weak nuclear charge of cesium, taking into account an explicit dependence on the LQ mass and coupling.

Moving to the heavier energy range,
we recast constraints from neutrino scattering off nuclei, namely those from the NuTeV experiment~\cite{NuTeV:2001whx}. This deep inelastic neutrino-nucleon scattering cross section has been measured with high accuracy by NuTeV, providing a test on the coupling of muon neutrinos to quarks. LQs can significantly modify the SM coupling and hence the neutrino-nucleon cross section. We then make use of the NSI framework to translate the results of Ref.~\cite{Escrihuela:2011cf}, that obtained constraints on the NSI coefficients from NuTeV data, into LQ masses and couplings.
Finally, in order to test the heavy mass and large energy regimes, we consider data from colliders. In particular we study how HERA~\cite{H1:2001ezk, ZEUS:2003uzd, ZEUS:2012pwm}, an electron-proton collider, LEP~\cite{L3:1991sow,OPAL:1991xaz, OPAL:1998gjo,OPAL:1998znn,L3:2000bql}, an electron-positron collider, SPS~\cite{UA2:1991ovi} and Tevatron~\cite{CDF:1993dmp,Norman:1994tp,D0:1997eun,Barfuss:2009zz}, both proton-antiproton colliders, and LHC~\cite{ATLAS:2020dsk, CMS:2018lab,CMS:2018ncu,ATLAS:2020yat, CMS:2021ctt,CMS:2021far}, a proton-proton collider, can shed some light on first- and second-generation LQs through their imprints on different processes. We 
recast bounds from HERA~\cite{ZEUS:2012pwm}, SPS~\cite{UA2:1991ovi}, Tevatron~\cite{Barfuss:2009zz} and LEP~\cite{OPAL:1998gjo,L3:1991sow,OPAL:1991xaz,OPAL:1998znn}, while we compute LHC constraints~\cite{ATLAS:2020dsk,CMS:2021ctt,CMS:2021far}
using Monte Carlo tools in order to obtain the most up-to-date limits from their searches.
We finally mention that during the completion of this work a study on LQs has been presented in Ref.~\cite{Kirk:2023fin}. Their analysis focuses on LQ interactions with third-generation leptons and second-generation quarks, and on their testability using neutrino telescopes, compared to other constraints including colliders. Their study complements the results presented here.\\

This paper is organised as follows. In Sec.~\ref{sec:LQ-formalism} we introduce the general LQ picture, particularly focusing on scalar LQs that couple to the first and second generation of leptons and to the first generation of quarks, which are hence accessible at \cevns~experiments. In Sec.~\ref{sec:current_constr} we briefly describe the \cevns~process, the associated LQ effects, and the COHERENT experiment. In the same section, we also detail our statistical analysis, that allows us to set stringent constraints on LQ properties. We then describe in Sec.~\ref{sec:other_constr} the effect of scalar LQs on other low-energy observables like APV, deep inelastic neutrino-nucleon scattering, and collider processes.  We provide a summary of all results and current constraints in Sec.~\ref{sec:results}. Next, we go back to \cevns~and present in Sec.~\ref{sec:future_sens} the expected sensitivities at future upgrades of COHERENT and at the ESS. Finally, we summarize and present our conclusions in Sec.~\ref{sec:concl}.

\section{Leptoquark formalism}
\label{sec:LQ-formalism}

As their name suggests, LQs are particles than can simultaneously couple to a lepton and a quark and, in general, they can be of either scalar or vector nature. These two properties allow us to add different terms to the SM Lagrangian that couple LQs with the SM fields while preserving the gauge structure of the SM. A list of all possible ways by which LQs can give rise to a lepton-quark interaction is long and has been studied, for instance, in Ref.~\cite{Dorsner:2016wpm}. However, since we are interested in signatures at \cevns~experiments, we will mainly focus on those interactions that can connect neutrinos and quarks. Respecting the SM symmetries, these interactions are possible through either a scalar or a vector LQ as long as they do not allow proton decay at tree level. Starting from the completely general LQ list given in~\cite{Dorsner:2016wpm}, we extract all the relevant operators involving neutrinos and we summarize them in Table~\ref{tab:operators}, indicating in each case the LQ quantum numbers under the SM gauge group. For each scenario, we follow the notation of~\cite{Dorsner:2016wpm} and we denote the SM lepton and quark doublets by $L$ and $Q$, respectively. We now briefly discuss the associated Lagrangian for each of the different scenarios. As seen from Table~\ref{tab:operators}, there are four possible scalar LQs giving rise to operators relevant for \cevns~(our main topic of interest): one $SU(2)_L$ singlet, two doublets, and one triplet. For simplicity, we will assume that LQs only interact with first-generation quarks and first- and second-generation leptons:

\begin{table*}[b]
\begin{center}
  \begin{tabular}{| c | c | c || c | c | c |}
  \hline
  \multicolumn{3}{|c ||}{\bf{Scalar}} & \multicolumn{3}{c|}{\bf{Vector}}\\
    \hline
      \hline
 LQ& Operator  & ($SU(3)_c$, $SU(2)_L$, $U(1)_Y$) & LQ & Operator & ($SU(3)_c$, $SU(2)_L$, $U(1)_Y$)  \\ 
   \hline
 $S_1$ & $QLS_1$& $(\boldsymbol{\bar{3},1},1/3)$ & $U_1$&$Q\gamma^\mu U_{1,\mu}L$& $(\boldsymbol{3,1},2/3)$ \\
 $R_2$ &$u_RLR_2$& $(\boldsymbol{3,2},7/6)$ &$V_2$&$d_R\gamma^\mu V_{2,\mu}L$& $(\boldsymbol{\bar{3},2},5/6)$ \\
 $\tilde{R}_2$ &$d_RL\tilde{R}_2$& $(\boldsymbol{3,2},1/6)$ &$\tilde{V}_2$&$u_R\gamma^\mu U_{1,\mu}L$& $(\boldsymbol{\bar{3},2},-1/6)$\\
 $S_3$ &$QLS_3$& $(\boldsymbol{\bar{3},3},1/3)$ &$U_3$&$Q\gamma^\mu \tilde{V}_{1,\mu}L$& $(\boldsymbol{3,3},2/3)$ \\
    \hline
  \end{tabular}
  \caption{Relevant LQ operators connecting neutrinos and quarks. Left/right column shows the possible operators when LQs are scalars/vectors.  The numbers between parenthesis denote the quantum numbers under $SU(3)_c, SU(2)_L$, and $U(1)_Y$, respectively.}
 \label{tab:operators}
\end{center}
\end{table*}

\begin{itemize}
    \item Leptoquark singlet $S_1=(\boldsymbol{\bar{3},1},1/3)$. 
    
Since this is a singlet under $SU(2)_L$, there is only one component under this symmetry, precisely denoted by $S_1$, which carries a charge $\mathcal{Q}=-1/3$. Then, the corresponding Lagrangian that adds to the SM reads 
\begin{eqnarray}
\mathcal{L}\subset \lambda_{ij}\bar{Q}_i^ci\tau_2 L_j S_1 + {\rm h.c.}\quad \to \quad \mathcal{L}\subset (\lambda_{1j}\bar{u}^cP_L\ell_j-\lambda_{1j}\bar{d}^cP_L\nu_j)S_1^{-1/3} + {\rm h.c.} \, ,
\label{eq:LQ}
\end{eqnarray}
where $\lambda_{ij}$ is in general a complex  matrix, $\tau_2$ is the indicated Pauli matrix, and $i,j=1,2,3$ are flavor indices. The right-hand side of the previous equation results from expanding the doublet terms in the Lagrangian. $P_L$ denotes the left-handed chirality operator and we just keep the $\lambda_{1j}$ ($i$ = 1) term to explicitly remark that we are interested in LQs coupling only to first-generation quarks, with $u_1 = u$ and $d_1 = d$. To illustrate how such an interaction in Eq.~\eqref{eq:LQ} can contribute to a neutral-current process as \cevns, we can see how the matrix element depends on the fermionic spinors involving neutrino reactions,
\begin{eqnarray}
    \mathcal{M}^{S_1}_{\mathrm{CE\nu NS}}\sim (\bar{d} P_L\nu_k)(\nu_j P_R d)=\frac{1}{2}(\bar{d}\gamma^\mu P_R d)(\bar{\nu}_j \gamma_\mu P_L\nu_k).
    \label{eq:MS1}
\end{eqnarray}

 Notice that to get the right-hand side of the previous equation we have performed a Fierz transformation (see Appendix~\ref{app:fierz}), giving the desired neutral current shape for the interaction. An important feature arising from Eq.~\eqref{eq:MS1} is that, when interacting with neutrinos, the scalar LQ $S_1$ couples only to down quarks.
    
    \item Leptoquark doublet $R_2=(\boldsymbol{{3},2},7/6)$.

Being a doublet under $SU(2)_L$, in this case there are two LQ components, which we denote $R_2 = (R_{2}^{5/3}, R_{2}^{2/3})^T$, where the superscript indicates the corresponding electric charge $\mathcal{Q}$. From Table~\ref{tab:operators}, we see that $R_2$ couples only to $u$-type quarks.  Then, the relevant Lagrangian for neutrino interactions is given by

\begin{eqnarray}
\mathcal{L}\subset \lambda_{ij}\bar{u}_i R_2^T i\tau_2 L_j + {\rm h.c.} \quad \to \quad \mathcal{L}\subset \lambda_{1j}(\bar{u}P_L\ell_j R_2^{5/3}-\bar{u}P_L\nu_j R_2^{2/3}) + {\rm h.c.} \, ,
\end{eqnarray}

where, again, $\lambda_{ij}$ is a complex matrix, and the right-hand side is obtained by expanding the different doublets. Assuming a degeneracy in the associated masses of the two states, the matrix element involved in neutrino interactions with matter is proportional to
\begin{eqnarray}
    \mathcal{M}^{R_2}_{\mathrm{CE\nu NS}}\sim (\bar{u} P_L\nu_k)(\nu_j P_R u)=\frac{1}{2}(\bar{u}\gamma^\mu P_R u)(\bar{\nu}_j \gamma_\mu P_L\nu_k),
    \label{eq:MR2}
\end{eqnarray}

where we have Fierz-transformed again the operator in order to get the expression on the right side of the equation.

    \item Leptoquark doublet $\tilde{R}_2=(\boldsymbol{{3},2},1/6)$.

As in the previous case, here we have a doublet under $SU(2)_L$, with components denoted by $\tilde{R}_2 = (\tilde{R}_2^{2/3}, \tilde{R}_2^{-1/3})^T$. Then, this LQ can only couple to $d$-type quarks, and the relevant Lagrangian for neutrino interactions reads
\begin{eqnarray}
\mathcal{L}\subset \lambda_{ij}\bar{d}_i \tilde{R}_2^T i\tau_2 L_j + {\rm h.c.} \quad \to \quad \mathcal{L}\subset \lambda_{1j}(\bar{d} P_L\ell_j \tilde{R}_2^{2/3} - \bar{d} P_L\nu_j\tilde{R}_2^{-1/3} )+ {\rm h.c.} \, .
\end{eqnarray}

Again we assume a degeneracy in mass between both states of the multiplet, so the matrix element reads
\begin{eqnarray}
    \mathcal{M}^{\tilde{R}_2}_{\mathrm{CE\nu NS}}\sim (\bar{d} P_L\nu_k)(\nu_j P_R d)=\frac{1}{2}(\bar{d}\gamma^\mu P_R d)(\bar{\nu}_j \gamma_\mu P_L\nu_k) \, .
    \label{eq:MRT2}
\end{eqnarray}

 By comparing with Eq.~\eqref{eq:MS1}, we can see that the relevant matrix element has the same structure as $S_1$, and $\tilde{R}_2$ couples only to down quarks.

    \item Leptoquark triplet $S_3=(\boldsymbol{\bar{3},3},1/3)$

    To finish our scalar LQ list, we now have a triplet under $SU(2)_L$, whose components are denoted as $S_3 = (S_3^{2/3}, S_3^{-1/3}, S_3^{-4/3} )$. Then, being $\tau = (\tau_1, \tau_2, \tau_3) $ the standard Pauli matrices, the associated Lagrangian involving neutrino interactions and respecting the SM symmetries reads
\begin{eqnarray}
\mathcal{L}\subset \lambda_{ij}\bar{Q}^c_i  i\tau_2 (\tau\cdot S_3)^\dagger L_j + {\rm h.c.} \quad \to \quad \mathcal{L}\subset \lambda_{1j}(\sqrt{2} \bar{u}^cP_L\nu_j S_3^{2/3*}-\bar{u}^cP_L\ell_jS_3^{-1/3}) \notag\\
-\lambda_{1j}(\bar{d}^c P_L\nu_j S_3^{-1/3*}+\sqrt{2}\bar{d}^cP_L\ell S_3^{-4/3*})+ {\rm h.c.}\, .
\end{eqnarray}
As in the previous cases we can see that the matrix element reads
\begin{eqnarray}
    \mathcal{M}^{S_3}_{\mathrm{CE\nu NS}}\sim  2(\bar{u} P_L\nu_k)(\nu_j P_R u) +(\bar{d} P_L\nu_k)(\nu_j P_R d)=\notag \\(\bar{u}\gamma^\mu P_R u)(\bar{\nu}_j \gamma_\mu P_L\nu_k)+\frac{1}{2}(\bar{d}\gamma^\mu P_R d)(\bar{\nu}_j \gamma_\mu P_L\nu_k) \, .
    \label{eq:MS3}
\end{eqnarray}

Interestingly, among all cases presented till now, only $S_3$ couples to both up- and down-type quarks. However, the strength of the corresponding coupling is not the same as there is a difference of a factor 1/2 between them. As we will see, this will result in an enhancement of the \cevns~cross section associated with the contribution of this type of LQ.

\end{itemize}

Notice that in Table~\ref{tab:operators} we have also listed operators involving vector-type LQs. In principle, these operators can also be studied, giving their corresponding contribution to the \cevns~cross section. However, if we perform a Fierz transformation the contribution turns out to have the same shape as the scalar cases listed above. To illustrate this we can take, for instance, the operator associated to the LQ denoted as $U_1$. Then, if we apply a Fierz transformation to the matrix element of neutrino interactions we have
\begin{eqnarray}
    \mathcal{M}^{U_1}_{\mathrm{CE\nu NS}}\sim  (\bar{u} \gamma_\mu \nu_k)(\nu_j \gamma^\mu  u) =(\bar{u}\gamma^\mu  u)(\bar{\nu}_j \gamma_\mu \nu_k) ,
    \label{eq:MU1}
\end{eqnarray}
that is the same Lorentz structure as $R_2$ with a factor 1/2 of difference. Given this similarity among Lorentz structures, in the following we will focus only on scalar LQs.\\

In the rest of the paper we will focus on scalar LQ interactions between first-generation quarks and first- and second-generation leptons. Hence the matrix structure of the parameters $\lambda_{ij}$ appearing in the Lagrangians of the different LQs considered here can be written as,
\begin{eqnarray}
    \lambda_{ij}=
    \begin{pmatrix}
        g & g & 0\\
        0 & 0 & 0\\
        0 & 0 & 0
    \end{pmatrix},
\end{eqnarray}
where $g$ is the strength of the interaction and we assume it to be the same for both lepton flavors. During the rest of this manuscript, we will consider this flavor structure and we will assume same strength of the coupling for the different LQ models under study.

\section{Current \cevns~data: COHERENT}
\label{sec:current_constr}
In this section, we investigate the potential of \cevns~to study LQs and we obtain constraints from COHERENT data in the parameter space of LQ masses and couplings. Within the SM, the \cevns~differential cross section, in terms of the nuclear recoil energy $E_\mathrm{nr}$, is given by~\cite{Freedman:1973yd}
\begin{equation}
\label{eq:xsec_CEvNS_SM}
\frac{d\sigma_{\nu_\ell \mathcal{N}}}{dE_\mathrm{nr}}\Big|_\mathrm{CE\nu NS} =\frac{G_F^2 m_N}{\pi}F_W(\qtransfer^2)\left({Q_W^\mathrm{SM}}\right)^2\left(1-\frac{m_N E_\mathrm{nr}}{2E_\nu^2}\right) \, ,
\end{equation}
where $G_F$ is the Fermi constant, $E_\nu$ denotes the incoming neutrino energy, $\ell$ indicates the neutrino flavor, and $m_N$ refers to the nuclear mass. Notice that at tree level, the \cevns~cross section is flavor independent, with small radiative corrections that are not relevant for present experimental sensitivities \cite{Tomalak:2020zfh}.
The SM weak charge $Q_W^\text{SM}$ is defined as
\begin{equation}
\label{eq:CEvNS_SM_Qw}
    Q_W^\text{SM} = g_V^p Z + g_V^n N \, ,
\end{equation}
where $g_V^{p,n}$ are the proton and neutron couplings, $g_V^p = 1/2 (1 - 4 \sin^2 \theta_W)$ and $ g_V^n = -1/2$.
The weak charge is the term which eventually encodes the typical $N^2$ dependence of the \cevns~cross section, and which gives rise to the relevant enhancement with respect to other neutrino processes. Notice that the proton contribution carries the dependence on the weak mixing angle, being this contribution subdominant due to an accidental cancellation generated by the SM value of the weak mixing angle at low energy.~\footnote{We take $\sin^2 \theta_W=0.23857(5)$~\cite{ParticleDataGroup:2020ssz}, from the RGE extrapolation in the minimal subtraction $ (\overline{\text{MS}})$ renormalization scheme.}

Nuclear-physics effects are encoded in the nuclear form factor $F_W(\qtransfer^2)$ appearing in Eq.~\eqref{eq:xsec_CEvNS_SM}. We adopt the Klein-Nystrand parametrization, which reads
\begin{equation}\label{eq:form-factor-kn}
    F_W(\qtransfer^2) = 3\, \dfrac{J_1(\qtransfer R_A)}{\qtransfer R_A (1 + a_k^2 \qtransfer^2)} \, ,
\end{equation}
where $J_1$ is the spherical Bessel function of order one, $\qtransfer \approx \sqrt{2 m_\mathrm{N} E_\mathrm{nr}}$ stands for the three-momentum transfer, $R_A = 1.23 \,A^{1/3}~\mathrm{fm}$ is the nuclear radius and $a_k = 0.7~\mathrm{fm}$ is the Yukawa potential range.\\

The weak charge term may be modified in the presence of new physics. In the specific case of the LQ scenarios of interest in this paper, we will compute the \cevns~events based on Eq.~\eqref{eq:xsec_CEvNS_SM} by changing

\begin{equation}
(Q_W^\text{SM})^2 \to (Q_{i}^\text{LQ})^2 = \left ( Q_W^\text{SM} + Q_{ii,\text{LQ}} \right )^2 + \sum_{i \neq j} Q^2_{ij, \text{LQ}}  \, ,
\end{equation}
where the first and second indexes in $Q_{ij,\textrm{LQ}}$ denote quark and lepton family, respectively, and $\text{LQ} = S_1, R_2, \tilde{R}_2, S_3$ stands for the LQ type. For simplicity, we assume that LQs couple with the same strength to electrons and muons, and to $u$ and $d$ quarks, with vanishing coupling to $\tau$ neutrinos and to the second- and third-generation quark families. Then, we denote $g^2 \equiv \lambda_{1i}\lambda_{1j}~  (i, j = e, \mu )$ and for the different models studied in Sec.~\ref{sec:LQ-formalism} we have
\begin{eqnarray}
    Q_{ij,S_1}=\frac{g^2}{4\sqrt{2}G_F}\frac{ZF_Z(q^2) + 2NF_N(q^2)}{q^2+m^2_{S_1}},
     \label{eq:s1:cross} \\
    Q_{ij,R_2}=\frac{g^2}{4\sqrt{2}G_F}\frac{2ZF_Z(q^2) + NF_N(q^2)}{q^2+m^2_{R_2}},
    \label{eq:r2:cross}
    \\
    Q_{ij,\tilde{R}_2}=\frac{g^2}{4\sqrt{2}G_F}\frac{ZF_Z(q^2) + 2NF_N(q^2)}{q^2+m^2_{\tilde{R}_2}},
    \label{eq:r2tilde:cross}\\
    Q_{ij,S_3}=\frac{g^2}{4\sqrt{2}G_F}\frac{5ZF_Z(q^2) + 4NF_N(q^2)}{q^2+m^2_{S_3}},
    \label{eq:s3:cross}
\end{eqnarray}
being $m_{\text{LQ}}$ the corresponding LQ mass. As it is clear from Eqs.~\eqref{eq:s1:cross} and~\eqref{eq:r2tilde:cross}, the impact of $S_1$ and $\tilde{R}_2$ on \cevns~is expected to be exactly the same. For the last case ($S_3$), we have assumed that the two LQs arising from the $SU(2)_L$ triplet have the same mass. In that case, one of the states couples to the $u$-type quarks, while the other one couples to the down quarks. However, given the parametrization used for the Lagrangian in Eq.~\eqref{eq:MS3}, they do not couple with the same strength.\\

\noindent
\begin{table}[t]
\centering
\begin{tabular}{l|c|c|c|c} 
\toprule
\textbf{Detector} & \textbf{Mass}  (kg) & \textbf{Baseline}  (m)& \textbf{Threshold} (keV$_\mathrm{nr}$) & $\mathbf{N_\mathrm{POT}}$\\ 
\midrule
COH-CsI & 14.6 & 19.3 & 4.2 &$1.38 \times 10^{23}$\\
COH-LAr & 24 & 27.5 & 20 & $1.38 \times 10^{23}$\\
\midrule
COH-CsI-700 & 700 & 19.3 & 1.4 &$5.18 \times 10^{23}$\\
COH-LAr-750 & 750 & 29 & 20 &$5.18 \times 10^{23}$ \\
\midrule
ESS-Si                & 1                  & 20                   & 0.16 & $2.8\times10^{23}$ \\
ESS-Xe                & 20                 & 20                    & 0.9  &    $2.8\times10^{23}$ \\
\bottomrule
\end{tabular}
\caption{Details of the \cevns~experiments considered in this paper.}
\label{tab:CEvNSexps}
\end{table} 

To set constraints on the LQ scenarios using \cevns~data, we rely on the most recent measurements of the COHERENT experiment, which were performed by using CsI~\cite{COHERENT:2021xmm} and LAr~\cite{COHERENT:2020ybo} detectors, whose specifications are summarized in the first two lines of Table~\ref{tab:CEvNSexps}. We perform a thorough analysis by including both energy and timing information together with all relevant systematic effects, for each detector, following Ref.~\cite{DeRomeri:2022twg}.\footnote{In this work, we do not include neutrino-electron (ES) scattering events given that in the SM the ES cross section is subdominant with respect to \cevns~and no new contributions appear due to the LQ exchange in the interaction.}  
The neutrino flux at COHERENT comes in three components, from $\pi$-DARs produced at the Spallation Neutron Source (SNS):
\begin{equation}
\begin{aligned} 
\frac{\d N_{\nu_\mu}}{\d E_\nu}(E_\nu) & = \eta \, \delta\left(E_\nu-\frac{m_{\pi}^{2}-m_{\mu}^{2}}{2 m_{\pi}}\right) \, , \\ 
\frac{\d N_{\bar{\nu}_\mu}}{\d E_\nu}(E_\nu) & = \eta \frac{64 E^{2}_\nu}{m_{\mu}^{3}}\left(\frac{3}{4}-\frac{E_\nu}{m_{\mu}}\right) \, ,\\ 
\frac{\d N_{\nu_e}}{\d E_\nu}(E_\nu) & = \eta \frac{192 E^{2}_\nu}{m_{\mu}^{3}}\left(\frac{1}{2}-\frac{E_\nu}{m_{\mu}}\right)  \, ,
\end{aligned}
\label{labor-nu}
\end{equation}
where $m_{\mu}, m_{\pi}$ denote the muon and pion masses, while $\eta = r N_{\mathrm{POT}}/4 \pi L^2$ is a normalization factor which depends on the number of neutrinos per flavor ($r$) produced for each proton on target (POT). We assume $r=0.0848\, (0.009)$ and $N_{\mathrm{POT}}=3.198\, (1.38) \times 10^{23}$ for the CsI (LAr) detector. Notice that the three different neutrino flux components come with different timing, the $\nu_\mu$ being prompt and the other two components delayed.

Next, we proceed to evaluate the expected number of events. We assume a detector mass $m_{\rm det} = 14.6 ~(24)$ kg located at a distance $L=19.3 ~(27.5)$~m from the SNS source, for the CsI (LAr) detector. The expected number of events, on a nuclear target $\mathcal{N}$, per neutrino flavor,  $\nu_\ell$, and in each nuclear recoil energy bin $i$ can be written as \cite{DeRomeri:2022twg,Candela:2023rvt}
\begin{align}
\label{eq:Nevents_CEvNS}
N_{i, \nu_\ell}(\mathcal{N})
= \nonumber
& \, N_\mathrm{target}
\int_{E_{\mathrm{nr}}^{i}}^{E_{\mathrm{nr}}^{i+1}}
\hspace{-0.3cm}
\d E_{\mathrm{nr}}\,
\epsilon_E(E_{\mathrm{nr}})
\int_{E^{\prime\mathrm{min}}_{\mathrm{nr}}}^{E^{\prime\text{max}}_{\text{nr}}}
\hspace{-0.3cm}
\d E'_{\text{nr}}
\,
\mathcal{R}(E_{\text{nr}},E'_{\text{nr}})  \\
& \times \int_{E_\nu^{\text{min}}(E'_{\text{nr}})}^{E_\nu^{\text{max}}}
\hspace{-0.3cm}
\d E_\nu \,
\frac{\d N_{\nu_\ell}}{\d E_\nu}(E_\nu)
\frac{\d \sigma_{\nu_\ell \mathcal{N}}}{\d E'_\mathrm{nr}}\Big|_\mathrm{CE\nu NS}(E_\nu, E'_{\mathrm{nr}})
,
\end{align}
where  $N_\mathrm{target} = N_{\mathrm{A}} m_{\mathrm{det}} / M_{\mathrm{\mathrm{target}}}$ is the number of target atoms in the detector, with $M_{\mathrm{\mathrm{target}}}$ the molar mass of the detector material, and  $N_{\mathrm{A}}$ the Avogadro's constant. 
Kinematically, the integration limits in Eq.~\eqref{eq:Nevents_CEvNS} are found to be $E_{\nu}^{\textrm{min}}(E_{\textrm{nr}}^{'}) = \sqrt{m_NE_{\textrm{nr}}/2}$ and $E_{\textrm{nr}}^{'\textrm{max}} = 2(E_{\nu}^{\textrm{max}})^2/m_N$, with $E_\nu^{\textrm{max}}$ the maximum incoming neutrino energy, which for SNS neutrinos is $\approx 52.8$ MeV.
Finally, the energy resolution function $\mathcal{R}(E_{\text{nr}},E'_{\text{nr}})$ appearing in Eq.~\eqref{eq:Nevents_CEvNS} associates the true nuclear recoil energy ($E'_{\text{nr}}$) with the reconstructed one ($E_{\text{nr}}$) and the $\epsilon_E(E_{\mathrm{nr}})$ is the energy-dependent detector efficiency. We refer the reader to Refs.~\cite{COHERENT:2021xmm,DeRomeri:2022twg,Candela:2023rvt} for more details.

In order to take into account the neutrino-flux timing information in our analysis, we distribute the predicted $N_{i, \nu_\ell}^{\mathrm{CE\nu NS}}(\mathcal{N})$ in each time bin $j$. At this scope, we rely on the time distributions $\mathcal{P}^{\nu_\ell}_T(t_{\mathrm{rec}})$ provided in~\cite{Picciau:2022xzi,COHERENT:2021xmm}, and we normalize them to 6 $\mathrm{\mu s}$~\cite{DeRomeri:2022twg,Candela:2023rvt}. The predicted event number, per observed nuclear recoil energy and time bins $i, j$ is finally obtained as
\begin{equation}
\label{eq:Nevents_DF_ij_CEvNS}
 N_{ij}(\mathcal{N}) = \sum_{\nu_\ell =\nu_{e}, \nu_{\mu}, \bar{\nu}_{\mu}}\int_{t_{\mathrm{rec}}^{j}}^{t_{\mathrm{rec}}^{j+1}} \d t_{\mathrm{rec}} \, \mathcal{P}^{\nu_\ell}_T(t_{\mathrm{rec}}, \alpha_6)\epsilon_T(t_{\mathrm{rec}}) N_{i, \nu_{\ell}}(\mathcal{N}),
\end{equation}
where $\epsilon_T (t_{\mathrm{rec}})$ is the time-dependent efficiency~\cite{COHERENT:2021xmm,DeRomeri:2022twg,Candela:2023rvt}. (We include an additional nuisance parameter on the beam timing, $\alpha_6$, see \cite{DeRomeri:2022twg,Candela:2023rvt}.)

To proceed with the statistical analysis of the COHERENT CsI data set we consider a Poissonian least-squares function~\cite{DeRomeri:2022twg, Candela:2023rvt}, expressed as 
\begin{equation}
\label{eq:chi2CsI}
	\chi^2_{\mathrm{CsI}}
	 =
	2
	\sum_{i=1}^{9}
	\sum_{j=1}^{11}
	\left[ N^\mathrm{th}_{ij}  -  N_{ij}^{\text{exp}} 
	 +  N_{ij}^{\text{exp}} \ln\left(\frac{N_{ij}^{\text{exp}}}{ N^\mathrm{th}_{ij}} \right)\right]\\
	+ \sum_{k=0}^{5}
	\left(
	\dfrac{ \alpha_{k} }{ \sigma_{k} }
	\right)^2  
	.
\end{equation}

The predicted number of events, which includes both SM and LQ \cevns~events, as well as backgrounds, depends on several nuisance parameters ($\alpha_i$) and reads 
\begin{align}\nonumber
N_{ij}^\mathrm{th} &= (1 + \alpha_{0} ) N_{ij} (\alpha_{4}, \alpha_{6}, \alpha_{7})
 + (1 + \alpha_{1}) N_{ij}^\mathrm{BRN}(\alpha_{6}) + (1 + \alpha_{2}) N_{ij}^\mathrm{NIN}(\alpha_{6}) 
  + (1 + \alpha_{3}) N_{ij}^\mathrm{SSB} \, .
	\label{eq:Nth_CsI_chi2}
\end{align}

The nuisances come together with their associated uncertainties $\sigma_i$~\cite{DeRomeri:2022twg,Candela:2023rvt}: $\sigma_{0} = 11\%$  (efficiency and flux uncertainties), $\sigma_{1} = 25\%$ (Beam Related Neutrons (BRN)), $\sigma_{2} = 35\%$ (Neutrino Induced Neutrons (NIN)) and $\sigma_{3} = 2.1\%$ (Steady State Background (SSB)),
$\sigma_{5} = 3.8\%$  (QF). The predicted number of events $N_{ij}^{\textrm{th}}$ also depends on three nuisance parameters: $\alpha_{4}$, which enters the nuclear form factor through the nuclear radius in Eq.~\eqref{eq:form-factor-kn}, via $R_A = 1.23 \, A^{1/3} (1+\alpha_4)$, with  $\sigma_{4} = 5\%$; $\alpha_{6}$ which accounts for the uncertainty in beam timing with no prior assigned; and $\alpha_{7}$ which allows for deviations of the uncertainty in the \cevns ~efficiency.\\

For the statistical analysis of the COHERENT-LAr data set we instead adopt the following Gaussian least-squares  approach based on~\cite{AtzoriCorona:2022moj,DeRomeri:2022twg,Candela:2023rvt} 
\begin{equation}
	\chi^2_{\mathrm{LAr}} = 
	\sum_{i=1}^{12}
	\sum_{j=1}^{10}
	\left( \frac{N^\mathrm{th}_{ij}  -  N_{ij}^{\text{exp} } }{\sigma_{ij}}  \right)^2  + \sum_{k=0,3,4,8}
	\left( 	\dfrac{ \beta_{k} }{ \sigma_{k} } 	\right)^2  +  \sum_{k=1,2,5,6,7} \left(\beta_{k}  	\right)^2 \,  .
\label{eq:chi2LAr}
\end{equation}

Here, the theoretical number of events is defined as
\begin{equation}
    \begin{aligned}
     N^\mathrm{th}_{ij} & =  (1 + \beta_0 +  \beta_1 \Delta_{\mathrm{CE\nu NS}}^{F_{90+}} + \beta_1 \Delta_{\mathrm{CE\nu NS}}^{F_{90-}} + \beta_2 \Delta_{\mathrm{CE\nu NS}}^\mathrm{t_{trig}}) N_{ij}\\   & 
     + (1 + \beta_4 + \beta_5 \Delta_\mathrm{pBRN}^{E_+} + \beta_5 \Delta_\mathrm{pBRN}^{E_-}
    + \beta_6 \Delta_\mathrm{pBRN}^{t_\text{trig}^+} + \beta_6 \Delta_\mathrm{pBRN}^{t_\text{trig}^-} + \beta_7 \Delta_\mathrm{pBRN}^{t_\text{trig}^\text{w}}) N_{ij}^\mathrm{pBRN}\\
    & + (1 + \beta_8) N_{ij}^\mathrm{dBRN} + (1 + \beta_3) N_{ij}^\mathrm{SSB}  \, .
    \end{aligned}
\end{equation}
and the experimental uncertainty is $\sigma_{ij}^2 = N_{ij}^\text{exp} +  N_{ij}^\mathrm{SSB}/5$.

The expected number of events depend on several nuisance parameters, dubbed $\beta_0,~\beta_3,~\beta_4$ and $\beta_8$, which account for the normalization uncertainties of \cevns, SS, prompt BRN (pBRN) and delayed BRN (dBRN) background rates respectively, with uncertainties $\{\sigma_{0},~\sigma_{3},~\sigma_{4},~\sigma_{8}\}$ = $\{0.13,~0.0079,~0.32,~1.0\}$~\cite{COHERENT:2020iec}. Let us notice that $\beta_0$ encodes multiple uncertainties, namely the flux (10\%), efficiency (3.6\%), energy calibration (0.8\%), the calibration of the pulse-shape discrimination parameter $F_{90}$  (7.8\%), QF (1\%), and nuclear form factor (2\%)~\cite{COHERENT:2020iec}.
The additional nuisance parameters $\beta_1, \beta_2, \beta_5, \beta_6$ and $\beta_7$ account for systematic effects affecting the shape uncertainties of the \cevns ~ and pBRN rates, namely the uncertainty on the \cevns ~shape due to existing systematic uncertainties on the $\pm 1\sigma$ energy distributions of the $F_{90}$ parameter ($\Delta_{\mathrm{CE\nu NS}}^{F_{90\pm}}$), due to the mean time to trigger distribution ($\Delta_{\mathrm{CE\nu NS}}^\mathrm{t_{trig}}$) or the pBRN shape uncertainty due to the corresponding uncertainty on the $\pm 1\sigma$ energy, time and trigger width distributions ($\Delta_\mathrm{pBRN}^{E_\pm}$, $\Delta_\mathrm{pBRN}^{t_\text{trig}^\pm}$ and $\Delta_\mathrm{pBRN}^{t_\text{trig}^\text{w}}$). These distributions are defined as departures from the central value (CV) ones~\cite{COHERENT:2020ybo} according to $
    \Delta_{\lambda}^{\xi_\lambda} = \frac{N_{ij}^{\lambda,\xi_\lambda} - N_{ij}^{\lambda,\mathrm{CV}}}{N_{ij}^{\lambda,\mathrm{CV}}}  ,
$
with $\lambda=$ \{CE$\nu$NS, pBRN\} and $\xi_\lambda$ referring to the different source uncertainties affecting the \cevns ~or  pBRN shapes.\\

\section{Other constraints}
\label{sec:other_constr}
In this section we proceed to discuss further current constraints on the LQ scenarios presented in Sec.~\ref{sec:LQ-formalism}. Following an increasing energy scale, we will start with APV, then proceed with deep inelastic neutrino-nucleon scattering, and finally move to collider searches.

\subsection{Atomic Parity Violation}

One very accurate determination of the weak mixing angle currently available in the low-energy regime comes from APV --- or parity nonconservation --- experiments on cesium atoms~\cite{Wood:1997zq,Guena:2004sq,Dzuba:2012kx}. 
It has been shown~\cite{Cadeddu:2018izq,Cadeddu:2021ijh,AtzoriCorona:2023ktl} that such a measurement can provide complementary information to \cevns,  also regarding nuclear physics parameters besides the weak mixing angle. 
Moreover, stringent APV bounds on LQ coupling to first-generation fermions have been obtained in the literature under the assumption of effective
four-fermion interactions and that only one contribution (from $u$ or $d$ quarks) is present at a given time~\cite{Barger:2000gv,Gresham:2012wc,Dorsner:2014axa,Dorsner:2016wpm,Gresham:2012wc}.
Here we want to exploit the low-energy measurement of the weak charge $Q_W$ of $^{133}$Cs from APV experiments to constrain the LQ scenarios proposed in Sec.~\ref{sec:LQ-formalism}, including model $S_3$ which simultaneously encodes couplings to both $u$ and $d$ quarks, and taking into account the explicit dependence on the LQ mass. In this subsection, we hence derive APV constraints on LQ through their effect on the weak charge. 
Including radiative corrections in the $\mathrm{\overline{MS}}$ scheme, the APV weak charge in the SM reads~\cite{Erler:2013xha,Cadeddu:2018izq,Cadeddu:2021ijh,Workman:2022ynf}

\begin{equation}
\label{eq:APV_SM_Qw_rad}
Q_W^{\mathrm{APV}}
(^{133}_{78}\mathrm{Cs}) \big |_\mathrm{SM}
=
- 2 [ Z (g_{A V}^{e p} + 0.00005) + N (g_{A V}^{e n} + 0.00006) ] 
\left( 1 - \dfrac{\alpha}{2 \pi} \right)
,
\end{equation}

where the couplings of electrons to 
nucleons are $g_{A V}^{e p} = 2 g_{A V}^{e u} + g_{A V}^{e d} = - 0.0357$ and $g_{A V}^{e n} = g_{A V}^{e u} + 2 g_{A V}^{e d} = - 0.495$, $\alpha$ is the fine-structure constant and $Z=55$, $N=78$ for cesium. The theoretical expected value is therefore $Q_W^{\mathrm{APV}}
\big |_\mathrm{SM}
= - 73.3 \pm 0.01$~\cite{Cadeddu:2018izq,Cadeddu:2021ijh}.
As anticipated, the presence of LQs will affect the value of the weak charge as following
\begin{equation}
Q_W^{\mathrm{APV}}
(^{133}_{78}\mathrm{Cs}) \big |_\mathrm{LQ} = \left ( Q_W^{\mathrm{APV}} \big |_\mathrm{SM} + Q_{e e,\mathrm{LQ}} (g, m_\mathrm{LQ}) \right )\, .
\end{equation}

The LQ charge $Q_{e e,\mathrm{LQ}}$ encodes the dependence on the free parameters $g$ and $m_\mathrm{LQ}$, and on the momentum transfer, which for APV is $\qtransfer^2 \simeq (2.4~\mathrm{MeV})^2$. 
The experimental value of $Q_W$
for cesium is extracted by measuring the ratio of the parity violating
amplitude $E_\mathrm{PNC}$ to the Stark vector transition polarizability, and by calculating theoretically $E_\mathrm{PNC}$ as a function of $Q_W$~\cite{Workman:2022ynf}. Taking into account small uncertainties associated with the atomic wave function calculations, most recent computations of the parity non-
conserving amplitude combined with the measurements~\cite{Guena:2004sq,Wood:1997zq} lead to~\cite{Workman:2022ynf}

\begin{equation}
Q_W^{\mathrm{APV}}(^{133}_{78}\mathrm{Cs})
\big |_\mathrm{exp} = -72.82 \pm (0.26)_\mathrm{exp} \pm (0.33)_\mathrm{th} .
\end{equation}

We evaluate the APV bound on the LQ scenarios by minimizing the following least-square function:

\begin{align}
	\chi^2_{\mathrm{APV}} 	=
    \left(
    \dfrac{
   Q_W^{\mathrm{APV}}(^{133}_{78}\mathrm{Cs})
\big |_\mathrm{LQ}
    -
   Q_W^{\mathrm{APV}}(^{133}_{78}\mathrm{Cs})
\big |_\mathrm{exp} }{\sigma_\mathrm{APV}}
    \right)^2
    \,
    ,
    \label{eq:chi2APV}
\end{align}

where $\sigma_{\rm APV} = 0.42$ is the total (experimental + theoretical) uncertainty.

\subsection{Deep inelastic neutrino-nucleon scattering (NuTeV)}
\label{sec:nutev}

The neutrino scattering off nuclei is a very accurate process by which the nature of the weak currents can be tested. Usually, neutrino scattering experiments make use of neutrino beams originating directly from colliders. This is the case for the NuTeV experiment that benefits from the Sign Selected Quadrupole Train (SSQT) beamline at the Fermilab Tevatron collider to obtain well-controlled muon neutrino beams and test the neutrino-nucleon cross section with iron targets. NuTeV measured the deep inelastic neutrino-nucleon  scattering cross section with high accuracy~\cite{NuTeV:2001whx}, improving upon its predecessors CDHS~\cite{Blondel:1989ev} and CHARM~\cite{CHARM:1987pwr}. Provided the target of neutrino experiments is isoscalar, contributions to the cross section from neutral and charged currents can be written as~\cite{Escrihuela:2011cf}
\begin{eqnarray}
    R^\nu =\frac{\sigma(\nu_\mu N\to \nu_\mu X)}{\sigma(\nu_\mu N\to \mu^- X)}= (g_\mu^L)^2 + r(g_\mu^R)^2, \quad R^{\bar{\nu}} =\frac{\sigma(\bar{\nu}_\mu N\to \bar{\nu}_\mu X)}{\sigma(\bar{\nu}_\mu N\to \mu^+ X)}= (g_\mu^L)^2 + \frac{1}{r}(g_\mu^R)^2,
\end{eqnarray}
where the coupling constants are $(g_\mu^{L,R})^2 = (g_\mu^{u(L,R)})^2+(g_\mu^{d(L,R)})^2$ and $r$ is defined as~\cite{Escrihuela:2011cf} 
\begin{eqnarray}
    r=\frac{\sigma(\bar{\nu}_\mu\to N)}{\sigma(\nu_\mu N\to \mu^-X)}.
\end{eqnarray}
By measuring these ratios one can measure the coupling of neutrinos to quarks. In the presence of new physics, these ratios will show a deviation from the SM predictions. One relevant example is the case of neutrino NSI.
In order to parameterize the presence of new physics giving rise to NSI, one can define the low-energy effective Lagrangian as~\cite{Farzan:2017xzy}
\begin{eqnarray}
    -\mathcal{L}_\text{NSI}^{\rm eff}=\sum_{\alpha\beta}\varepsilon_{\alpha\beta}^{fP}2\sqrt{2}G_F(\bar{\nu}_\alpha \gamma_\mu L \nu_{\beta})(\bar{f}\gamma^{\mu}P f),
    \label{eq:LNSI}
\end{eqnarray}
where $P=L,R$ are the chiral projectors and $\varepsilon_{\alpha\beta}^{fP}$ are the coefficients that parameterize the NSI, where $\alpha,\beta$ run over the different neutrino flavors, and $f$ are the charged fermions of the SM. Still focusing on NuTeV, we can relate the NSI coefficients to the couplings $(g_\mu^{L,R})$ in the following way~\cite{Escrihuela:2011cf}
\begin{eqnarray}
    (\tilde{g}^L_\mu)^2=(g_\mu^{uL} + \varepsilon_{\mu\mu}^{uL})^2 + (g_\mu^{dL} + \varepsilon_{\mu\mu}^{dL})^2,\quad (\tilde{g}^R_\mu)^2=(g_\mu^{uR} + \varepsilon_{\mu\mu}^{uR})^2 + (g_\mu^{dR} + \varepsilon_{\mu\mu}^{dR})^2,
\end{eqnarray}
where we set $\alpha=\beta=\mu$.

The introduction of LQs may alter the interaction between neutrinos and quarks, since they couple directly to them, for this reason the presence of LQs may induce non-zero NSI coefficients.
For instance, in the case of $S_1$, for large masses we can integrate out the LQ degrees of freedom, thus obtaining
\begin{eqnarray}
    \mathcal{L}_{\rm eff}^{S_1}\subset \frac{\lambda_{1j}\lambda_{1k}}{m_{S_1}^2}
    (\bar{d} P_L\nu_k)(\nu_j P_R d)=\frac{g^2}{2m_{S_1}^2}(\bar{d}\gamma^\mu P_R d)(\bar{\nu}_j \gamma_\mu P_L\nu_k),
    \label{eq:effLS1}
\end{eqnarray}
where we have performed a Fierz transformation to get the right-hand side of the equation. Analogously, for the other LQ scenarios we have
\begin{eqnarray}
    \label{eq:effLR2}
    \mathcal{L}_{\rm eff}^{R_2}\subset \frac{\lambda_{1j}\lambda_{1k}}{m_{R_2}^2}
    (\bar{u} P_L\nu_k)(\nu_j P_R u)=\frac{g^2}{2m_{R_2}^2}(\bar{u}\gamma^\mu P_R u)(\bar{\nu}_j\gamma_\mu P_L\nu_k)\, ,
\end{eqnarray}
\begin{eqnarray}
\label{eq:effLRT2}
    \mathcal{L}_{\rm eff}^{\tilde{R}_2}\subset \frac{\lambda_{1j}\lambda_{1k}}{m_{\tilde{R}_2}^2}
    (\bar{d} P_L\nu_k)(\nu_j P_R d)=\frac{g^2}{2m_{\tilde{R}_2}^2}(\bar{d}\gamma^\mu P_R d)(\bar{\nu}_j \gamma_\mu P_L\nu_k) \, ,
\end{eqnarray}
\begin{eqnarray}
    \mathcal{L}_{\rm eff}^{S_3}\subset \frac{\lambda_{1j}\lambda_{1k}}{m_{S_3}^2}[2(\bar{u}P_L\nu_k)(\bar{\nu}_jP_R u) + (\bar{d}P_L\nu_k)(\bar{\nu}_j P_R d)] =\notag \\
    \frac{g^2}{2m_{S_3}^2}[2(\bar{u}\gamma^\mu P_R u)(\bar{\nu}_j\gamma_\mu P_L \nu_k) + (\bar{d}\gamma^\mu P_R d)(\bar{\nu}_j \gamma_\mu P_L\nu_k)].
    \label{eq:effLS3}
\end{eqnarray}
Then, we can compare Eq.~\eqref{eq:LNSI} with Eqs.~\eqref{eq:effLS1}, \eqref{eq:effLR2}, \eqref{eq:effLRT2}, and \eqref{eq:effLS3} to obtain the relations between NSI coefficients and LQ parameters~\cite{Davidson:1993qk,Barranco:2007tz,Billard:2018jnl}
\begin{eqnarray}
    S_1: \varepsilon_{\mu\mu}^{dV}=\frac{\sqrt{2}g^2}{4 G_F m_{S_1}^2}\quad R_2: \varepsilon_{\mu\mu}^{uV}=\frac{\sqrt{2}g^2}{4G_F m_{R_2}^2}\\
    S_3: \varepsilon_{\mu\mu}^{dV}=\frac{\sqrt{2}g^2}{2G_F m_{S_3}^2},\quad \varepsilon_{\mu\mu}^{uV}=\frac{\sqrt{2}g^2}{4G_F m_{S_3}^2}.
\end{eqnarray}
It is important to note that given the Lagrangian of $\tilde{R}_2$, the coefficient for this state is the same as the one for $S_1$.

NuTeV data have been studied in Ref.~\cite{Escrihuela:2011cf} where a detailed analysis was done to translate the experimental measurements into the NSI parameter space. We have recast their results to obtain limits on the LQ mass-coupling plane through the NSI coefficients. It is important to note that due to the fact that the NSI parametrization is given in terms of an effective interaction, NuTeV bounds apply for LQ masses larger than 10 GeV, where the effective theory holds.

\subsection{Collider data}
\label{sec:colliders}

As LQs couple to both quarks and leptons, they are very likely to be produced in lepton or hadron colliders. Furthermore, given the nature of their couplings, LQs may give rise to a very interesting set of signatures  \cite{Dorsner:2016wpm,Schmaltz:2018nls}. Depending on the collider nature, the production of LQs may differ and also the signatures obtained at the detectors. Since we focus on LQ interactions with neutrinos, it is reasonable to expect that they will also interact with charged leptons. This interaction could be mediated by the same LQ state, as it happens for $S_1$, by the other charged state from the LQ multiplet, as for $R_2$ and $\tilde{R}_2$, or eventually by both at the same time, as it occurs in the $S_3$ scenario. 
Given that we are considering the different states from the multiplets to be mass-degenerate, we will translate the bounds on the masses to the state that couples to neutrinos. Moreover, since the couplings are the same before decomposing the multiplets, limits on the LQ couplings apply to all states. In this subsection we will focus on different kind of colliders regarding the nature of their collisions. 
First, we will recast data from HERA (an $e^-p$ collider), LEP ($e^+e^-$), and both SPS and Tevatron ($p\bar{p}$). Then, we will compute bounds from the LHC proton-proton collider.

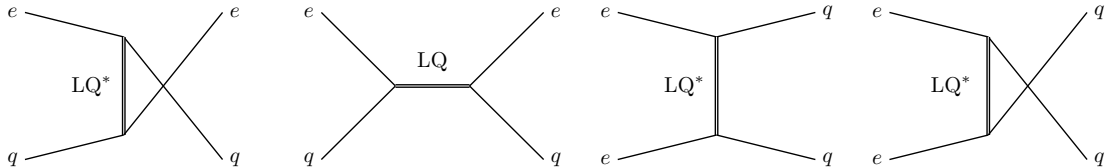
\begin{figure}
  \begin{center}
    \scalebox{0.65}{
      \begin{tikzpicture}
	\begin{scope}[thick] 
    \draw[thick, -] (0, 8.5)--(2, 8);
    \draw[thick, -] (2, 8)--(4, 5.5);
    \draw[thick, double] (2, 8)--(2, 6);
    \draw[thick, -] (2, 6)--(4, 8.5);
    \draw[thick, -] (0, 5.5)--(2, 6);
    \node[black,scale=1.25] at (1.35,7) {{$\rm{LQ}^*$}};
    \node[black,scale=1.25] at (4.25,8.5) {{$e$}};
    \node[black,scale=1.25] at (-0.25,8.5) {{$e$}};
    \node[black,scale=1.25] at (4.25,5.5) {{$q$}};
    \node[black,scale=1.25] at (-0.25,5.5) {{$q$}};

    \draw[thick, -] (6, 8.5)--(7.5, 7);
    \draw[thick, -] (6, 5.5)--(7.5, 7);
    \draw[thick, double] (7.5, 7)--(9, 7);
    \draw[thick, -] (9, 7)--(10.5, 8.5);
    \draw[thick, -] (9, 7)--(10.5, 5.5);
    \node[black,scale=1.25] at (8.25,7.5) {{$\rm{LQ}$}};
    \node[black,scale=1.25] at (5.75,8.5) {{$e$}};
    \node[black,scale=1.25] at (5.75,5.5) {{$q$}};
    \node[black,scale=1.25] at (10.75,8.5) {{$e$}};
    \node[black,scale=1.25] at (10.75,5.5) {{$q$}};

    \draw[thick, -] (12, 8.5)--(14, 8);
    \draw[thick, -] (14, 8)--(16, 8.5);
    \draw[thick, double] (14, 8)--(14, 6);
    \draw[thick, -] (14, 6)--(16, 5.5);
    \draw[thick, -] (12, 5.5)--(14, 6);
    \node[black,scale=1.25] at (13.35,7) {{$\rm{LQ}^*$}};
    \node[black,scale=1.25] at (11.75,8.5) {{$e$}};
    \node[black,scale=1.25] at (16.25,8.5) {{$q$}};
    \node[black,scale=1.25] at (11.75,5.5) {{$e$}};
    \node[black,scale=1.25] at (16.25,5.5) {{$q$}};

    \draw[thick, -] (17.5, 8.5)--(19.5, 8);
    \draw[thick, -] (19.5, 8)--(21.5, 5.5);
    \draw[thick, double] (19.5, 8)--(19.5, 6);
    \draw[thick, -] (19.5, 6)--(21.5, 8.5);
    \draw[thick, -] (17.5, 5.5)--(19.5, 6);
    \node[black,scale=1.25] at (18.75,7) {{$\rm{LQ}^*$}};
    \node[black,scale=1.25] at (17.25,8.5) {{$e$}};
    \node[black,scale=1.25] at (21.75,8.5) {{$q$}};
    \node[black,scale=1.25] at (17.25,5.5) {{$e$}};
    \node[black,scale=1.25] at (21.75,5.5) {{$q$}};

	\end{scope}
      \end{tikzpicture}
    }
  \end{center}
  \caption{Relevant LQ diagrams in $ep$ and $ee$ colliders. The first two diagrams correspond to the LQ contribution to electron jet production in HERA while the last two represent the LQ contributions to the dijet production in LEP.
    \label{fig:HERALEP}}
\end{figure}

HERA was an electron-proton collider that operated at center-of mass energies up to 320 GeV in the regime of deep inelastic scattering. In this regime, a better understanding of the nature of the proton was reached. However, other results were obtained using the deep inelastic scattering of electrons and protons. For example, at those energies LQs can contribute to the total electron jet cross section in both $s$ and $t$ channel, as we can see in the first two diagrams of Fig.~\ref{fig:HERALEP}. It is important to notice that LQs are only resonantly produced in $ep$ colliders. For that reason, the experiments H1 and ZEUS performed searches for LQs in electron plus jet final states~\cite{H1:2001ezk, ZEUS:2003uzd, ZEUS:2012pwm}. To set limits in the mass-coupling plane, we recast the bounds from Ref.~\cite{ZEUS:2012pwm}. This search focused on the production of LQs of first generation leading to an electron and a jet signature at the ZEUS experiment with a luminosity of 498 pb$^{-1}$. Using this search, the ZEUS collaboration could set constraints on the LQ production in the 150 GeV to 1 TeV mass range. To recast these results we have taken into account the coupling structure of the models presented in Sec.~\ref{sec:LQ-formalism} and weighted the data with the corresponding branching ratios.

In electron-positron colliders it is also possible to measure the presence of LQs. As we can see in the last two diagrams of Fig.~\ref{fig:HERALEP}, LQs can contribute through $t$ and $u$ channels to the total dijet cross section. The L3 and OPAL experiments from LEP performed searches in the dijet cross section looking for new physics~\cite{L3:2000bql, OPAL:1998gjo}. We have recasted the search of OPAL~\cite{OPAL:1998gjo} that looked for constraints on new physics in the dijet cross section. This search can be translated into limits on LQ masses from 100 to 400 GeV. We have weighted their data according to the characteristics of our models in order to recast the search. Furthermore, experiments at LEP have also looked for LQs produced in decays of an on-shell $Z$ boson~\cite{L3:1991sow,OPAL:1991xaz}. Since no positive results were found, this can be translated into a lower limit of $m_{\text{LQ}}>45$ GeV. Finally, OPAL at LEP has also searched for charged long-lived LQ production~\cite{OPAL:1998znn}. In order to recast this search we have to verify the range of validity where the LQs are long-lived. The decay length of a long-lived particle is given by $L=\gamma\beta c \tau$, where $\gamma\beta$ is the boost factor and $\tau$ is the proper lifetime of that particle. For scalar LQs, assuming that the decay products have smaller masses than the initial particle, we find
\begin{eqnarray}
    \Gamma \sim \frac{g^2}{16\pi} m_\text{LQ},
\end{eqnarray}
where $g$ is the LQ coupling and $m_\text{LQ}$ is its mass. As the proper lifetime of a particle is given by $\tau=\hbar /\Gamma$ we can then obtain an expression for $g$ where the long-lived regime starts (assuming that a long-lived particle can be identified when it has a decay length of 0.1 cm~\cite{Dreiner:2023bvs})
\begin{eqnarray}
g^2 < \gamma\beta\frac{10^{-12}\, \rm{GeV} }{m_\text{LQ}}.
\label{eq:coupllp}
\end{eqnarray}
Using the information provided in~\cite{OPAL:1998znn} to compute the $\gamma$ and $\beta$ factors for different masses and center-of-mass energies, this gives us a range of validity for the long-lived LQ search of  $g<2.0 \times 10^{-7}$.

On the other hand, proton-antiproton colliders have also been used to look for LQ signals. In particular, the UA2 experiment at SPS has performed a search for double production of LQs that decay into a charged lepton plus jet or into a neutrino plus jet~\cite{UA2:1991ovi}. Double LQ production in a proton-antiproton collider does not depend on the specific LQ scenario, however the identification of the different LQ decays relies on the different branching ratios of the produced LQ, and hence from the model under scrutiny. We have recast the results from~\cite{UA2:1991ovi} according to our specific benchmark models, obtaining a mass limit of $m_{\text{LQ}}<50$ GeV.\\
Other LQ searches have been done at Tevatron, where double pair production is the leading production mechanism that allows to constrain the LQ mass. Both CDF and D\O\,  experiments at Tevatron have performed such a search~\cite{CDF:1993dmp,Norman:1994tp,D0:1997eun,UA2:1991ovi,Barfuss:2009zz} for different luminosities, being the production of LQs independent of the LQ coupling, as in the case of the UA2 bounds. However, the limits derived by the collaboration are dependent on the branching ratios. We have recast such limits into our scenarios and we have found that for our branching ratios these bounds are less powerful than the one obtained by UA2~\cite{UA2:1991ovi}. In the case of Ref.~\cite{Barfuss:2009zz} the experimental results do not show the results for the specific branching ratios typical of the benchmark points considered in our study (i.e. BR $\lesssim 25\%$ for each channel). However, we have decided to take as a reference the limits of the most similar case (BR = $50\%$), even if in this case the resulting exclusion region is overestimated. Taking all that into account, we have found a window in tension with Tevatron data~\cite{Barfuss:2009zz} that lies in the mass range of $150<m_{\text{LQ}}<260$ GeV.

\begin{figure}
  \begin{center}
    \scalebox{0.6}{
      \begin{tikzpicture}
	\begin{scope}[thick] 
    \draw[thick, gluon] (0, 8.5)--(2, 8);
    \draw[thick, double] (2, 8)--(3.5, 8);
    \draw[thick, double] (2, 8)--(2, 6);
    \draw[thick, double] (2, 6)--(3.5, 6);
    \draw[thick, gluon] (0, 5.5)--(2, 6);
    \draw[thick, -] (3.5, 6)--(4.5,6.5);
    \draw[thick, -] (3.5, 8)--(4.5,8.5);
    \draw[thick, -] (3.5, 6)--(4.5,5.5);
    \draw[thick, -] (3.5, 8)--(4.5,7.5);
    \node[black,scale=1.25] at (2.75,8.5) {{$\rm{LQ}$}};
    \node[black,scale=1.25] at (2.75,5.5) {{$\rm{LQ}$}};
    \node[black,scale=1.25] at (4.95,8.5) {{$\ell /\nu$}};
    \node[black,scale=1.25] at (4.95,7.5) {{$q$}};
    \node[black,scale=1.25] at (4.95,6.5) {{$\ell /\nu$}};
    \node[black,scale=1.25] at (4.95,5.5) {{$q$}};

    \draw[thick, gluon] (6, 8.5)--(7, 7);
    \draw[thick, gluon] (6, 5.5)--(7, 7);
    \draw[thick, gluon] (7, 7)--(8, 7);
    \draw[thick, double] (8, 7)--(9, 8);
    \draw[thick, double] (8, 7)--(9, 6);
    \draw[thick, -] (9, 8)--(10, 8.5);
    \draw[thick, -] (9, 8)--(10, 7.5);
    \draw[thick, -] (9, 6)--(10, 6.5);
    \draw[thick, -] (9, 6)--(10, 5.5);
    \node[black,scale=1.25] at (8.5,8.5) {{$\rm{LQ}$}};
    \node[black,scale=1.25] at (8.5,5.5) {{$\rm{LQ}$}};
    \node[black,scale=1.25] at (10.45,8.5) {{$\ell /\nu$}};
    \node[black,scale=1.25] at (10.45,7.5) {{$q$}};
    \node[black,scale=1.25] at (10.45,6.5) {{$\ell /\nu$}};
    \node[black,scale=1.25] at (10.45,5.5) {{$q$}};

    \draw[thick, gluon] (11.5, 8.5)--(12.5, 7);
    \draw[thick, gluon] (11.5, 5.5)--(12.5, 7);
    \draw[thick, double] (12.5, 7)--(13.5, 8);
    \draw[thick, double] (12.5, 7)--(13.5, 6);
    \draw[thick, -] (13.5, 8)--(14.5, 8.5);
    \draw[thick, -] (13.5, 8)--(14.5, 7.5);
    \draw[thick, -] (13.5, 6)--(14.5, 6.5);
    \draw[thick, -] (13.5, 6)--(14.5, 5.5);
    \node[black,scale=1.25] at (13,8.5) {{$\rm{LQ}$}};
    \node[black,scale=1.25] at (13,5.5) {{$\rm{LQ}$}};
    \node[black,scale=1.25] at (14.95,8.5) {{$\ell /\nu$}};
    \node[black,scale=1.25] at (14.95,7.5) {{$q$}};
    \node[black,scale=1.25] at (14.95,6.5) {{$\ell /\nu$}};
    \node[black,scale=1.25] at (14.95,5.5) {{$q$}};

    \draw[thick, -] (16, 8.5)--(17, 7);
    \draw[thick, -] (16, 5.5)--(17, 7);
    \draw[thick, gluon] (17, 7)--(18, 7);
    \draw[thick, double] (18, 7)--(19, 8);
    \draw[thick, double] (18, 7)--(19, 6);
    \draw[thick, -] (19, 8)--(20, 8.5);
    \draw[thick, -] (19, 8)--(20, 7.5);
    \draw[thick, -] (19, 6)--(20, 6.5);
    \draw[thick, -] (19, 6)--(20, 5.5);
    \node[black,scale=1.25] at (18.5,8.5) {{$\rm{LQ}$}};
    \node[black,scale=1.25] at (18.5,5.5) {{$\rm{LQ}$}};
    \node[black,scale=1.25] at (20.45,8.5) {{$\ell /\nu$}};
    \node[black,scale=1.25] at (20.45,7.5) {{$q$}};
    \node[black,scale=1.25] at (20.45,6.5) {{$\ell /\nu$}};
    \node[black,scale=1.25] at (20.45,5.5) {{$q$}};

    \draw[thick, -] (21.5, 8.5)--(23, 8);
    \draw[thick, -] (21.5, 5.5)--(23, 6);
    \draw[thick, -] (23, 6)--(23, 8);
    \draw[thick, double] (23, 8)--(24.5, 8);
    \draw[thick, double] (23, 6)--(24.5, 6);
    \draw[thick, -] (24.5, 6)--(25.5,6.5);
    \draw[thick, -] (24.5, 8)--(25.5,8.5);
    \draw[thick, -] (24.5, 6)--(25.5,5.5);
    \draw[thick, -] (24.5, 8)--(25.5,7.5);
    \node[black,scale=1.25] at (23.5,8.5) {{$\rm{LQ}$}};
    \node[black,scale=1.25] at (23.5,5.5) {{$\rm{LQ}$}};
    \node[black,scale=1.25] at (25.95,8.5) {{$\ell /\nu$}};
    \node[black,scale=1.25] at (25.95,7.5) {{$q$}};
    \node[black,scale=1.25] at (25.95,6.5) {{$\ell /\nu$}};
    \node[black,scale=1.25] at (25.95,5.5) {{$q$}};
    \node[black,scale=1.25] at (23.4,7) {{$\ell^*$}};

	\end{scope}
      \end{tikzpicture}
    }
  \end{center}
  \caption{Diagrams contributing to LQ double production at the LHC.
    \label{fig:DP}}
\end{figure}
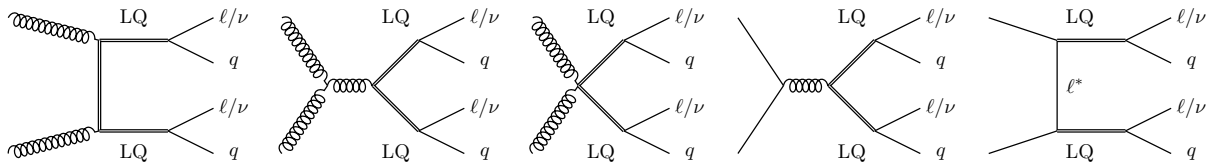

Finally, LQs can also be produced in multiple ways at the LHC. For this reason there are different searches from ATLAS and CMS that look for LQs taking into account their multiple production channels and decays. One of these searches is the double production of LQs. Being LQs color triplets, they can be created in the LHC in events initiated by gluons and quarks. Fig.~\ref{fig:DP} shows all the diagrams that contribute to the LQ double production. As we can see, the first four diagrams are initiated by the strong coupling, so they scale as $\alpha_s^2$, which is independent on the LQ coupling. The last diagram is the only one that depends on the LQ coupling,  $g^4$. Because of this, we expect this kind of searches to be independent of LQ couplings when their value is small, and the limits will only display their dependence on $g$ when they reach values of $\mathcal{O}(1)$ \cite{Schmaltz:2018nls}. For that reason, this search can exclude LQ masses  independently of the LQ coupling. ATLAS and CMS have performed several searches for double production of LQs in multiple final states, mainly the presence of two quarks and the combination of charged and neutral leptons, $qq\ell\ell$, $qq\ell\nu$ and $qq\nu\nu$, where, depending on the LQ generation, signals may vary into different flavors of leptons. Nonetheless, given the purposes of our work, we are only interested in those searches that contain first- and second-generation leptons in the final state \cite{ATLAS:2020dsk, CMS:2018lab,CMS:2018ncu}. In order to set limits from LQ double production we have recast the search from Ref.~\cite{ATLAS:2020dsk} using the recommendations of Refs.~\cite{Diaz:2017lit,Schmaltz:2018nls}. To simulate the LQ double production we have made use of \texttt{MadGraph5\_aMC@NLO-v3.5.0}~\cite{Alwall:2014hca,Frederix:2018nkq} using the codes and recommendations from Refs.~\cite{Borschensky:2020hot,Borschensky:2021hbo,urllq}. We have then compared the production cross sections weighted by their corresponding branching fractions with those from Ref.~\cite{ATLAS:2020dsk} to obtain the limits. Finally, it is important to note that even if the limits from LQ pair production are independent of the LQ coupling, $g$, when this is small ($g \ll 1$), there is an actual limit to the constraints on this coupling. This is given by the fact that the LQs could be long-lived due to the smallness of the coupling. Hence we can use the same procedure as we used above, following Eq.~\eqref{eq:coupllp}, in order to infer an estimate of the reliability of the double production bounds. Using this method we obtain that this search is valid for values of the coupling $g> \mathcal{O}(10^{-6}-10^{-7})$.

\begin{figure}
  \begin{center}
    \scalebox{0.65}{
      \begin{tikzpicture}
	\begin{scope}[thick] 
    \draw[thick, -] (0, 8.5)--(2, 8);
    \draw[thick, -] (2, 8)--(4, 8.5);
    \draw[thick, double] (2, 8)--(2, 6);
    \draw[thick, -] (2, 6)--(4, 5.5);
    \draw[thick, -] (0, 5.5)--(2, 6);
    \node[black,scale=1.25] at (2.75,7) {{$\rm{LQ}^*$}};
    \node[black,scale=1.25] at (4.25,8.5) {{$\ell$}};
    \node[black,scale=1.25] at (-0.25,8.5) {{$q$}};
    \node[black,scale=1.25] at (4.25,5.5) {{$\ell$}};
    \node[black,scale=1.25] at (-0.25,5.5) {{$q$}};

    \draw[thick, -] (6, 8.5)--(7, 7);
    \draw[thick, gluon] (6, 5.5)--(7, 7);
    \draw[thick, -] (7, 7)--(8, 7);
    \draw[thick, -] (8, 7)--(9, 8);
    \draw[thick, double] (8, 7)--(9, 6);
    \draw[thick, -] (9, 6)--(10, 6.5);
    \draw[thick, -] (9, 6)--(10, 5.5);
    \node[black,scale=1.25] at (8.0,6.25) {{$\rm{LQ}$}};
    \node[black,scale=1.25] at (9.45,8.5) {{$\ell /\nu$}};
    \node[black,scale=1.25] at (5.65,8.5) {{$q$}};
    \node[black,scale=1.25] at (10.45,6.5) {{$\ell /\nu$}};
    \node[black,scale=1.25] at (10.45,5.5) {{$q$}};
    \node[black,scale=1.25] at (5.65,5.5) {{$g$}};

    \draw[thick, -] (12, 8.5)--(14, 8);
    \draw[thick, gluon] (12, 5.5)--(14, 6);
    \draw[thick, double] (14, 8)--(14, 6);
    \draw[thick, -] (14, 8)--(16, 8.5);
    \draw[thick, double] (14, 6)--(16, 6);
    \draw[thick, -] (16, 6)--(17, 6.5);
    \draw[thick, -] (16, 6)--(17, 5.5);
    \node[black,scale=1.25] at (17.45,6.5) {{$\ell /\nu$}};
    \node[black,scale=1.25] at (17.45,5.5) {{$q$}};
    \node[black,scale=1.25] at (17.45,8.5) {{$\ell /\nu$}};
    \node[black,scale=1.25] at (11.5,8.5) {{$q$}};
    \node[black,scale=1.25] at (11.5,5.5) {{$g$}};
    \node[black,scale=1.25] at (14.75,7) {{$\rm{LQ}^*$}};
    \node[black,scale=1.25] at (15.0,5.25) {{$\rm{LQ}$}};

	\end{scope}
      \end{tikzpicture}
    }
  \end{center}
  \caption{Diagrams contributing to Drell Yan production and single LQ production. The first diagram corresponds to the LQ contribution to Drell Yan while the last ones are the responsible for LQ single production.
    \label{fig:DYSP}}
\end{figure}
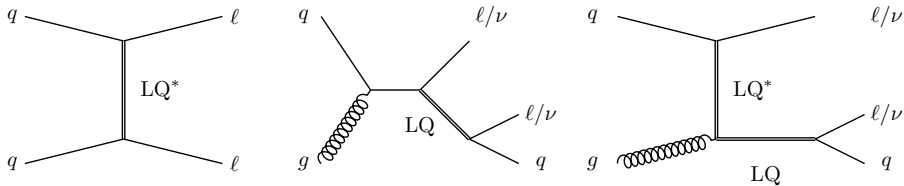

Another imprint of LQs at LHC is through  dilepton production. As LQs couple to both quarks and leptons they can contribute to the Drell Yan cross section as it is shown in the first diagram of Fig.~\ref{fig:DYSP}. The LQ appears in the $t$-channel of a process initiated by a pair of quarks giving as a final results two leptons. The presence of LQs in this process can interfere with the SM processes mediated by $\gamma/Z$. ATLAS and CMS experiments have performed several searches studying the dilepton cross section \cite{ATLAS:2020yat, CMS:2021ctt}. As we can see from the first diagram in Fig.~\ref{fig:DYSP}, this process is coupling-dependent, contrary to what we had for the double production mechanism. To understand how strong these limits are, we simulate the LQ cross section using \texttt{MadGraph5\_aMC@NLO-v3.5.0}~\cite{Alwall:2014hca,Frederix:2018nkq} and again, using the codes from  \cite{Borschensky:2020hot,Borschensky:2021hbo,urllq} we compare it against data from Ref.~\cite{CMS:2021ctt}. Concerning this last step, we follow the prescriptions of  Ref.~\cite{Schmaltz:2018nls} to obtain the cross section with the cuts from the search and compare the results against actual data. Another important channel to consider in hadron colliders is the single LQ production. LQs can be produced together with a lepton in processes initiated by a quark and a gluon, as it is shown in the last two diagrams of Fig.~\ref{fig:DYSP}. Once the LQ is produced it will subsequently decay into a quark and a lepton, leading to different final states containing either one jet and two charged leptons, or one jet, one charged lepton and missing transverse energy, or simply one jet and missing energy. Several searches from ATLAS and CMS have been looking for these signals: we choose to recast the CMS search for energetic jets and missing transverse energy~\cite{CMS:2021far} at a center-of-mass energy of $\sqrt{s}=13$ TeV and a luminosity of $\mathcal{L}=$101 fb$^{-1}$ since it proves directly the coupling of LQ to neutrinos. These limits are also computed using \texttt{MadGraph5\_aMC@NLO-v3.5.0}~\cite{Alwall:2014hca,Frederix:2018nkq}, benefiting from the codes of  \cite{Borschensky:2020hot,Borschensky:2021hbo,urllq} and compared with the data obtained from Ref.~\cite{CMS:2021far}.

\section{Results}
\label{sec:results}
\begin{figure}[!tb]
\centering
\includegraphics[width=0.49\textwidth]{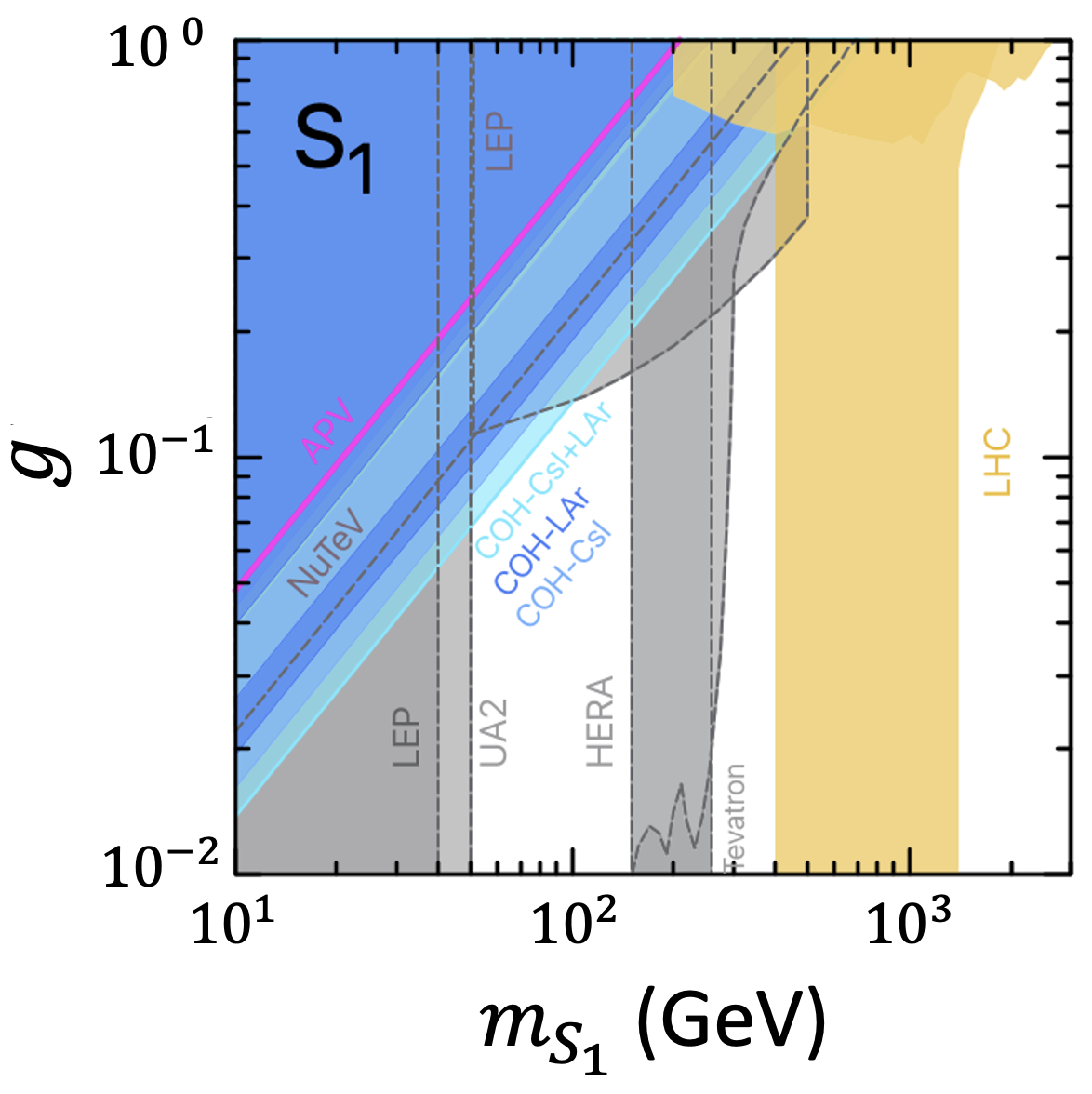}
\includegraphics[width=0.49\textwidth]{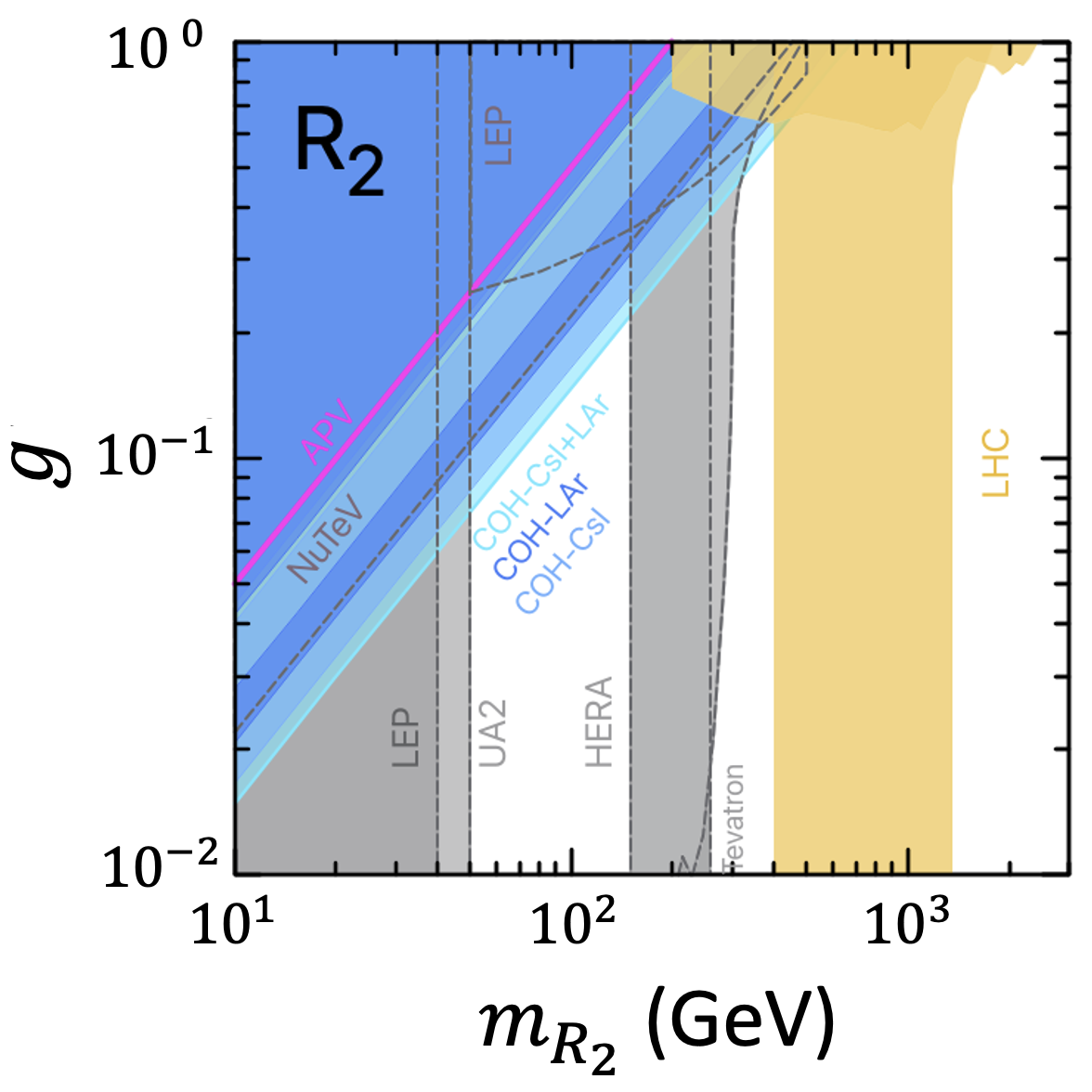}
\includegraphics[width=0.5\textwidth]{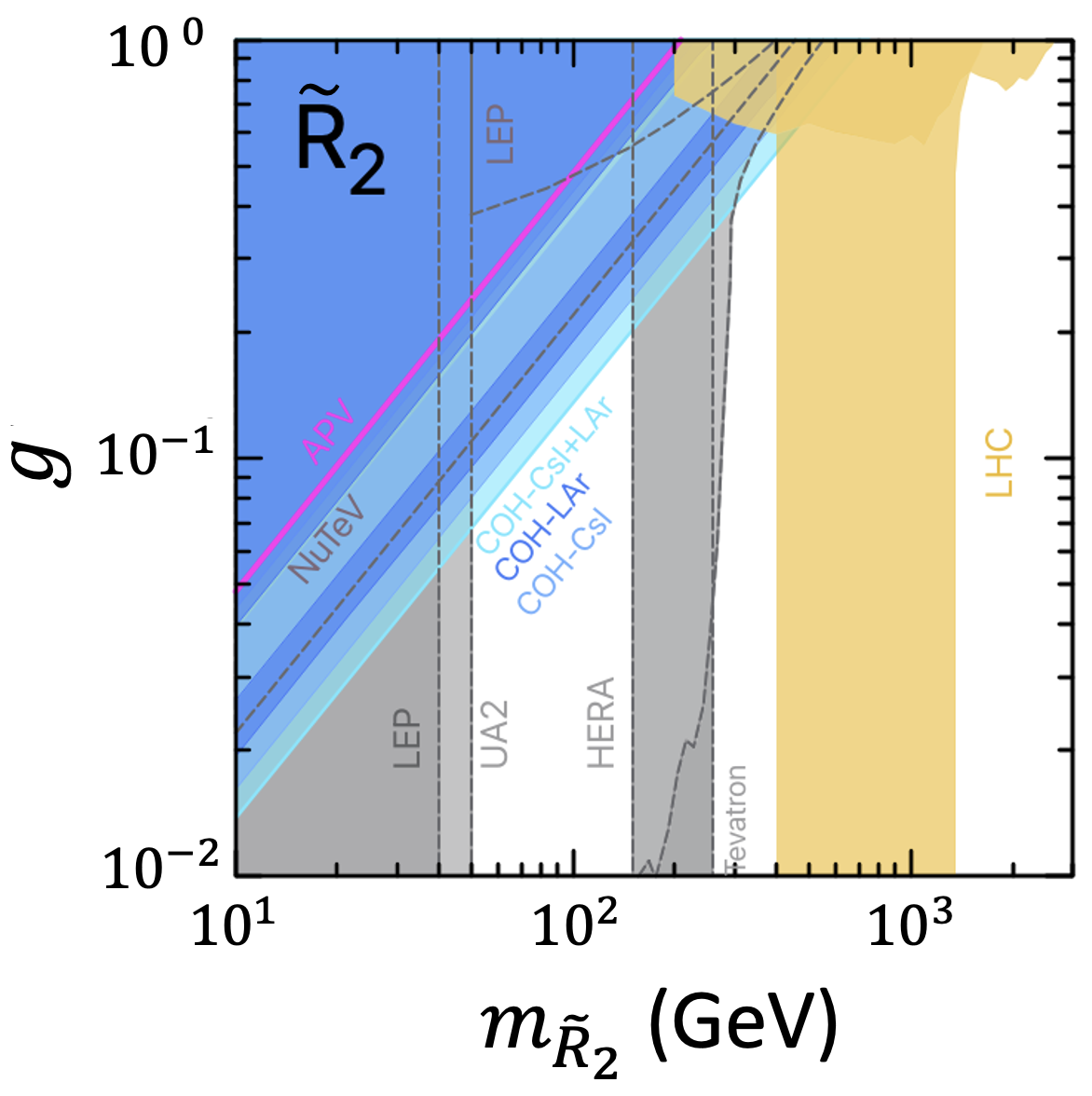}
\includegraphics[width=0.49\textwidth]{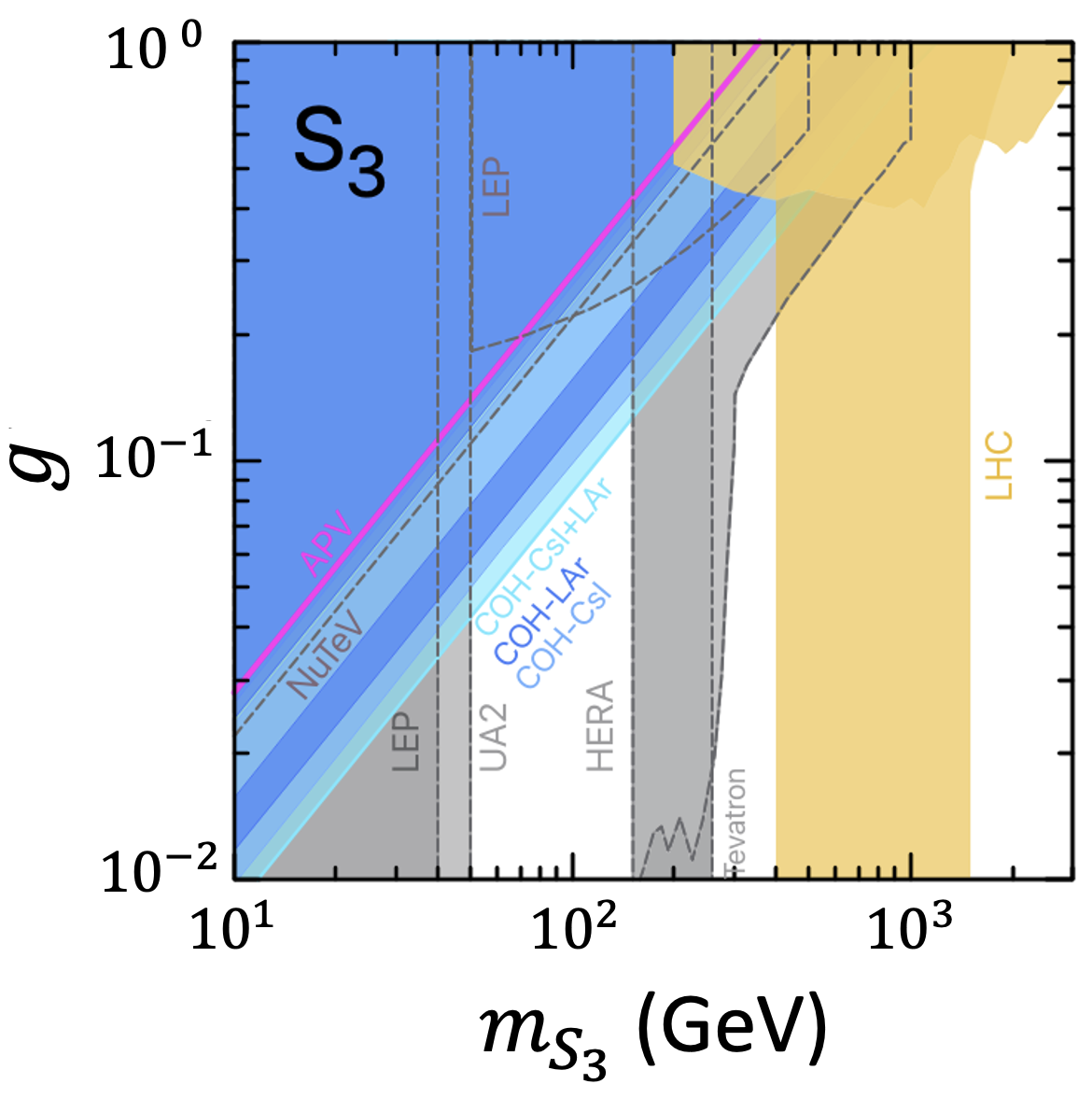}
\caption{$90 \%$ C.L. excluded regions, in the ($m_\textrm{LQ}, g$) plane, on different LQ scenarios, $S_1, R_2, \tilde{R}_2,$ and $S_3$. Colored contours and filled areas denote new upper bounds obtained in this work: APV (magenta line); \cevns~data from the COHERENT-CsI (2021)~\cite{COHERENT:2021xmm} and COHERENT-LAr (2020)~\cite{COHERENT:2020iec} data sets (different shades of blue); single production, double production and Drell Yan processes at LHC (yellow region).  For comparison, we also show previously obtained limits in the literature and recast them here into the LQ scenarios under scrutiny (grey-shaded regions): NuTeV~\cite{Escrihuela:2011cf}, ZEUS at HERA~\cite{ZEUS:2012pwm}, UA2 at SPS~\cite{UA2:1991ovi}, CDF and D0 at Tevatron~\cite{Barfuss:2009zz}, and OPAL at LEP~\cite{OPAL:1991xaz,OPAL:1998gjo}. See main text for more details.  }
\label{fig:current:results}
\end{figure}

In this section, we summarize the results for the  constraints obtained with all different processes discussed in the previous section, each covering a different range of validity in the LQ parameter space ($m_\textrm{LQ},g$). Notice that, for all our computations, we have assumed the same coupling constant $g$ for all leptons except for taus, for which we assume a vanishing coupling. The main results for each LQ scenario, in terms of the LQ mass $m_\text{LQ}$ and coupling $g$, are shown in the different panels of Fig.~\ref{fig:current:results}, where we show the $90 \%$ C.L. exclusion limits from each case. 
The top-left panel in the figure shows the constraints obtained for the scalar scenario listed as $S_1$ in Table~\ref{tab:operators}. From the figure, we can  distinguish the different mass ranges that each experiment is able to test, and hence, we can infer the complementarity between different observables and facilities to constrain a wide region of parameter space. 
Colored regions in the figure indicate new results computed in this work, while grey-shaded regions correspond to previous limits found in the literature, including those from NuTeV~\cite{NuTeV:2001whx} and colliders such as ZEUS at HERA~\cite{ZEUS:2012pwm}, OPAL at LEP~\cite{OPAL:1991xaz,OPAL:1998gjo} and UA2 at SPS~\cite{UA2:1991ovi}, which we have recast for the LQ scenarios of our interest.

 Starting from low-energy observables, APV upper limits obtained from Eq.~\eqref{eq:chi2APV} are indicated as magenta contours,\footnote{We do not color-fill them not to overcrow the figures.} while COHERENT excluded regions (see Eqs.~\eqref{eq:chi2CsI} and \eqref{eq:chi2LAr}) are shown as different shades of blue. Dark blue indicates the limits from the COHERENT-LAr (2020) detector alone, light blue stands for the COHERENT-CsI data set (2021), and cyan is used for their combined analysis. We can notice that, when considered individually, the excluded region for COHERENT-LAr is not continuous, as it contains a tiny allowed band. This degeneracy is a consequence of a destructive interference between the SM and the LQ contributions to the \cevns~cross section, resulting in a combination of non-zero parameters that can mimic the SM solution. Although not visible, the situation is similar for the COHERENT-CsI analysis. However, when combining the results from the two detectors, the degeneracy is lifted, resulting in an excluded cyan region which is now continuous. We refer the reader to Appendix~\ref{app:int} for a detailed discussion on the origin and effects of this interference. Moreover, when comparing to previous results that use the older CsI data set~\cite{Calabrese:2022mnp}, we can see that new data allow to constrain slightly lower couplings and their combination with the LAr result allows to remove the SM degeneracy, as just discussed.

Still referring to the top-left panel in the figure, and moving to the heavy mass (and energy) regime, the yellow region in the figure corresponds to LHC constraints that have been obtained including channels like single production, double production and Drell Yan (see discussion in Sec.~\ref{sec:colliders}). Regarding LHC constraints, it is worth mentioning that, within their validity range, double pair production limits are mass independent for $g \lesssim 3\times10^{-1}$. This is shown as a vertical yellow band that extends for LQ masses in the range $400\,\rm{GeV} \lesssim m_{LQ} \lesssim 1400$ GeV.
On the other hand, we further show as grey-shaded regions those excluded by LEP, UA2, HERA, Tevatron and NuTeV, which we recast from pre-existing analyses (see Secs.~\ref{sec:nutev} and \ref{sec:colliders}). In general, these bounds apply to heavy LQ masses, and some of them have been obtained under effective-theory assumptions, except for the LEP search, which probes LQs from an on-shell $Z$ boson decay, and can be extended to very low masses. 
For the displayed mass range, we see that LQ searches in $Z$ decays from LEP are able to constrain masses below $m_\text{LQ} \lesssim 40-45$ GeV, this result being overtaken by the double production of LQ at UA2, which excludes masses up to $m_\text{LQ} \lesssim 50$ GeV.
Let us remark that the LQ double production channel is actually independent on the LQ coupling. However, the LEP and UA2 bounds that we derived here depend upon the flavor structure of their couplings and therefore upon their branching ratios, being these searches less sensitive to smaller branching fractions. As we work under the assumption that LQs couple to both the first and second lepton families scenarios, the best limit we get from LEP and UA2 double production searches is $m_\text{LQ} \lesssim 50$ while leaving heavier masses unconstrained.  
As a result, we see that COHERENT data lead to the most stringent constraints in one small region in the parameter space that goes from $m_\text{LQ} \sim 50$~GeV and up to $m_\text{LQ} \sim 100-150$~GeV. In this small window, we see that COHERENT clearly overtakes former bounds from NuTeV, APV and LEP. For masses greater than $m_\text{LQ} \gtrsim 100-150$~GeV, LEP, HERA and Tevatron become the most stringent bounds, being the latter two the most powerful constraints in terms of the LQ coupling due to its nature as an $ep$ collider that can produce on-shell LQ for the case of HERA, and due to the double LQ production in Tevatron. As mentioned before, it is important to note that the bounds imposed by Tevatron searches are over-estimated due to different assumptions on the branching ratios, and the actual limits would lie within the area set by HERA. However, the strength of the bounds weakens for masses around $m_\text{LQ} \sim 400$~GeV. From this LQ mass and on, LHC constraints dominate thanks to the double pair production, which leads to coupling-independent bounds on the LQ mass. As anticipated in Sec.~\ref{sec:colliders}, while covering the whole $g$ parameter space in these panels, the LHC-excluded yellow band is expected to extend down to $\sim 10^{-6} - 10^{-7}$ due to LQs lifetime considerations. The small region at $g \sim 0.8$ bounded at $m_\text{LQ} \sim 2$ TeV comes from Drell Yan processes. All in all, we can safely conclude that for $m_\text{LQ} \lesssim 0.1$ TeV the dominant exclusion process is \cevns, thus complementing the strong collider bounds which instead dominate above $0.1$ TeV. It is important to note that existing long-lived charged particles searches, such as~\cite{OPAL:1998znn} by LEP, constrain the LQ parameter space for couplings below $g\lesssim 10^{-7}$, however these constraints lie outside the parameter range shown in these plots.

The top-right and lower panels in Fig.~\ref{fig:current:results} show the corresponding results for other LQ scenarios: $R_2$ (top right), $\tilde{R}_2,$ (bottom left) and $S_3$ (bottom right). 
Overall, we see a similar behaviour for all cases, being \cevns~the dominant channel to constrain LQ masses  in one small window that goes from  $50~\text{GeV} \lesssim m_\text{LQ} \lesssim 150$ GeV for the $S_3$ scenario, and for two small windows that go from $50~\text{GeV} \lesssim m_\text{LQ} \lesssim 150$ GeV  and $300~\text{GeV} \lesssim  m_\text{LQ} \lesssim 400$ GeV in the case of $R_2$ and $\tilde{R}_2$. Above these mass ranges, LHC  data have  a major ability in setting constraints on the LQ coupling $g$. When comparing the different panels, notice that the most stringent \cevns~constraint is found for $S_3$. This was expected from  the modified weak charge defined in Eq.~\eqref{eq:s3:cross}, where we  see that the \cevns~cross section in this case is effectively enhanced with a factor $(4N)^2$ for fixed $m_\text{LQ}$ and $g$. On the other hand, for $R_2$ the \cevns~constraint is less robust because of a factor $2Z$ in the cross section given in Eq.~\eqref{eq:r2:cross} which, given the different relative sign between $g_V^p$ and $g_V^n$, results into a smaller cross section, and hence a lower number of events expected in the statistical analysis. Another interesting feature is that, when coupling to neutrinos, scenarios $S_1$ and $\tilde{R}_2$ are indistinguishable for \cevns (see Sec.~\ref{sec:LQ-formalism}), and in consequence, the excluded blue-shaded regions in the two left panels are the same.
However, this is not the case for collider observables since, when coupling to charged leptons, $S_1$ couples only to up quarks while $\tilde{R}_2$ couples only to down quarks. Then, because of the ratio between up and down quarks within the proton, this results in different excluded yellow regions in the top and bottom left panels of Fig.~\ref{fig:current:results}.

\section{Future sensitivities}
\label{sec:future_sens}
After having analyzed the current picture of LQ constraints in the parameter space ($m_\text{LQ},g$), we now turn our attention to sensitivities that can be reached at future \cevns~experiments. We consider upcoming upgrades of both the CsI and LAr detectors planned by the COHERENT collaboration, as well as two of the different detectors from a proposal at the ESS discussed in \cite{Baxter:2019mcx}. We discuss these prospects in the following.

\subsection{\cevns~data (COH-CsI-700 and COH-LAr-750)}
\label{subsec:COH-future}

The intense experimental program of the COHERENT collaboration envisages, among others, upgrades of current detectors, namely a 700-kg cryogenic CsI scintillator and a tonne-scale LAr time-projection
chamber detector~\cite{Asaadi:2022ojm,Akimov:2022oyb}. Moreover, planned up-scales of the SNS proton beam foresee an upgrade of the proton energy 
$E_p = 0.984 \to 1.3$ GeV
and of the beam power $P = 1.4 \to 2$ MW. By assuming a data-taking time of 5000 hr per year, this leads to 
$N_\mathrm{POT} = 5.18 \times 10^{23}$ (for three years)~\cite{Asaadi:2022ojm,Akimov:2022oyb,AtzoriCorona:2023ktl}, and to a predicted number of neutrinos per flavor produced for each POT $r = 0.0848 \to 0.13$ ~\cite{COHERENT:2021yvp}.
We estimate the future sensitivities for the COH-LAr-750 and COH-CsI-700 updates of the COHERENT detectors, assuming the technical upgrades summarized in Table~\ref{tab:CEvNSexps}, and a detector mass of 750 and 700 kg, respectively. We perform a statistical analysis in energy and time following that done for current data and previously detailed in Sec.~\ref{sec:current_constr}.
In the case of COH-CsI-700 we take into account the expected improvement in energy sensitivity, by  considering a threshold of $1.4$ keV$_\mathrm{nr}$~\cite{Akimov:2022oyb,AtzoriCorona:2023ktl} while keeping the
shape of the energy efficiency unaltered. Pragmatically, we add in the statistical analysis an extra bin in energy, from 1 to 4 PE. Concerning backgrounds, keeping in mind that the collaboration anticipates that the cryogenic technology will allow to reduce them (in particular to remove the Cherenkov radiation background), we choose to be conservative and re-scale the current BRN, NIN and SSB backgrounds to the new detector's mass. Moreover, again assuming a conservative approach, we fix the numbers of background events in the first, new energy bin ([1-4] photoelectrons) to be exactly the same as in the second one ([4-8] photoelectrons).\\

\begin{figure}
\centering
\includegraphics[width=0.49\textwidth]{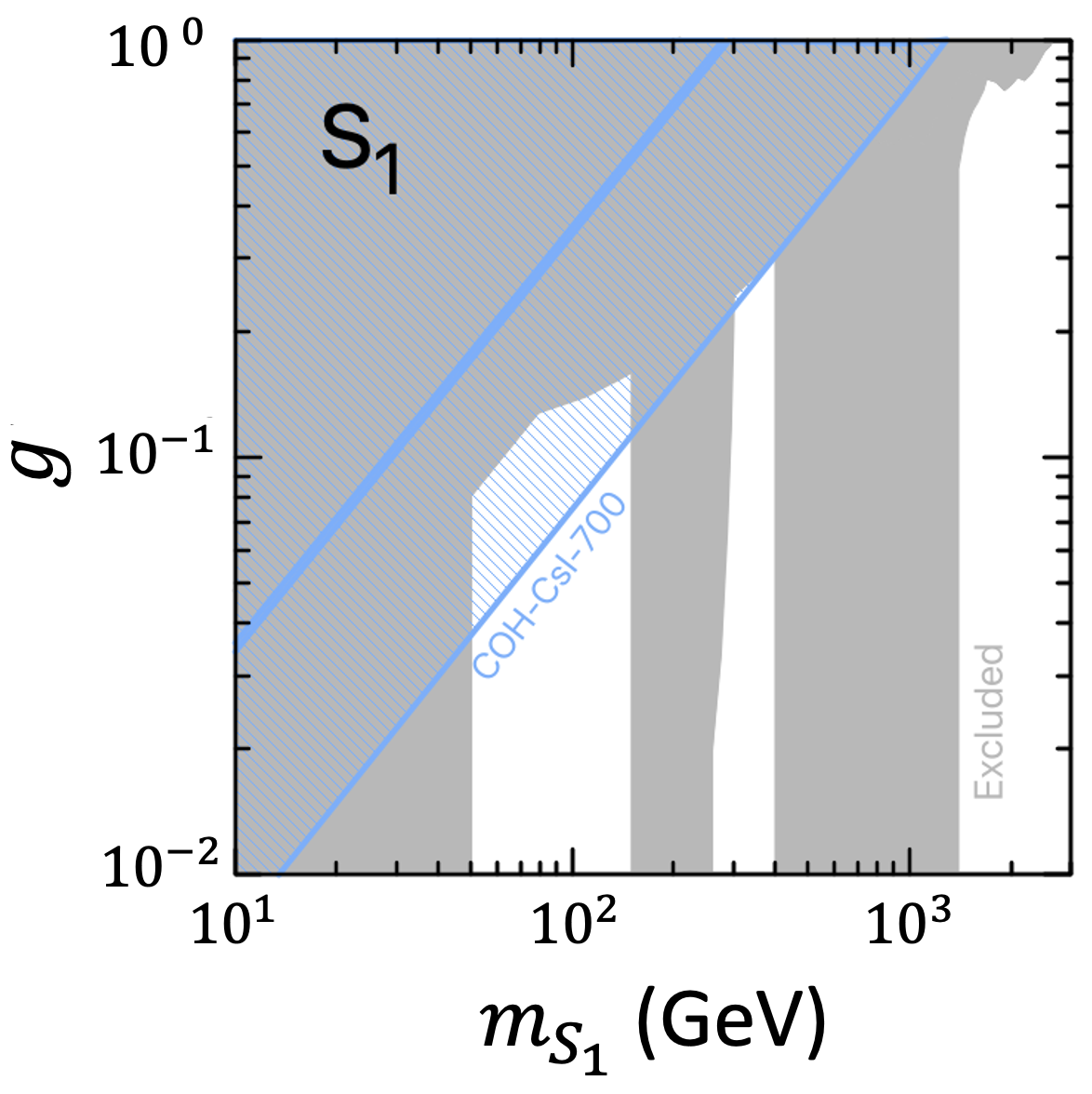}
\includegraphics[width=0.49\textwidth]{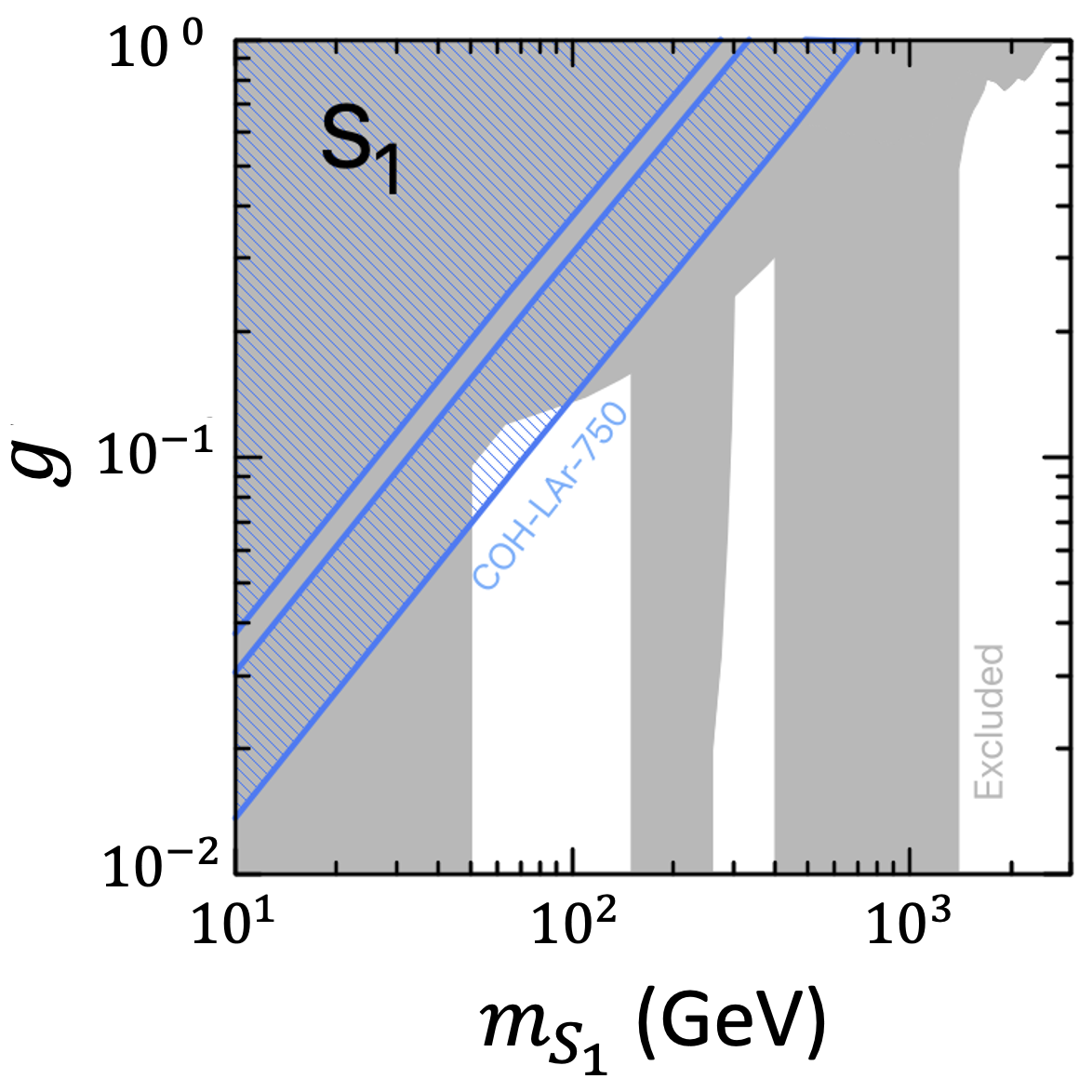}
\caption{Expected $90 \%$ C.L. sensitivities, in the ($m_\textrm{LQ}, g$) plane, obtained for the  COH-CsI-700 (left panel) and  COH-LAr-750 (right) detectors and assuming model $S_1$. These bounds apply also to model $\tilde{R}_2$. The grey-shaded regions refer to current limits previously presented in Fig.~\ref{fig:current:results}.  In the case of current \cevns~bounds, only the constraint obtained with the corresponding target is shown. See text for more details.  }
\label{fig:future:SNS}
\end{figure}

\begin{figure}
\centering
\includegraphics[width=0.49
\textwidth]{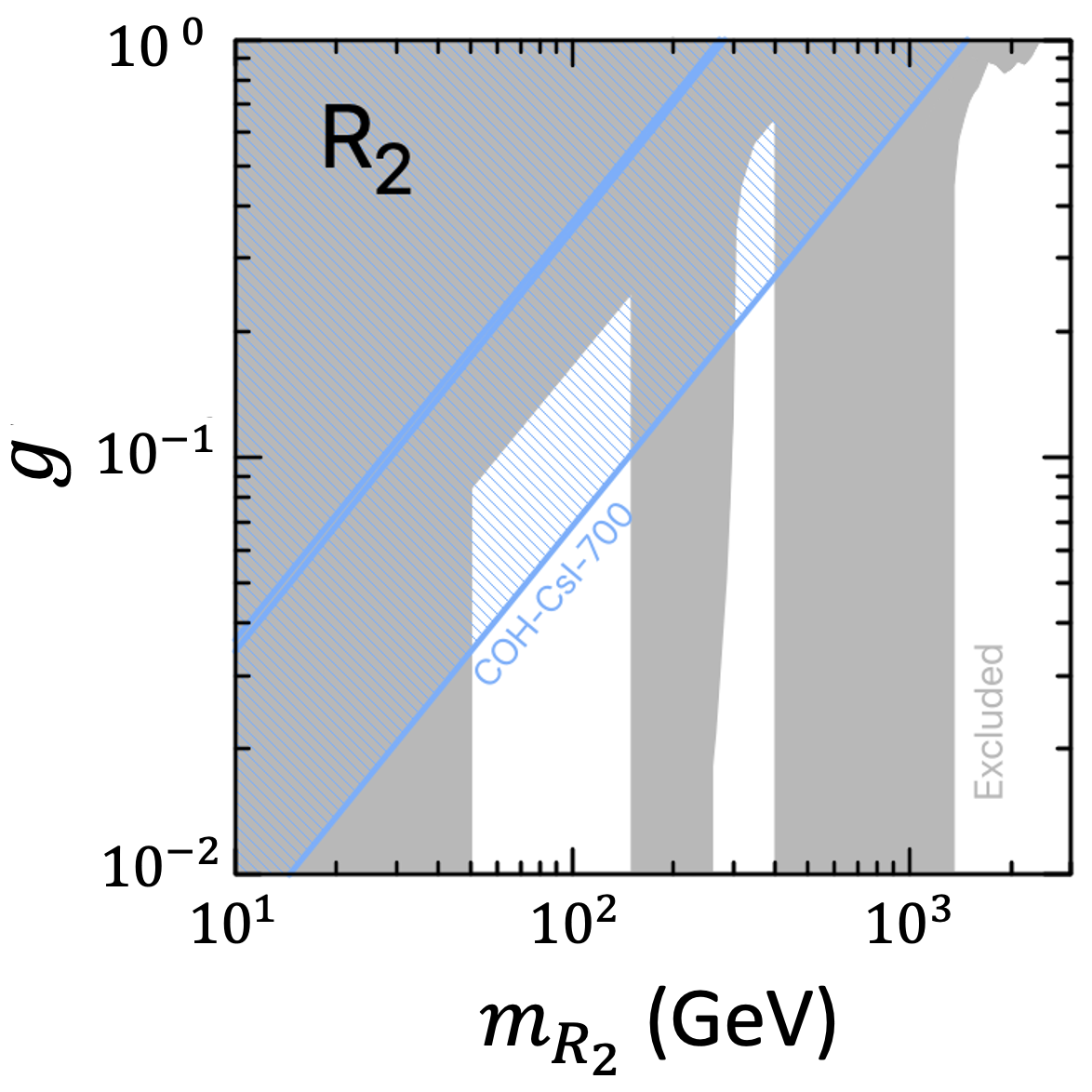}
\includegraphics[width=0.49
\textwidth]{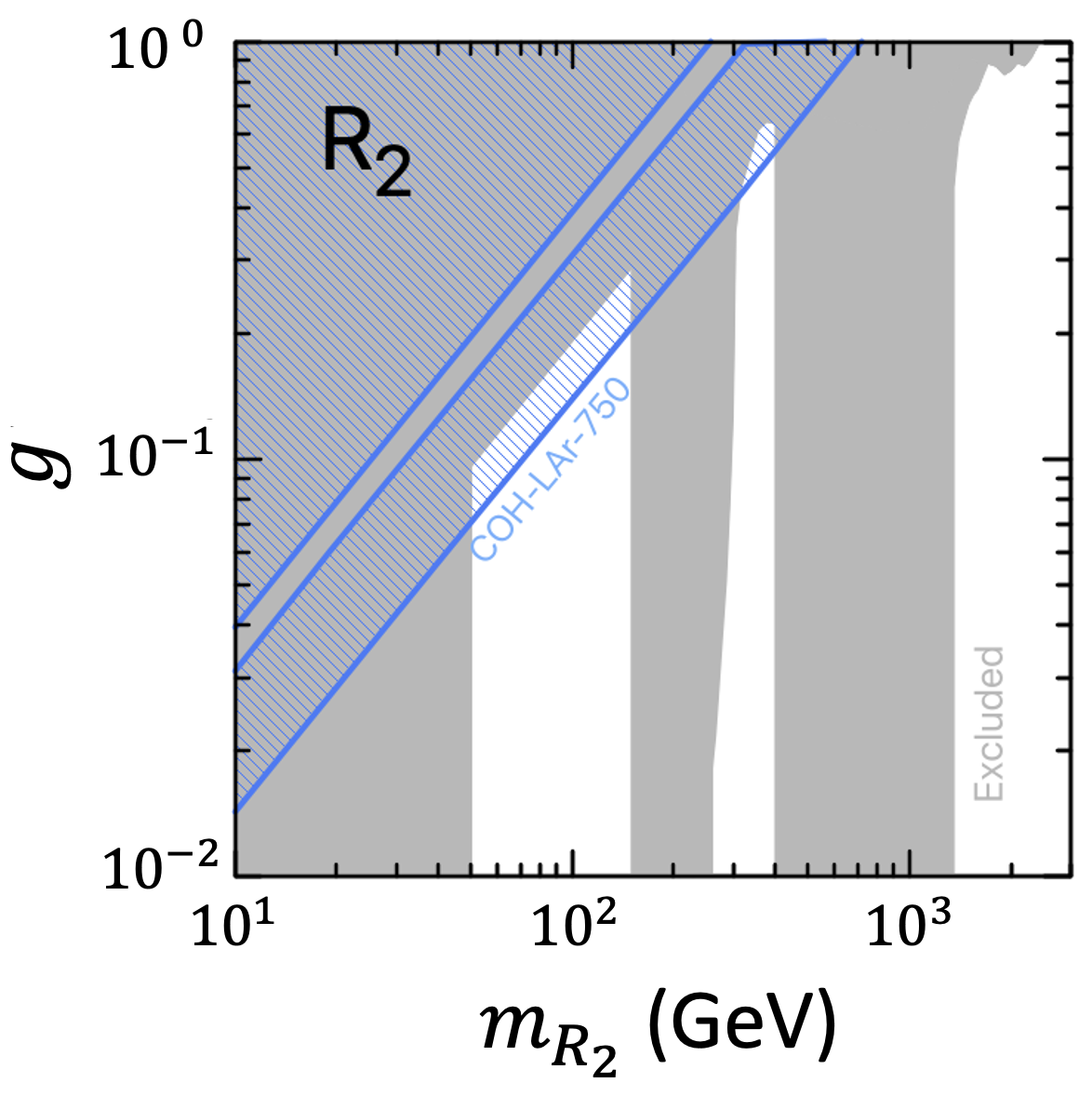}
\caption{Expected $90 \%$ C.L. sensitivities, in the ($m_\textrm{LQ}, g$) plane, obtained for the  COH-CsI-700 (left panel) and  COH-LAr-750 (right) detectors and assuming model $R_2$. The grey-shaded regions refer to current limits previously presented in Fig.~\ref{fig:current:results}.  In the case of current \cevns~bounds, only the constraint obtained with the corresponding target is shown. See text for more details.  }
\label{fig:future:SNS2}
\end{figure}

\begin{figure}
\centering
\includegraphics[width=0.49\textwidth]{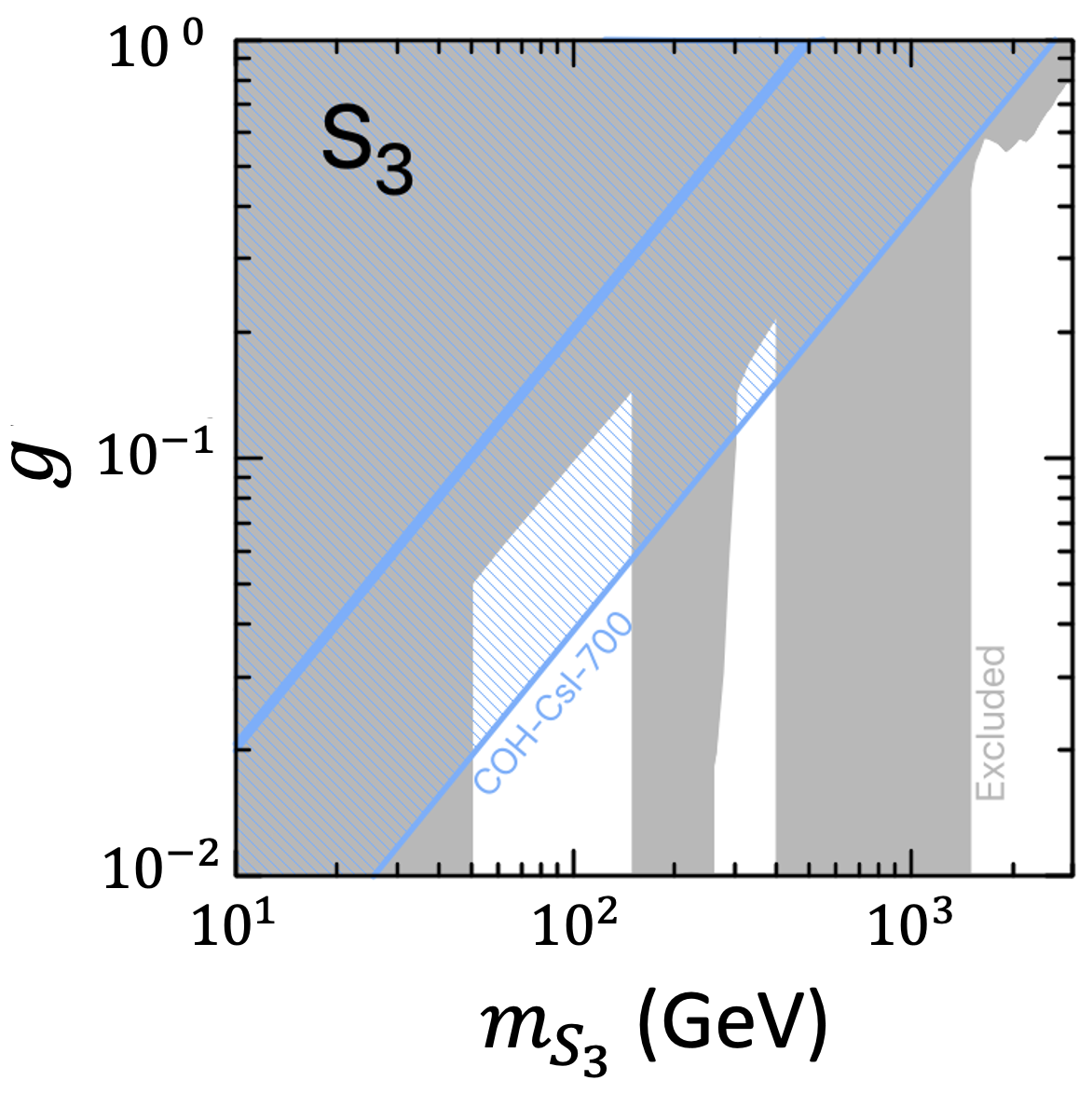}
\includegraphics[width=0.49\textwidth]{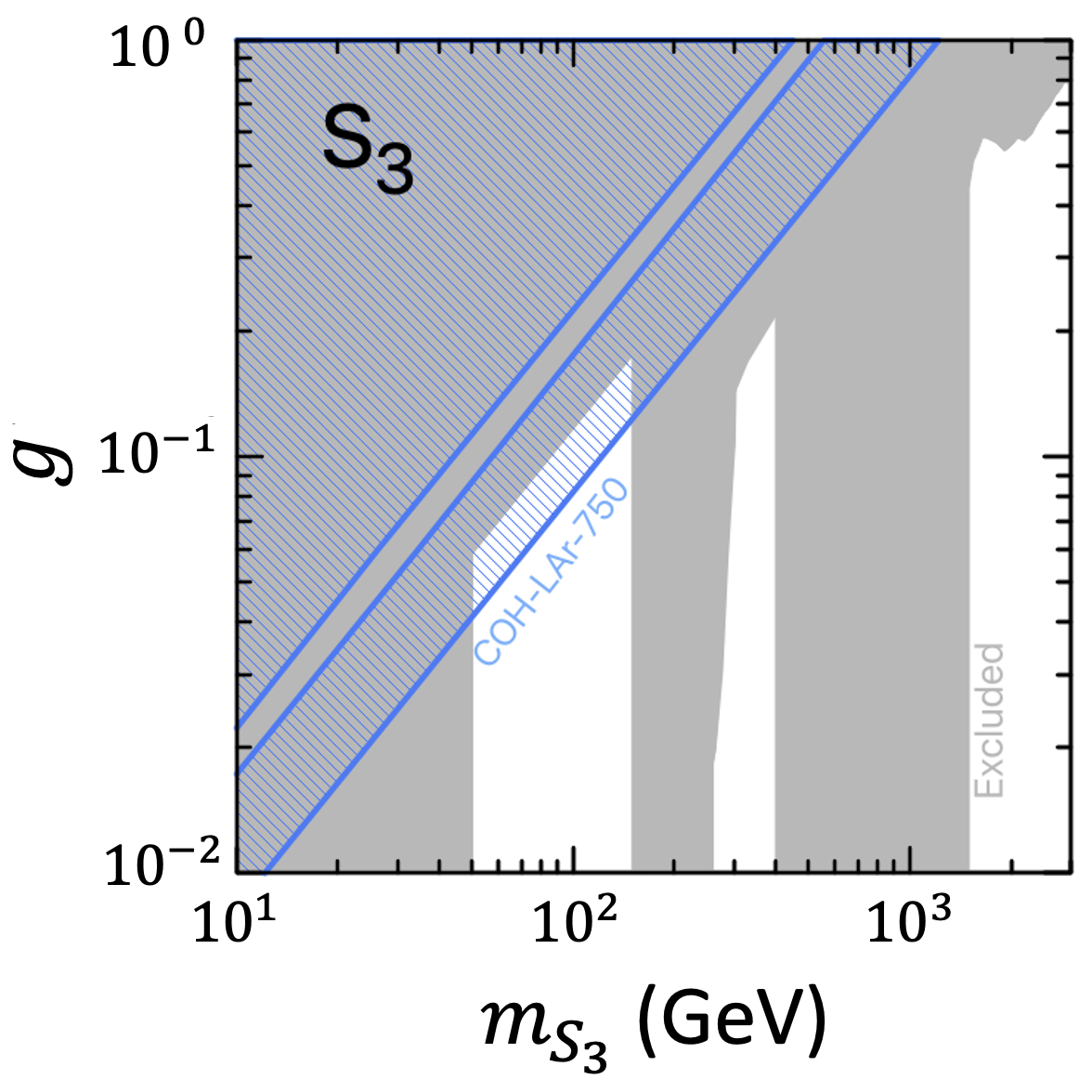}
\caption{Expected $90 \%$ C.L. sensitivities, in the ($m_\textrm{LQ}, g$) plane, obtained for the  COH-CsI-700 (left panel) and  COH-LAr-750 (right) detectors and assuming model $S_3$. The grey-shaded regions refer to current limits previously presented in Fig.~\ref{fig:current:results}. In the case of current \cevns~bounds, only the constraint obtained with the corresponding target is shown. See text for more details. }
\label{fig:future:SNS3}
\end{figure}

The expected sensitivities for these two future COHERENT detectors are shown in Fig.~\ref{fig:future:SNS}, where we give the results for the $S_1$ scenario. The left and right panels in the figure correspond to the COH-CsI-700 and COH-LAr-750 upgrades, respectively. The region within the colored blue lines in each panel indicates the expected excluded values for masses and coupling constants. Similarly to what already observed for current (individual) sensitivites in Fig.~\ref{fig:current:results}, we see the presence of an allowed band within each of the excluded regions, which corresponds  to combinations of LQ masses and parameters that allow for a destructive interference with the SM (see Appendix~\ref{app:int}).  
The grey-shaded regions in each panel indicate current constraints as obtained in Fig.~\ref{fig:current:results} and that include colliders, APV, and current COHERENT data, as discussed in Sec.~\ref{sec:current_constr}. Regarding \cevns~current bounds shown in Fig.~\ref{fig:future:SNS}, 
we show in each case only the constraint obtained assuming the corresponding 
detector and not the (more stringent) combined CsI+LAr result. In such a way we allow for an easier comparison that indicates how much future upgrades are expected to improve upon current detectors.

The corresponding results for scenarios $R_2$ and $S_3$ are shown in Fig.~\ref{fig:future:SNS2} and Fig.~\ref{fig:future:SNS3}, respectively, showing a similar qualitative behaviour to $S_1$. (We recall that model $\tilde{R}_2$ is equivalent to $S_1$ from the \cevns~point of view.)  Overall, we notice that a future LAr detector is expected to enhance current constraints of up to around $50\%$ when compared to the constraints obtained through its (current) predecessor, while CsI will be able to improve by almost one order of magnitude in some regions of the parameter space. 

\subsection{\cevns~data (ESS)}
In addition to the COHERENT program, there are other collaborations aiming at performing new \cevns~measurements. Here we consider the particular case of the ESS, a facility that will be located at Lund, Sweden, and that at full power will become the most intense neutron beam source in the world.  
 The physics potential of the ESS within the context of particle physics is  summarized in Ref.~\cite{Abele:2022iml}. Furthermore, a proposal of measuring \cevns~at the ESS was presented in Ref.~\cite{Baxter:2019mcx}, and  different analyses have explored its sensitivity to new physics, particularly within the context of NSI \cite{Baxter:2019mcx, Chatterjee:2022mmu} and electromagnetic properties of neutrinos \cite{Baxter:2019mcx}. 

Here we explore the sensitivity of the ESS to scalar LQ models by analyzing two of the proposed detection technologies~\cite{Baxter:2019mcx}, namely silicon and xenon, characterized by having a very different ratio of protons to neutrons. In contrast to CsI and LAr, for these detectors we compute the expected number of events simply through Eq.~\eqref{eq:Nevents_CEvNS}, by separating the data in nuclear recoil energy bins as done in~\cite{Chatterjee:2022mmu}.

\begin{figure}
\centering
\includegraphics[width=0.49
\textwidth]{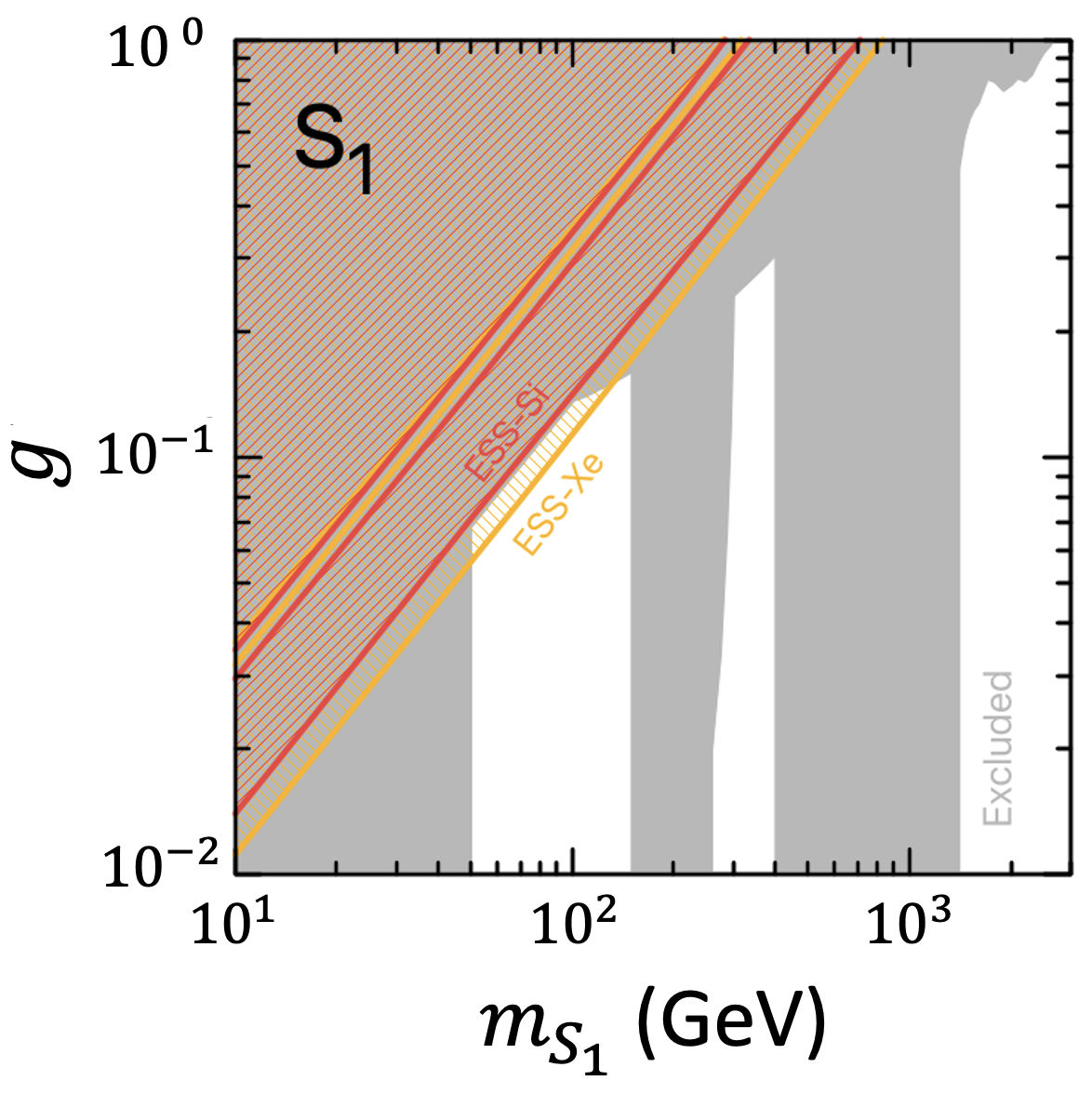}
\includegraphics[width=0.49\textwidth]{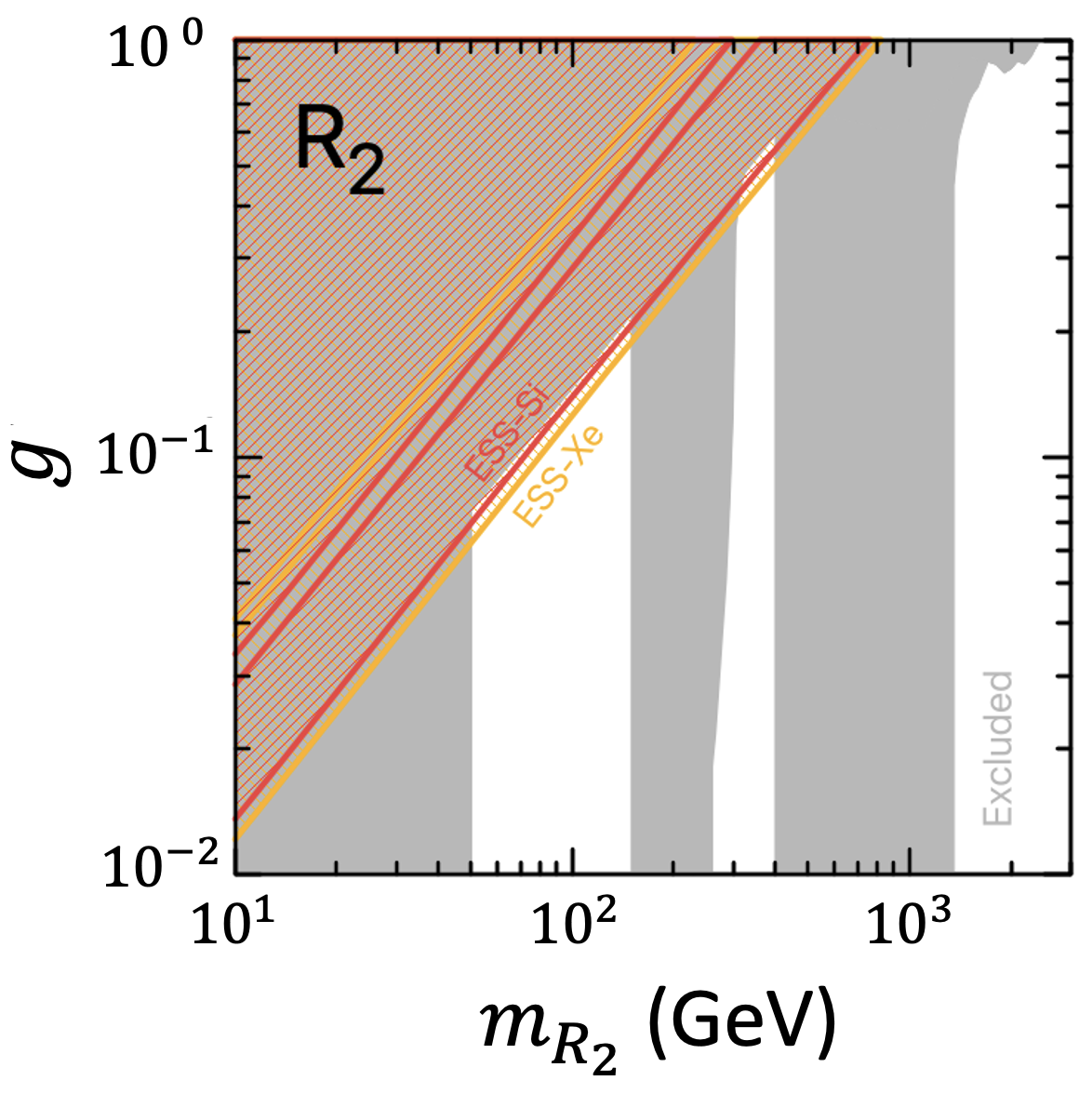}
\includegraphics[width=0.49\textwidth]{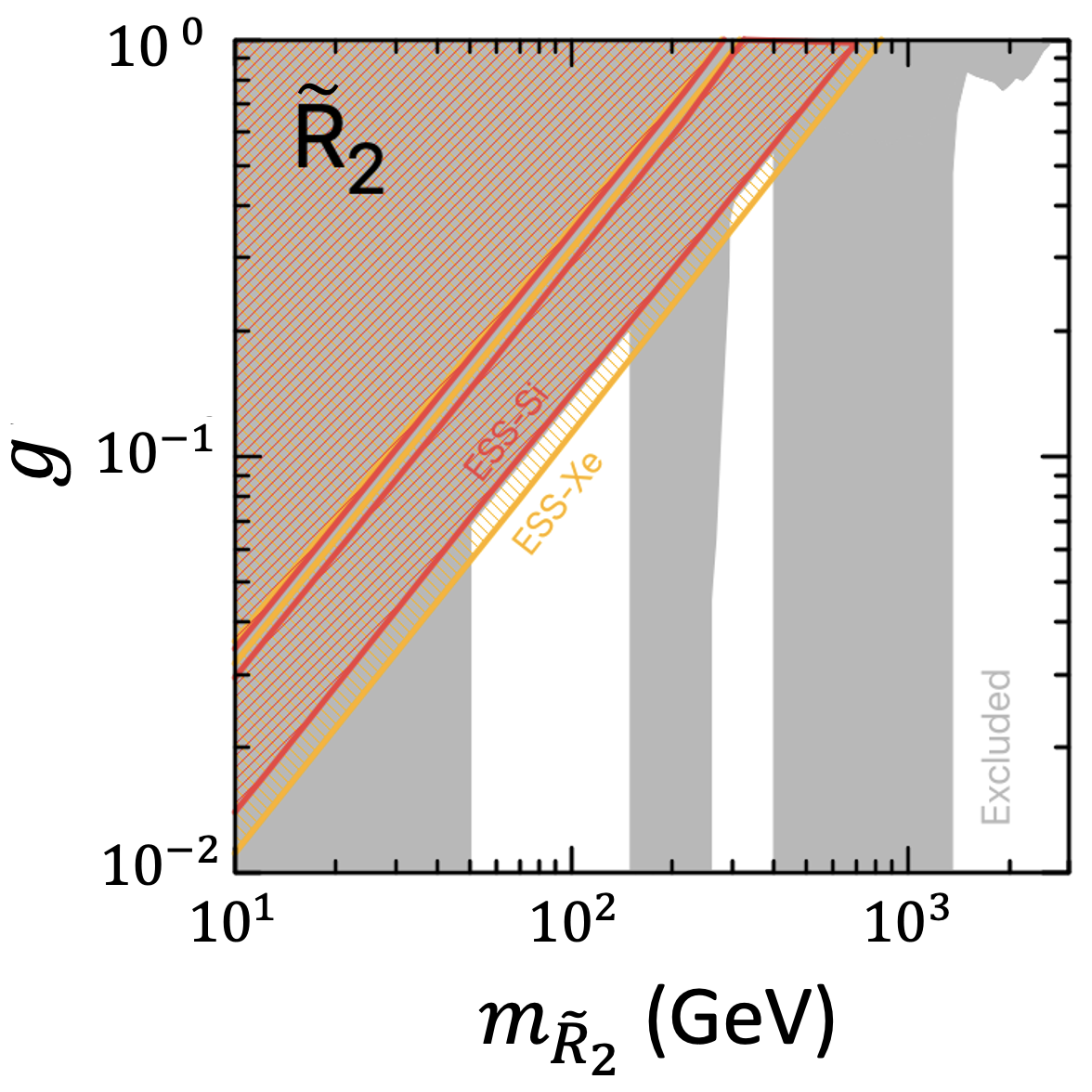}
\includegraphics[width=0.49\textwidth]{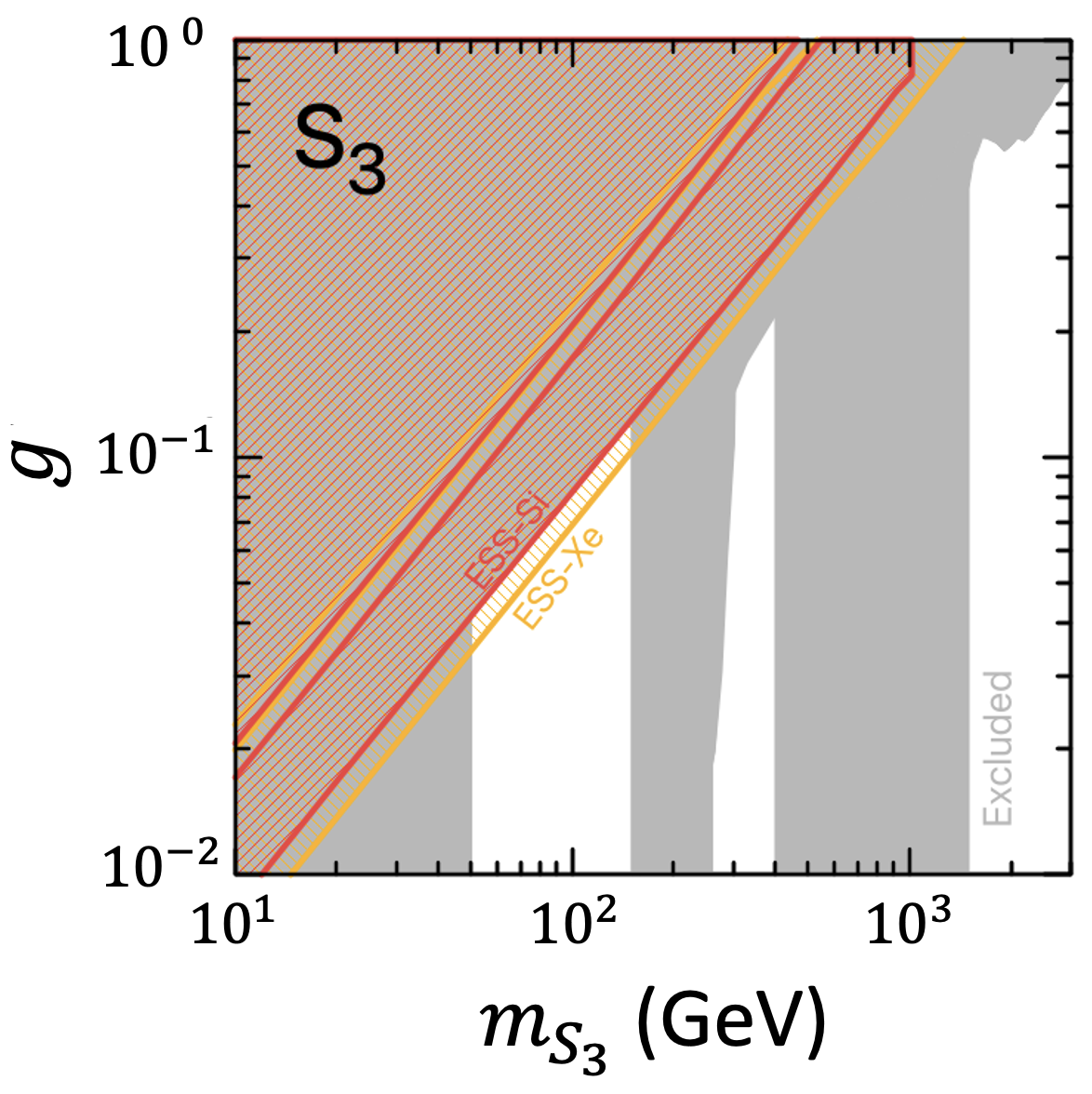}
\caption{Expected $90 \%$ C.L. sensitivities, in the ($m_\textrm{LQ}, g$) plane, obtained for the Si (red contour) and Xe (yellow) detectors at the ESS, assuming different LQ models. The grey-shaded regions refer to current limits previously presented in Fig.~\ref{fig:current:results}. In the case of current \cevns~bounds, we show the combined COHERENT CsI+LAr result. See text for more details. }
\label{fig:future:ESS}
\end{figure}  

Being a spallation source, the total neutrino flux at the ESS will also have the contributions from prompt and delayed neutrinos as given in Eq.~\eqref{labor-nu}. At full capacity, this facility will operate at a beam energy of 2 GeV and a beam power of 5 MW, resulting on an $N_\textrm{POT}$ of $2.8\times10^{23}$ per calendar year of operations ($\approx 5000$ hours), with a number of released neutrinos per flavor of $r = 0.3$.
We assume these values for our analysis. As a result, the ESS will provide larger statistics when compared to current measurements at the SNS, the upshot being a smaller beam frequency pulse that disfavors background discrimination in particular from SSB, expected to be the dominant contribution among all backgrounds. Regarding detectors' characteristics, the considered mass, baseline, and threshold for Xe and Si are given in Table~\ref{tab:CEvNSexps}. In addition,  we follow the procedure proposed in Ref. \cite{Baxter:2019mcx} and for the analysis we consider a Gaussian smearing distribution with a resolution $\sigma = \sigma_0\sqrt{T_\textrm{Th}E_{\textrm{nr}}}$, with $\sigma_0 = 0.40~(0.60)$ for Xe (Si), $T_\textrm{Th}$ being the energy threshold.

Given the absence, at present, of timing information for this proposal, to infer LQ sensitivities for this experiment, we minimize the following Poissionan $\chi^2$ function
\begin{equation}
\label{eq:chi2ESS}
	\chi^2_{\mathrm{ESS}}
	 =
	2
	\sum_{i=1}
	\left[ N^\mathrm{th}_{i}  -  N_{i}^{\text{exp}} 
	 +  N_{i}^{\text{exp}} \ln\left(\frac{N_{i}^{\text{exp}}}{ N^\mathrm{th}_{i}} \right)\right]\\
	+ 
	\left(
	\dfrac{ \alpha }{ \sigma_\alpha }
	\right)^2 
 + 
	\left(
	\dfrac{ \beta }{ \sigma_{\beta} }
	\right)^2,
\end{equation}
where the index $i$ runs over the recoil energy bins. The predicted number of events in this case is given by 
\begin{equation}
    N^{\textrm{th}}_i = (1+\alpha) N^{\textrm{\cevns}}_i +  (1+\beta)N^{\textrm{SSB}}_i ,
\end{equation}
where $N_i^{\textrm{\cevns}}$ stands for the expected number of \cevns~events as a function of the LQ model parameters under study, and $N_i^{\textrm{bckg}}$ is the number of background events. As discussed above, the large pulse shape at the ESS makes it more difficult to discriminate background events from SSB contributions. Being this the dominant background component, we model it as expected counts per keV per kilogram per day (ckkd), as also done in Ref. \cite{Baxter:2019mcx}, where it is assumed a value of 10 ckkd (1~ckkd) for Xe (Si). Going back to Eq.~\eqref{eq:chi2ESS}, $N_i^{\textrm{exp}}$ is the experimental number of events, which we assume as the SM prediction. To perform the analysis, the $\chi^2$ function in Eq.~\eqref{eq:chi2ESS} is minimized with respect to the nuisance parameters $\alpha$ and $\beta$, which are associated to the predicted \cevns~and background events, respectively, each with its corresponding uncertainty taken as $\sigma_\alpha = 10\%$ and
$\sigma_\beta = 1\%$~\cite{Baxter:2019mcx}.

The expected sensitivities for the described ESS detectors, at 90\% C.L., are shown in Fig.~\ref{fig:future:ESS} for all LQ models. Colored lines in the figure represent the exclusion regions for Si (red) and Xe (yellow). Notice again the presence of an allowed band within the region, whose position and width depend not only on the specific LQ scenario considered, but also on the ratio of protons to neutrons of the target material (see Appendix~\ref{app:int}). The grey-shaded regions correspond to current excluded limits from colliders, DIS, APV, and the \cevns~bounds obtained by the combination of COHERENT CsI and LAr detectors. Given the larger nuclear mass, we see that, for all models, better sensitivies are expected for Xe when compared to Si, getting a better improvement for the $S_1$ and $S_3$ cases when compared to current bounds.

\section{Conclusions}
\label{sec:concl}

In this work we have explored the potential of \cevns~in probing scalar leptoquarks. We have considered four different models, each of them giving rise to a different contribution to the weak charge. First we have analyzed current COHERENT data, from the CsI (2021) and the LAr (2020) detectors. By means of a detailed statistical analysis, which took into account timing information and all experimental uncertainties, we obtained stringent constraints on the LQ mass and couplings. We further obtained upper limits on the LQ parameter space from atomic parity violation experiments, which turned out to be comparable (although slightly less stringent) to COHERENT bounds. Next, we have obtained bounds on the same LQ models from LHC data, considering different processes and production mechanisms: single production, double
production and Drell Yan. These strong collider bounds lead to an exclusion region in the mass range $0.4 \lesssim m_\textrm{LQ} \lesssim 1.5$ TeV, independent of the LQ coupling. To complete the picture on the LQ parameter space we have also recast bounds from deep inelastic neutrino-nucleon scattering (NuTeV) and older colliders (HERA, SPS, LEP and Tevatron). Among them, UA2 at SPS and OPAL at LEP set strong constraints on $m_\text{LQ} \lesssim 50$ GeV, while HERA and Tevatron disfavor a thin region around 0.2 TeV.
However, 
we have identified two regions in parameter space where \cevns~data may improve upon existing constraints and provide a complementary probe, at $50~\text{GeV} \lesssim m_\text{LQ} \lesssim 150$ GeV and $300~\text{GeV} \lesssim  m_\text{LQ} \lesssim 400$ GeV,  depending on the LQ scenario.

Additionally, we have computed sensitivities at future upgrades of the COHERENT CsI and LAr detectors, and at the European Spallation Source. We have found that these future facilities, thanks to their larger exposures and exquisitely low thresholds, will allow to improve upon current bounds by up to  a factor 3 at $m_\textrm{LQ} \sim 100$ GeV.
Let us mention that we analyzed \cevns~data using $\pi-$DAR neutrinos motivated by current available measurements, however there is a vast array of experiments using reactor neutrinos \cite{CONNIE:2021ggh,CONUS:2021dwh,nuGeN:2022bmg,MINER:2016igy,Billard:2016giu,Strauss:2017cuu,Wong:2015kgl,Fernandez-Moroni:2020yyl,Akimov:2022xvr,NEON:2022hbk,NEWS-G:2021mhf,SBC:2021yal,Colaresi:2022obx} that can also provide valuable information on LQ scenarios and are therefore worth studying in a future work.

As a last remark, it is important to note that the present Run 3 of LHC will soon allow to explore the LQ parameter space involving large masses. With increasing luminosity, searches like double pair production will be able to set coupling-independent limits on LQ masses that lie in the TeV range. Furthermore, single LQ production and Drell Yan will be able to cover even higher masses, imposing constraints on couplings of the order of $\mathcal{O}(10^{-1})$. In addition, there are new searches in the literature that are specific for LQ signatures and that are not yet exploited by the experiments. One example is the single-lepton channel initiated using the lepton content in the proton \cite{Dreiner:2021ext}. This search is more sensitive than others listed before, in particular being more powerful than Drell Yan up to LQ masses $m_\textrm{LQ}\sim$ 4 TeV, so it could be decisive in the search for high mass LQs.

\section*{Acknowledgments}
This work has been supported by the Spanish grants PID2020-113775GB-I00 (MCIN/AEI/ 10.13039/501100011033) and CIPROM/2021/054 (Generalitat Valenciana).
V.D.R. acknowledges financial support by the CIDEXG/2022/20 grant (project ``D'AMAGAT'') funded by
Generalitat Valenciana. 
V.M.L.\ acknowledges the financial support by Ministerio de Universidades and “European Union-NextGenerationEU/PRTR” under the grant María Zambrano UP2021-044 (ZA2021-081). G.S.G. acknowledges financial support by the CIAPOS/2022/254 grant funded by
Generalitat Valenciana.

\noindent

\appendix
 \section{Fierz transformations}
\label{app:fierz}

Starting from a scalar interaction, we can apply a Fierz transformation to obtain the four fermion operator~\cite{Husek:2021isa}
 \begin{eqnarray}
     (\bar{a}P_Lb)(\bar{c}P_Rd)=-\frac{1}{2}(\bar{a}\gamma^\mu P_R d)(\bar{c}\gamma_\mu P_L b),
 \end{eqnarray}
that reads as a vector interaction.

For the case of a vector interaction, if we apply a Fierz transformation we obtain
 \begin{eqnarray}
     (\bar{a}\gamma^\mu P_{L,R}b)(\bar{c}\gamma_\mu P_{L,R}d)=(\bar{a}\gamma^\mu P_{L,R} d)(\bar{c}\gamma_\mu P_{L,R} b),
 \end{eqnarray}
that is the same as we had for the scalar interaction after the Fierz transformation. For this reason scalar and vector LQs would give rise to the same results in terms of the \cevns~cross section.

\section{Interference with the SM in \cevns}
\label{app:int}

\begin{figure}
\centering
\includegraphics[width=0.45\textwidth]
{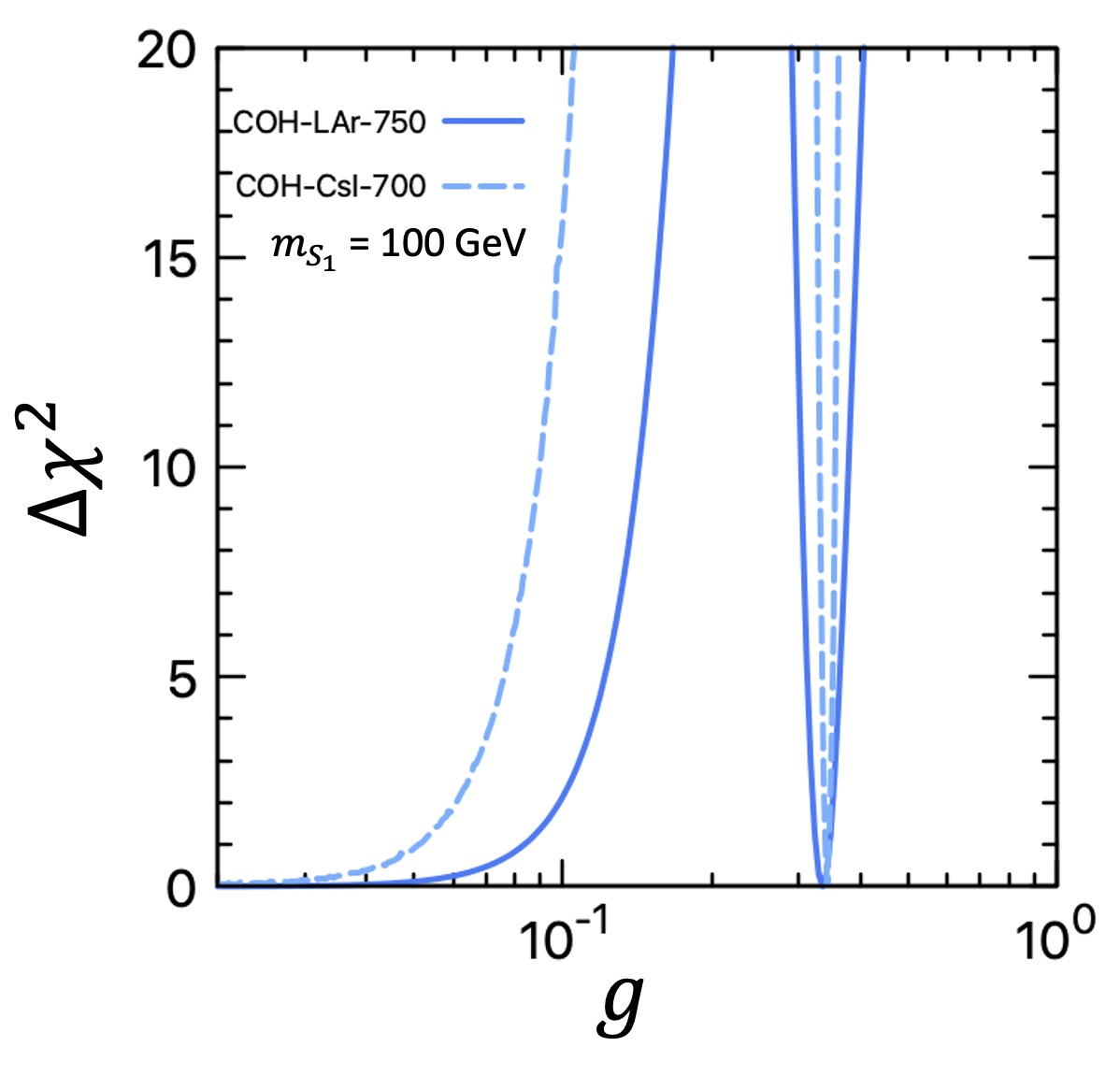}
\includegraphics[width=0.45\textwidth]
{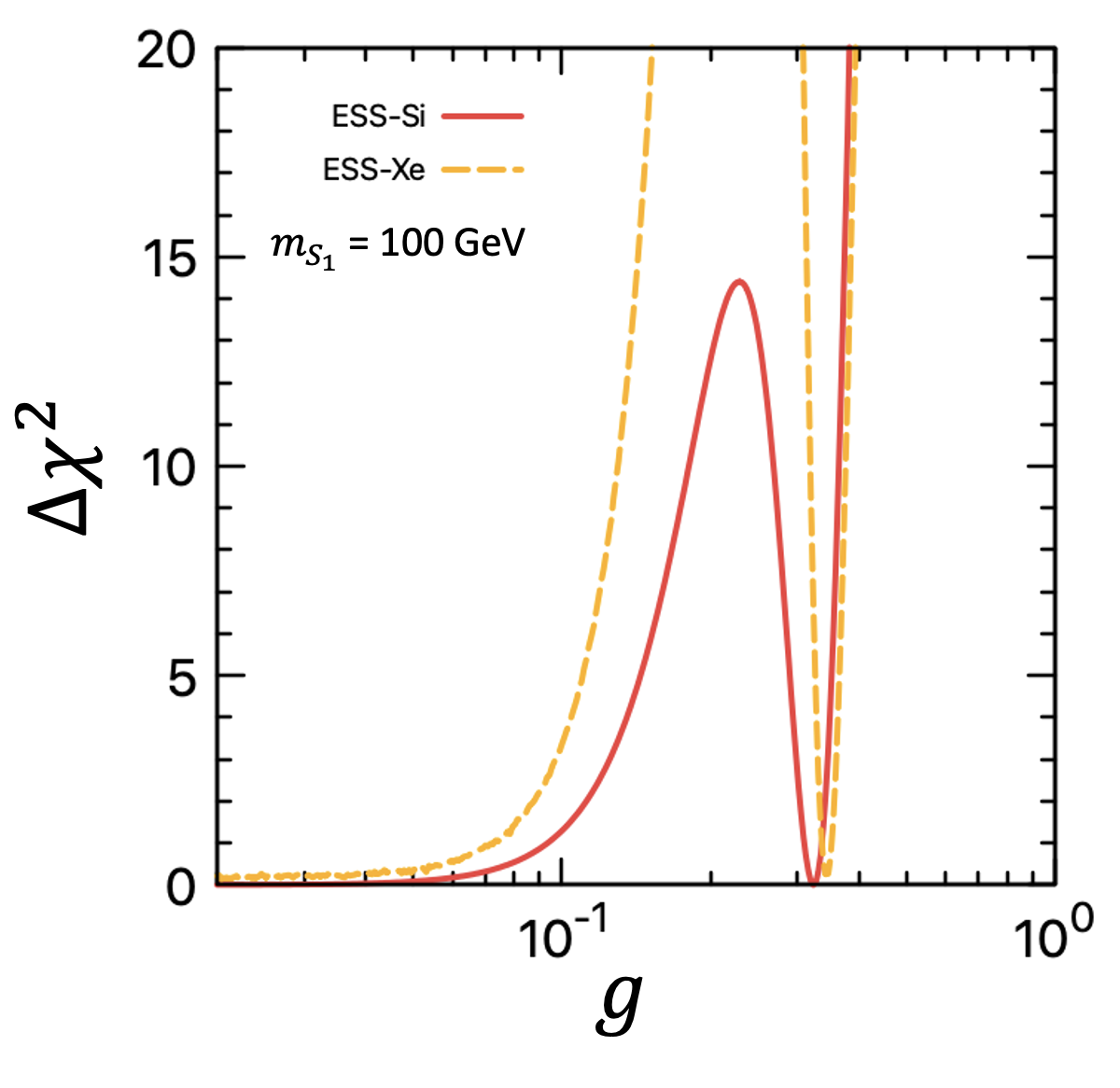}
\caption{Left panel: Reduced $\chi^2$ profile as a function of the LQ coupling $g$ for two future detectors proposed at COHERENT, CsI-700 (light blue, dashed) and LAr-750 (blue). We fix $m_\textrm{LQ} = 100~$GeV and we consider the $S_1$ model.
Right panel: Same as left panel but for two future detectors proposed at the ESS, Si (red) and Xe (yellow, dashed).}
\label{fig:chi:profile}
\end{figure}

Regardless of the LQ scenario under study, we have seen that its impact on the \cevns~cross section results on a shift of the SM weak charge, ending up with
\begin{equation}
Q_{i,\textrm{LQ}}^2 = \left ( Q_W^{\textrm{SM}} + Q_{ii,\textrm{LQ}} \right )^2 + \sum_{j\neq i} Q_{ij,\textrm{LQ}}^2.
\end{equation}
This shift gives rise to an interesting feature through which, within the excluded regions obtained for the different analyses, there is always an allowed band for each detector. This can be seen, for instance, from the color contours in Figs.~\ref{fig:future:SNS} to \ref{fig:future:ESS}. This particular behaviour can be easily explained for the scenarios studied in this work where we assume $Q_{ii,\textrm{LQ}} = Q_{ij,\textrm{LQ}}$, leading in general  to
\begin{equation}
Q_{ii,\textrm{LQ}} = Q_{ij,\textrm{LQ}}= \frac{g^2}{4\sqrt{2}G_F}\frac{\mathcal{C}_1 Z+ \mathcal{C}_2 N}{q^2+ m_{\textrm{LQ}}^2} \, ,
\label{eq:app:LQ}
\end{equation}
with $\mathcal{C}_1$ and $\mathcal{C}_2 \in \mathbb{N}$. 
The shape of the weak charge in Eq.~\eqref{eq:app:LQ} allows for different pairs of $g$ and $m_\textrm{LQ}$ to reproduce the SM cross section, and hence the SM prediction for the number of events, giving as a result a degeneracy in the parameter space, which corresponds to an allowed band. 
For instance, in the case of the future sensitivities studied in Sec.~\ref{sec:future_sens},  if we want to reproduce the SM number of events (for which $\Delta\chi^2$ is minimum), under the assumption in Eq.~\eqref{eq:app:LQ}, we need
\begin{equation}
    (Q_W^{\textrm{SM}}+Q_{ii,\textrm{LQ}})^2 + Q_{ii,\textrm{LQ}}^2 = (Q_W^{\textrm{SM}})^2 \, , 
\end{equation}
which can be satisfied when $ Q_{ii,\textrm{LQ}} = -Q_W^{\textrm{SM}}$ or, in other words, when the parameters $g$ and $m_\textrm{LQ}$ are such that 
\begin{equation}
    g^2 = \frac{4\sqrt{2}G_F\left ( g_V^pZ + g_V^nN \right )\left ( 2m_NE_{\textrm{nr}} + m_\textrm{LQ}^2 \right )}{\mathcal{C}_1 Z + \mathcal{C}_2 N} \, .
\end{equation}
Notice that for $m_\textrm{LQ} \gg 2m_NE_{\textrm{nr}}$ the needed value of $g^2$ is energy-independent and we have  
\begin{equation}
    g^2 \approx \frac{4\sqrt{2}G_F\left ( g_V^pZ + g_V^nN \right ) m_\textrm{LQ}^2 }{\mathcal{C}_1 Z + \mathcal{C}_2 N} \, .
    \label{eq:simp:g}
\end{equation}
For SNS and ESS neutrinos, the condition $m_\textrm{LQ}^2 \gg 2m_NE_{\textrm{nr}}$ can be easily satisfied with a relatively large LQ mass, $m_\textrm{LQ}$. For instance, let us consider the $S_1$ LQ scenario, for which $ \mathcal{C}_1 = 1$ and $ \mathcal{C}_2 = 2$. Then, regardless of the target material, for a mass of $m_{\textrm{LQ}} = 100$ GeV, we have $2m_NE_{\textrm{nr}}^{\textrm{max}}/m_\textrm{LQ}^2 \approx 1\times 10^{-6}$ and Eq.~\eqref{eq:simp:g} safely applies. Then, in this case it is possible to find a solution in the parameter space that reproduces the SM solution and the $\chi^2$ reaches a minimum, as we can see in the two panels of Fig.~\ref{fig:chi:profile}, where we show the $\Delta\chi^2$ profile as a function of $g$ for $m_{\textrm{LQ}} = 100$ GeV. The profiles are shown for the COH-LAr-750 and COH-CsI-700 detectors in the left panel, and for ESS-Si and ESS-Xe in the right panel. These results are consistent with the allowed bands observed in Figs.~\ref{fig:future:SNS} and  \ref{fig:future:ESS} at $m_{\textrm{LQ}} = 100$ GeV. The situation would be different for low LQ masses, when the terms $2m_NE_{\textrm{nr}}$ and $ m_\textrm{LQ}^2$ are comparable and Eq.~\eqref{eq:simp:g} does not hold. However, we are not interested in such low masses given the LEP and UA2 constraints shown in Fig.~\ref{fig:current:results}.

\bibliographystyle{utphys}
\bibliography{bibliography}  

\providecommand{\href}[2]{#2}\begingroup\raggedright\begin{thebibliography}{100}

\bibitem{Pati:1973uk}
J.~C. Pati and A.~Salam, ``{Unified Lepton-Hadron Symmetry and a Gauge Theory
  of the Basic Interactions},''
  \href{http://dx.doi.org/10.1103/PhysRevD.8.1240}{{\em Phys. Rev. D}
  {\bfseries 8} (1973) 1240--1251}.

\bibitem{Georgi:1974sy}
H.~Georgi and S.~L. Glashow, ``{Unity of All Elementary Particle Forces},''
  \href{http://dx.doi.org/10.1103/PhysRevLett.32.438}{{\em Phys. Rev. Lett.}
  {\bfseries 32} (1974) 438--441}.

\bibitem{Georgi:1974my}
H.~Georgi, ``{The State of the Art\textemdash{}Gauge Theories},''
  \href{http://dx.doi.org/10.1063/1.2947450}{{\em AIP Conf. Proc.} {\bfseries
  23} (1975) 575--582}.

\bibitem{Fritzsch:1974nn}
H.~Fritzsch and P.~Minkowski, ``{Unified Interactions of Leptons and
  Hadrons},'' \href{http://dx.doi.org/10.1016/0003-4916(75)90211-0}{{\em Annals
  Phys.} {\bfseries 93} (1975) 193--266}.

\bibitem{Nath:2006ut}
P.~Nath and P.~Fileviez~Perez, ``{Proton stability in grand unified theories,
  in strings and in branes},''
  \href{http://dx.doi.org/10.1016/j.physrep.2007.02.010}{{\em Phys. Rept.}
  {\bfseries 441} (2007) 191--317},
  \href{http://arxiv.org/abs/hep-ph/0601023}{{\ttfamily arXiv:hep-ph/0601023}}.

\bibitem{Buchmuller:1986zs}
W.~Buchmuller, R.~Ruckl, and D.~Wyler, ``{Leptoquarks in Lepton - Quark
  Collisions},'' \href{http://dx.doi.org/10.1016/0370-2693(87)90637-X}{{\em
  Phys. Lett. B} {\bfseries 191} (1987) 442--448}. [Erratum: Phys.Lett.B 448,
  320--320 (1999)].

\bibitem{Belyaev:2005ew}
A.~Belyaev, C.~Leroy, R.~Mehdiyev, and A.~Pukhov, ``{Leptoquark single and pair
  production at LHC with CalcHEP/CompHEP in the complete model},''
  \href{http://dx.doi.org/10.1088/1126-6708/2005/09/005}{{\em JHEP} {\bfseries
  09} (2005) 005}, \href{http://arxiv.org/abs/hep-ph/0502067}{{\ttfamily
  arXiv:hep-ph/0502067}}.

\bibitem{Dorsner:2005fq}
I.~Dorsner and P.~Fileviez~Perez, ``{Unification without supersymmetry:
  Neutrino mass, proton decay and light leptoquarks},''
  \href{http://dx.doi.org/10.1016/j.nuclphysb.2005.06.016}{{\em Nucl. Phys. B}
  {\bfseries 723} (2005) 53--76},
  \href{http://arxiv.org/abs/hep-ph/0504276}{{\ttfamily arXiv:hep-ph/0504276}}.

\bibitem{Arnold:2013cva}
J.~M. Arnold, B.~Fornal, and M.~B. Wise, ``{Phenomenology of scalar
  leptoquarks},'' \href{http://dx.doi.org/10.1103/PhysRevD.88.035009}{{\em
  Phys. Rev. D} {\bfseries 88} (2013) 035009},
  \href{http://arxiv.org/abs/1304.6119}{{\ttfamily arXiv:1304.6119 [hep-ph]}}.

\bibitem{FileviezPerez:2013zmv}
P.~Fileviez~Perez and M.~B. Wise, ``{Low Scale Quark-Lepton Unification},''
  \href{http://dx.doi.org/10.1103/PhysRevD.88.057703}{{\em Phys. Rev. D}
  {\bfseries 88} (2013) 057703},
  \href{http://arxiv.org/abs/1307.6213}{{\ttfamily arXiv:1307.6213 [hep-ph]}}.

\bibitem{Dorsner:2016wpm}
I.~Dor\v{s}ner, S.~Fajfer, A.~Greljo, J.~F. Kamenik, and N.~Ko\v{s}nik,
  ``{Physics of leptoquarks in precision experiments and at particle
  colliders},'' \href{http://dx.doi.org/10.1016/j.physrep.2016.06.001}{{\em
  Phys. Rept.} {\bfseries 641} (2016) 1--68},
  \href{http://arxiv.org/abs/1603.04993}{{\ttfamily arXiv:1603.04993
  [hep-ph]}}.

\bibitem{Tanaka:2012nw}
M.~Tanaka and R.~Watanabe, ``{New physics in the weak interaction of $\bar B\to
  D^{(*)}\tau\bar\nu$},''
  \href{http://dx.doi.org/10.1103/PhysRevD.87.034028}{{\em Phys. Rev. D}
  {\bfseries 87} no.~3, (2013) 034028},
  \href{http://arxiv.org/abs/1212.1878}{{\ttfamily arXiv:1212.1878 [hep-ph]}}.

\bibitem{Sakaki:2013bfa}
Y.~Sakaki, M.~Tanaka, A.~Tayduganov, and R.~Watanabe, ``{Testing leptoquark
  models in $\bar B \to D^{(*)} \tau \bar\nu$},''
  \href{http://dx.doi.org/10.1103/PhysRevD.88.094012}{{\em Phys. Rev. D}
  {\bfseries 88} no.~9, (2013) 094012},
  \href{http://arxiv.org/abs/1309.0301}{{\ttfamily arXiv:1309.0301 [hep-ph]}}.

\bibitem{Dorsner:2013tla}
I.~Dor\v{s}ner, S.~Fajfer, N.~Ko\v{s}nik, and I.~Ni\v{s}and\v{z}i\'c,
  ``{Minimally flavored colored scalar in $\bar B \to D^{(*)} \tau \bar \nu$
  and the mass matrices constraints},''
  \href{http://dx.doi.org/10.1007/JHEP11(2013)084}{{\em JHEP} {\bfseries 11}
  (2013) 084}, \href{http://arxiv.org/abs/1306.6493}{{\ttfamily arXiv:1306.6493
  [hep-ph]}}.

\bibitem{Gripaios:2014tna}
B.~Gripaios, M.~Nardecchia, and S.~A. Renner, ``{Composite leptoquarks and
  anomalies in $B$-meson decays},''
  \href{http://dx.doi.org/10.1007/JHEP05(2015)006}{{\em JHEP} {\bfseries 05}
  (2015) 006}, \href{http://arxiv.org/abs/1412.1791}{{\ttfamily arXiv:1412.1791
  [hep-ph]}}.

\bibitem{Bauer:2015knc}
M.~Bauer and M.~Neubert, ``{Minimal Leptoquark Explanation for the
  $R_{D^{(*)}}$ , $R_K$ , and $(g-2)_\mu$ Anomalies},''
  \href{http://dx.doi.org/10.1103/PhysRevLett.116.141802}{{\em Phys. Rev.
  Lett.} {\bfseries 116} no.~14, (2016) 141802},
  \href{http://arxiv.org/abs/1511.01900}{{\ttfamily arXiv:1511.01900
  [hep-ph]}}.

\bibitem{Buttazzo:2017ixm}
D.~Buttazzo, A.~Greljo, G.~Isidori, and D.~Marzocca, ``{B-physics anomalies: a
  guide to combined explanations},''
  \href{http://dx.doi.org/10.1007/JHEP11(2017)044}{{\em JHEP} {\bfseries 11}
  (2017) 044}, \href{http://arxiv.org/abs/1706.07808}{{\ttfamily
  arXiv:1706.07808 [hep-ph]}}.

\bibitem{Dorsner:2018ynv}
I.~Dor\v{s}ner and A.~Greljo, ``{Leptoquark toolbox for precision collider
  studies},'' \href{http://dx.doi.org/10.1007/JHEP05(2018)126}{{\em JHEP}
  {\bfseries 05} (2018) 126}, \href{http://arxiv.org/abs/1801.07641}{{\ttfamily
  arXiv:1801.07641 [hep-ph]}}.

\bibitem{Hiller:2014yaa}
G.~Hiller and M.~Schmaltz, ``{$R_K$ and future $b \to s \ell \ell$ physics
  beyond the standard model opportunities},''
  \href{http://dx.doi.org/10.1103/PhysRevD.90.054014}{{\em Phys. Rev. D}
  {\bfseries 90} (2014) 054014},
  \href{http://arxiv.org/abs/1408.1627}{{\ttfamily arXiv:1408.1627 [hep-ph]}}.

\bibitem{ColuccioLeskow:2016dox}
E.~Coluccio~Leskow, G.~D'Ambrosio, A.~Crivellin, and D.~M\"uller,
  ``{$(g-2)\mu$, lepton flavor violation, and $Z$ decays with leptoquarks:
  Correlations and future prospects},''
  \href{http://dx.doi.org/10.1103/PhysRevD.95.055018}{{\em Phys. Rev. D}
  {\bfseries 95} no.~5, (2017) 055018},
  \href{http://arxiv.org/abs/1612.06858}{{\ttfamily arXiv:1612.06858
  [hep-ph]}}.

\bibitem{Crivellin:2017zlb}
A.~Crivellin, D.~M\"uller, and T.~Ota, ``{Simultaneous explanation of
  $R_{D^{(*)}}$ and $b\to s \mu^+ \mu^-$: the last scalar leptoquarks
  standing},'' \href{http://dx.doi.org/10.1007/JHEP09(2017)040}{{\em JHEP}
  {\bfseries 09} (2017) 040}, \href{http://arxiv.org/abs/1703.09226}{{\ttfamily
  arXiv:1703.09226 [hep-ph]}}.

\bibitem{Hiller:2018wbv}
G.~Hiller, D.~Loose, and I.~Ni\v{s}and\v{z}i\'c, ``{Flavorful leptoquarks at
  hadron colliders},'' \href{http://dx.doi.org/10.1103/PhysRevD.97.075004}{{\em
  Phys. Rev. D} {\bfseries 97} no.~7, (2018) 075004},
  \href{http://arxiv.org/abs/1801.09399}{{\ttfamily arXiv:1801.09399
  [hep-ph]}}.

\bibitem{Crivellin:2020tsz}
A.~Crivellin, D.~Mueller, and F.~Saturnino, ``{Correlating $h\to\mu^+\mu^-$ to
  the Anomalous Magnetic Moment of the Muon via Leptoquarks},''
  \href{http://dx.doi.org/10.1103/PhysRevLett.127.021801}{{\em Phys. Rev.
  Lett.} {\bfseries 127} no.~2, (2021) 021801},
  \href{http://arxiv.org/abs/2008.02643}{{\ttfamily arXiv:2008.02643
  [hep-ph]}}.

\bibitem{Angelescu:2021lln}
A.~Angelescu, D.~Be\v{c}irevi\'c, D.~A. Faroughy, F.~Jaffredo, and
  O.~Sumensari, ``{Single leptoquark solutions to the B-physics anomalies},''
  \href{http://dx.doi.org/10.1103/PhysRevD.104.055017}{{\em Phys. Rev. D}
  {\bfseries 104} no.~5, (2021) 055017},
  \href{http://arxiv.org/abs/2103.12504}{{\ttfamily arXiv:2103.12504
  [hep-ph]}}.

\bibitem{Nomura:2021oeu}
T.~Nomura and H.~Okada, ``{Explanations for anomalies of muon anomalous
  magnetic dipole moment, $b\to s\mu^+\mu^-$, and radiative neutrino masses in
  a leptoquark model},''
  \href{http://dx.doi.org/10.1103/PhysRevD.104.035042}{{\em Phys. Rev. D}
  {\bfseries 104} no.~3, (2021) 035042},
  \href{http://arxiv.org/abs/2104.03248}{{\ttfamily arXiv:2104.03248
  [hep-ph]}}.

\bibitem{Marzocca:2021azj}
D.~Marzocca and S.~Trifinopoulos, ``{Minimal Explanation of Flavor Anomalies:
  B-Meson Decays, Muon Magnetic Moment, and the Cabibbo Angle},''
  \href{http://dx.doi.org/10.1103/PhysRevLett.127.061803}{{\em Phys. Rev.
  Lett.} {\bfseries 127} no.~6, (2021) 061803},
  \href{http://arxiv.org/abs/2104.05730}{{\ttfamily arXiv:2104.05730
  [hep-ph]}}.

\bibitem{FileviezPerez:2021lkq}
P.~Fileviez~Perez, C.~Murgui, and A.~D. Plascencia, ``{Leptoquarks and matter
  unification: Flavor anomalies and the muon g-2},''
  \href{http://dx.doi.org/10.1103/PhysRevD.104.035041}{{\em Phys. Rev. D}
  {\bfseries 104} no.~3, (2021) 035041},
  \href{http://arxiv.org/abs/2104.11229}{{\ttfamily arXiv:2104.11229
  [hep-ph]}}.

\bibitem{Murgui:2021bdy}
C.~Murgui and M.~B. Wise, ``{Scalar leptoquarks, baryon number violation, and
  Pati-Salam symmetry},''
  \href{http://dx.doi.org/10.1103/PhysRevD.104.035017}{{\em Phys. Rev. D}
  {\bfseries 104} no.~3, (2021) 035017},
  \href{http://arxiv.org/abs/2105.14029}{{\ttfamily arXiv:2105.14029
  [hep-ph]}}.

\bibitem{Singirala:2021gok}
S.~Singirala, S.~Sahoo, and R.~Mohanta, ``{Light dark matter, rare B decays
  with missing energy in L\ensuremath{\mu}-L\ensuremath{\tau} model with a
  scalar leptoquark},''
  \href{http://dx.doi.org/10.1103/PhysRevD.105.015033}{{\em Phys. Rev. D}
  {\bfseries 105} no.~1, (2022) 015033},
  \href{http://arxiv.org/abs/2106.03735}{{\ttfamily arXiv:2106.03735
  [hep-ph]}}.

\bibitem{Crivellin:2022mff}
A.~Crivellin, B.~Fuks, and L.~Schnell, ``{Explaining the hints for lepton
  flavour universality violation with three S$_{2}$ leptoquark generations},''
  \href{http://dx.doi.org/10.1007/JHEP06(2022)169}{{\em JHEP} {\bfseries 06}
  (2022) 169}, \href{http://arxiv.org/abs/2203.10111}{{\ttfamily
  arXiv:2203.10111 [hep-ph]}}.

\bibitem{Abdullah:2022zue}
M.~Abdullah {\em et~al.}, ``{Coherent elastic neutrino-nucleus scattering:
  Terrestrial and astrophysical applications},''
  \href{http://arxiv.org/abs/2203.07361}{{\ttfamily arXiv:2203.07361
  [hep-ph]}}.

\bibitem{COHERENT:2017ipa}
{\bfseries COHERENT} Collaboration, D.~Akimov {\em et~al.}, ``{Observation of
  Coherent Elastic Neutrino-Nucleus Scattering},''
  \href{http://dx.doi.org/10.1126/science.aao0990}{{\em Science} {\bfseries
  357} (2017) 1123--1126}, \href{http://arxiv.org/abs/1708.01294}{{\ttfamily
  arXiv:1708.01294 [nucl-ex]}}.

\bibitem{Freedman:1973yd}
D.~Z. Freedman, ``{Coherent Neutrino Nucleus Scattering as a Probe of the Weak
  Neutral Current},'' \href{http://dx.doi.org/10.1103/PhysRevD.9.1389}{{\em
  Phys. Rev. D} {\bfseries 9} (1974) 1389--1392}.

\bibitem{COHERENT:2020iec}
{\bfseries COHERENT} Collaboration, D.~Akimov {\em et~al.}, ``{First
  Measurement of Coherent Elastic Neutrino-Nucleus Scattering on Argon},''
  \href{http://dx.doi.org/10.1103/PhysRevLett.126.012002}{{\em Phys. Rev.
  Lett.} {\bfseries 126} no.~1, (2021) 012002},
  \href{http://arxiv.org/abs/2003.10630}{{\ttfamily arXiv:2003.10630
  [nucl-ex]}}.

\bibitem{COHERENT:2021xmm}
{\bfseries COHERENT} Collaboration, D.~Akimov {\em et~al.}, ``{Measurement of
  the Coherent Elastic Neutrino-Nucleus Scattering Cross Section on CsI by
  COHERENT},'' \href{http://dx.doi.org/10.1103/PhysRevLett.129.081801}{{\em
  Phys. Rev. Lett.} {\bfseries 129} (2022) 081801},
  \href{http://arxiv.org/abs/2110.07730}{{\ttfamily arXiv:2110.07730
  [hep-ex]}}.

\bibitem{CONNIE:2021ggh}
{\bfseries CONNIE} Collaboration, A.~Aguilar-Arevalo {\em et~al.}, ``{Search
  for coherent elastic neutrino-nucleus scattering at a nuclear reactor with
  CONNIE 2019 data},'' \href{http://dx.doi.org/10.1007/JHEP05(2022)017}{{\em
  JHEP} {\bfseries 05} (2022) 017},
  \href{http://arxiv.org/abs/2110.13033}{{\ttfamily arXiv:2110.13033
  [hep-ex]}}.

\bibitem{CONUS:2021dwh}
{\bfseries CONUS} Collaboration, H.~Bonet {\em et~al.}, ``{Novel constraints on
  neutrino physics beyond the standard model from the CONUS experiment},''
  \href{http://dx.doi.org/10.1007/JHEP05(2022)085}{{\em JHEP} {\bfseries 05}
  (2022) 085}, \href{http://arxiv.org/abs/2110.02174}{{\ttfamily
  arXiv:2110.02174 [hep-ph]}}.

\bibitem{nuGeN:2022bmg}
{\bfseries {\ensuremath{\nu}}GeN} Collaboration, I.~Alekseev {\em et~al.},
  ``{First results of the \ensuremath{\nu}GeN experiment on coherent elastic
  neutrino-nucleus scattering},''
  \href{http://dx.doi.org/10.1103/PhysRevD.106.L051101}{{\em Phys.Rev.D}
  {\bfseries 106} (2022) L051101},
  \href{http://arxiv.org/abs/2205.04305}{{\ttfamily arXiv:2205.04305
  [nucl-ex]}}.

\bibitem{MINER:2016igy}
{\bfseries MINER} Collaboration, G.~Agnolet {\em et~al.}, ``{Background Studies
  for the MINER Coherent Neutrino Scattering Reactor Experiment},''
  \href{http://dx.doi.org/10.1016/j.nima.2017.02.024}{{\em Nucl. Instrum. Meth.
  A} {\bfseries 853} (2017) 53--60},
  \href{http://arxiv.org/abs/1609.02066}{{\ttfamily arXiv:1609.02066
  [physics.ins-det]}}.

\bibitem{Billard:2016giu}
J.~Billard {\em et~al.}, ``{Coherent Neutrino Scattering with Low Temperature
  Bolometers at Chooz Reactor Complex},''
  \href{http://dx.doi.org/10.1088/1361-6471/aa83d0}{{\em J. Phys. G} {\bfseries
  44} no.~10, (2017) 105101}, \href{http://arxiv.org/abs/1612.09035}{{\ttfamily
  arXiv:1612.09035 [physics.ins-det]}}.

\bibitem{Strauss:2017cuu}
R.~Strauss {\em et~al.}, ``{The $\nu$-cleus experiment: A gram-scale
  fiducial-volume cryogenic detector for the first detection of coherent
  neutrino-nucleus scattering},''
  \href{http://dx.doi.org/10.1140/epjc/s10052-017-5068-2}{{\em Eur. Phys. J. C}
  {\bfseries 77} (2017) 506}, \href{http://arxiv.org/abs/1704.04320}{{\ttfamily
  arXiv:1704.04320 [physics.ins-det]}}.

\bibitem{Wong:2015kgl}
H.~T.-K. Wong, ``{Taiwan EXperiment On NeutrinO \textemdash{} History and
  Prospects},'' \href{http://dx.doi.org/10.1142/S0217751X18300144}{{\em The
  Universe} {\bfseries 3} (2015) 22--37},
  \href{http://arxiv.org/abs/1608.00306}{{\ttfamily arXiv:1608.00306
  [hep-ex]}}.

\bibitem{Fernandez-Moroni:2020yyl}
G.~Fernandez-Moroni, P.~A.~N. Machado, I.~Martinez-Soler, Y.~F. Perez-Gonzalez,
  D.~Rodrigues, and S.~Rosauro-Alcaraz, ``{The physics potential of a reactor
  neutrino experiment with Skipper CCDs: Measuring the weak mixing angle},''
  \href{http://dx.doi.org/10.1007/JHEP03(2021)186}{{\em JHEP} {\bfseries 03}
  (2021) 186}, \href{http://arxiv.org/abs/2009.10741}{{\ttfamily
  arXiv:2009.10741 [hep-ph]}}.

\bibitem{Akimov:2022xvr}
D.~Y. Akimov {\em et~al.}, ``{The RED-100 experiment},''
  \href{http://dx.doi.org/10.1088/1748-0221/17/11/T11011}{{\em JINST}
  {\bfseries 17} (2022) T11011},
  \href{http://arxiv.org/abs/2209.15516}{{\ttfamily arXiv:2209.15516
  [physics.ins-det]}}.

\bibitem{NEON:2022hbk}
{\bfseries NEON} Collaboration, J.~J. Choi {\em et~al.}, ``{Exploring coherent
  elastic neutrino-nucleus scattering using reactor electron antineutrinos in
  the NEON experiment},''
  \href{http://dx.doi.org/10.1140/epjc/s10052-023-11352-x}{{\em Eur. Phys. J.
  C} {\bfseries 83} (2023) 226},
  \href{http://arxiv.org/abs/2204.06318}{{\ttfamily arXiv:2204.06318
  [hep-ex]}}.

\bibitem{NEWS-G:2021mhf}
{\bfseries NEWS-G} Collaboration, L.~Balogh {\em et~al.}, ``{Quenching factor
  measurements of neon nuclei in neon gas},''
  \href{http://dx.doi.org/10.1103/PhysRevD.105.052004}{{\em Phys. Rev. D}
  {\bfseries 105} (2022) 052004},
  \href{http://arxiv.org/abs/2109.01055}{{\ttfamily arXiv:2109.01055
  [physics.ins-det]}}.

\bibitem{SBC:2021yal}
{\bfseries SBC, CE\ensuremath{\nu}NS Theory Group at IF-UNAM} Collaboration,
  L.~J. Flores {\em et~al.}, ``{Physics reach of a low threshold scintillating
  argon bubble chamber in coherent elastic neutrino-nucleus scattering reactor
  experiments},'' \href{http://dx.doi.org/10.1103/PhysRevD.103.L091301}{{\em
  Phys. Rev. D} {\bfseries 103} (2021) L091301},
  \href{http://arxiv.org/abs/2101.08785}{{\ttfamily arXiv:2101.08785
  [hep-ex]}}.

\bibitem{Colaresi:2022obx}
J.~Colaresi, J.~I. Collar, T.~W. Hossbach, C.~M. Lewis, and K.~M. Yocum,
  ``{Measurement of Coherent Elastic Neutrino-Nucleus Scattering from Reactor
  Antineutrinos},''
  \href{http://dx.doi.org/10.1103/PhysRevLett.129.211802}{{\em Phys. Rev.
  Lett.} {\bfseries 129} no.~21, (2022) 211802},
  \href{http://arxiv.org/abs/2202.09672}{{\ttfamily arXiv:2202.09672
  [hep-ex]}}.

\bibitem{CCM:2021leg}
{\bfseries CCM} Collaboration, A.~A. Aguilar-Arevalo {\em et~al.}, ``{First
  Dark Matter Search Results From Coherent CAPTAIN-Mills},''
  \href{http://arxiv.org/abs/2105.14020}{{\ttfamily arXiv:2105.14020
  [hep-ex]}}.

\bibitem{Baxter:2019mcx}
D.~Baxter {\em et~al.}, ``{Coherent Elastic Neutrino-Nucleus Scattering at the
  European Spallation Source},''
  \href{http://dx.doi.org/10.1007/JHEP02(2020)123}{{\em JHEP} {\bfseries 02}
  (2020) 123}, \href{http://arxiv.org/abs/1911.00762}{{\ttfamily
  arXiv:1911.00762 [physics.ins-det]}}.

\bibitem{AristizabalSierra:2021uob}
D.~Aristizabal~Sierra, B.~Dutta, D.~Kim, D.~Snowden-Ifft, and L.~E. Strigari,
  ``{Coherent elastic neutrino-nucleus scattering with the
  \ensuremath{\nu}BDX-DRIFT directional detector at next generation neutrino
  facilities},'' \href{http://dx.doi.org/10.1103/PhysRevD.104.033004}{{\em
  Phys. Rev. D} {\bfseries 104} no.~3, (2021) 033004},
  \href{http://arxiv.org/abs/2103.10857}{{\ttfamily arXiv:2103.10857
  [hep-ph]}}.

\bibitem{AristizabalSierra:2022jgg}
{\bfseries \ensuremath{\nu}BDX-DRIFT} Collaboration, D.~Aristizabal~Sierra,
  J.~L. Barrow, B.~Dutta, D.~Kim, D.~Snowden-Ifft, L.~Strigari, and M.~H. Wood,
  ``{Rock neutron backgrounds from FNAL neutrino beamlines in the
  \ensuremath{\nu}BDX-DRIFT detector},''
  \href{http://dx.doi.org/10.1103/PhysRevD.107.013003}{{\em Phys. Rev. D}
  {\bfseries 107} no.~1, (2023) 013003},
  \href{http://arxiv.org/abs/2210.08612}{{\ttfamily arXiv:2210.08612
  [hep-ex]}}.

\bibitem{Papoulias:2019lfi}
D.~K. Papoulias, T.~S. Kosmas, R.~Sahu, V.~K.~B. Kota, and M.~Hota,
  ``{Constraining nuclear physics parameters with current and future COHERENT
  data},'' \href{http://dx.doi.org/10.1016/j.physletb.2019.135133}{{\em Phys.
  Lett. B} {\bfseries 800} (2020) 135133},
  \href{http://arxiv.org/abs/1903.03722}{{\ttfamily arXiv:1903.03722
  [hep-ph]}}.

\bibitem{Coloma:2020nhf}
P.~Coloma, I.~Esteban, M.~C. Gonzalez-Garcia, and J.~Menendez, ``{Determining
  the nuclear neutron distribution from Coherent Elastic neutrino-Nucleus
  Scattering: current results and future prospects},''
  \href{http://dx.doi.org/10.1007/JHEP08(2020)030}{{\em JHEP} {\bfseries 08}
  no.~08, (2020) 030}, \href{http://arxiv.org/abs/2006.08624}{{\ttfamily
  arXiv:2006.08624 [hep-ph]}}.

\bibitem{Cadeddu:2021ijh}
M.~Cadeddu, N.~Cargioli, F.~Dordei, C.~Giunti, Y.~F. Li, E.~Picciau, C.~A.
  Ternes, and Y.~Y. Zhang, ``{New insights into nuclear physics and weak mixing
  angle using electroweak probes},''
  \href{http://dx.doi.org/10.1103/PhysRevC.104.065502}{{\em Phys. Rev. C}
  {\bfseries 104} no.~6, (2021) 065502},
  \href{http://arxiv.org/abs/2102.06153}{{\ttfamily arXiv:2102.06153
  [hep-ph]}}.

\bibitem{Majumdar:2022nby}
A.~Majumdar, D.~K. Papoulias, R.~Srivastava, and J.~W.~F. Valle, ``{Physics
  implications of recent Dresden-II reactor data},''
  \href{http://dx.doi.org/10.1103/PhysRevD.106.093010}{{\em Phys. Rev. D}
  {\bfseries 106} no.~9, (2022) 093010},
  \href{http://arxiv.org/abs/2208.13262}{{\ttfamily arXiv:2208.13262
  [hep-ph]}}.

\bibitem{AristizabalSierra:2022axl}
D.~Aristizabal~Sierra, V.~De~Romeri, and D.~K. Papoulias, ``{Consequences of
  the Dresden-II reactor data for the weak mixing angle and new physics},''
  \href{http://dx.doi.org/10.1007/JHEP09(2022)076}{{\em JHEP} {\bfseries 09}
  (2022) 076}, \href{http://arxiv.org/abs/2203.02414}{{\ttfamily
  arXiv:2203.02414 [hep-ph]}}.

\bibitem{DeRomeri:2022twg}
V.~De~Romeri, O.~G. Miranda, D.~K. Papoulias, G.~Sanchez~Garcia, M.~T\'ortola,
  and J.~W.~F. Valle, ``{Physics implications of a combined analysis of
  COHERENT CsI and LAr data},''
  \href{http://dx.doi.org/10.1007/JHEP04(2023)035}{{\em JHEP} {\bfseries 04}
  (2023) 035}, \href{http://arxiv.org/abs/2211.11905}{{\ttfamily
  arXiv:2211.11905 [hep-ph]}}.

\bibitem{Sierra:2023pnf}
D.~A. Sierra, ``{Extraction of neutron density distributions from
  high-statistics coherent elastic neutrino-nucleus scattering data},''
  \href{http://arxiv.org/abs/2301.13249}{{\ttfamily arXiv:2301.13249
  [hep-ph]}}.

\bibitem{AtzoriCorona:2023ktl}
M.~Atzori~Corona, M.~Cadeddu, N.~Cargioli, F.~Dordei, C.~Giunti, and G.~Masia,
  ``{Nuclear neutron radius and weak mixing angle measurements from latest
  COHERENT CsI and atomic parity violation Cs data},''
  \href{http://arxiv.org/abs/2303.09360}{{\ttfamily arXiv:2303.09360
  [nucl-ex]}}.

\bibitem{Barranco:2005yy}
J.~Barranco, O.~G. Miranda, and T.~I. Rashba, ``{Probing new physics with
  coherent neutrino scattering off nuclei},''
  \href{http://dx.doi.org/10.1088/1126-6708/2005/12/021}{{\em JHEP} {\bfseries
  12} (2005) 021}, \href{http://arxiv.org/abs/hep-ph/0508299}{{\ttfamily
  arXiv:hep-ph/0508299}}.

\bibitem{Ohlsson:2012kf}
T.~Ohlsson, ``{Status of non-standard neutrino interactions},''
  \href{http://dx.doi.org/10.1088/0034-4885/76/4/044201}{{\em Rept. Prog.
  Phys.} {\bfseries 76} (2013) 044201},
  \href{http://arxiv.org/abs/1209.2710}{{\ttfamily arXiv:1209.2710 [hep-ph]}}.

\bibitem{Miranda:2015dra}
O.~G. Miranda and H.~Nunokawa, ``{Non standard neutrino interactions: current
  status and future prospects},''
  \href{http://dx.doi.org/10.1088/1367-2630/17/9/095002}{{\em New J. Phys.}
  {\bfseries 17} (2015) 095002},
  \href{http://arxiv.org/abs/1505.06254}{{\ttfamily arXiv:1505.06254
  [hep-ph]}}.

\bibitem{Dent:2016wcr}
J.~B. Dent, B.~Dutta, S.~Liao, J.~L. Newstead, L.~E. Strigari, and J.~W.
  Walker, ``{Probing light mediators at ultralow threshold energies with
  coherent elastic neutrino-nucleus scattering},''
  \href{http://dx.doi.org/10.1103/PhysRevD.96.095007}{{\em Phys. Rev. D}
  {\bfseries 96} (2017) 095007},
  \href{http://arxiv.org/abs/1612.06350}{{\ttfamily arXiv:1612.06350
  [hep-ph]}}.

\bibitem{Farzan:2017xzy}
Y.~Farzan and M.~Tortola, ``{Neutrino oscillations and Non-Standard
  Interactions},'' \href{http://dx.doi.org/10.3389/fphy.2018.00010}{{\em Front.
  in Phys.} {\bfseries 6} (2018) 10},
  \href{http://arxiv.org/abs/1710.09360}{{\ttfamily arXiv:1710.09360
  [hep-ph]}}.

\bibitem{Liao:2017uzy}
J.~Liao and D.~Marfatia, ``{COHERENT constraints on nonstandard neutrino
  interactions},'' \href{http://dx.doi.org/10.1016/j.physletb.2017.10.046}{{\em
  Phys. Lett. B} {\bfseries 775} (2017) 54--57},
  \href{http://arxiv.org/abs/1708.04255}{{\ttfamily arXiv:1708.04255
  [hep-ph]}}.

\bibitem{AristizabalSierra:2018eqm}
D.~Aristizabal~Sierra, V.~De~Romeri, and N.~Rojas, ``{COHERENT analysis of
  neutrino generalized interactions},''
  \href{http://dx.doi.org/10.1103/PhysRevD.98.075018}{{\em Phys. Rev. D}
  {\bfseries 98} (2018) 075018},
  \href{http://arxiv.org/abs/1806.07424}{{\ttfamily arXiv:1806.07424
  [hep-ph]}}.

\bibitem{Abdullah:2018ykz}
M.~Abdullah, J.~B. Dent, B.~Dutta, G.~L. Kane, S.~Liao, and L.~E. Strigari,
  ``{Coherent elastic neutrino nucleus scattering as a probe of a Z' through
  kinetic and mass mixing effects},''
  \href{http://dx.doi.org/10.1103/PhysRevD.98.015005}{{\em Phys. Rev. D}
  {\bfseries 98} (2018) 015005},
  \href{http://arxiv.org/abs/1803.01224}{{\ttfamily arXiv:1803.01224
  [hep-ph]}}.

\bibitem{Billard:2018jnl}
J.~Billard, J.~Johnston, and B.~J. Kavanagh, ``{Prospects for exploring New
  Physics in Coherent Elastic Neutrino-Nucleus Scattering},''
  \href{http://dx.doi.org/10.1088/1475-7516/2018/11/016}{{\em JCAP} {\bfseries
  11} (2018) 016}, \href{http://arxiv.org/abs/1805.01798}{{\ttfamily
  arXiv:1805.01798 [hep-ph]}}.

\bibitem{Denton:2018xmq}
P.~B. Denton, Y.~Farzan, and I.~M. Shoemaker, ``{Testing large non-standard
  neutrino interactions with arbitrary mediator mass after COHERENT data},''
  \href{http://dx.doi.org/10.1007/JHEP07(2018)037}{{\em JHEP} {\bfseries 07}
  (2018) 037}, \href{http://arxiv.org/abs/1804.03660}{{\ttfamily
  arXiv:1804.03660 [hep-ph]}}.

\bibitem{Han:2019zkz}
T.~Han, J.~Liao, H.~Liu, and D.~Marfatia, ``{Nonstandard neutrino interactions
  at COHERENT, DUNE, T2HK and LHC},''
  \href{http://dx.doi.org/10.1007/JHEP11(2019)028}{{\em JHEP} {\bfseries 11}
  (2019) 028}, \href{http://arxiv.org/abs/1910.03272}{{\ttfamily
  arXiv:1910.03272 [hep-ph]}}.

\bibitem{Giunti:2019xpr}
C.~Giunti, ``{General COHERENT constraints on neutrino nonstandard
  interactions},'' \href{http://dx.doi.org/10.1103/PhysRevD.101.035039}{{\em
  Phys. Rev. D} {\bfseries 101} no.~3, (2020) 035039},
  \href{http://arxiv.org/abs/1909.00466}{{\ttfamily arXiv:1909.00466
  [hep-ph]}}.

\bibitem{AristizabalSierra:2019ykk}
D.~Aristizabal~Sierra, B.~Dutta, S.~Liao, and L.~E. Strigari, ``{Coherent
  elastic neutrino-nucleus scattering in multi-ton scale dark matter
  experiments: Classification of vector and scalar interactions new physics
  signals},'' \href{http://dx.doi.org/10.1007/JHEP12(2019)124}{{\em JHEP}
  {\bfseries 12} (2019) 124}, \href{http://arxiv.org/abs/1910.12437}{{\ttfamily
  arXiv:1910.12437 [hep-ph]}}.

\bibitem{AristizabalSierra:2019ufd}
D.~Aristizabal~Sierra, V.~De~Romeri, and N.~Rojas, ``{CP violating effects in
  coherent elastic neutrino-nucleus scattering processes},''
  \href{http://dx.doi.org/10.1007/JHEP09(2019)069}{{\em JHEP} {\bfseries 09}
  (2019) 069}, \href{http://arxiv.org/abs/1906.01156}{{\ttfamily
  arXiv:1906.01156 [hep-ph]}}.

\bibitem{Denton:2020hop}
P.~B. Denton and J.~Gehrlein, ``{A Statistical Analysis of the COHERENT Data
  and Applications to New Physics},''
  \href{http://dx.doi.org/10.1007/JHEP04(2021)266}{{\em JHEP} {\bfseries 04}
  (2021) 266}, \href{http://arxiv.org/abs/2008.06062}{{\ttfamily
  arXiv:2008.06062 [hep-ph]}}.

\bibitem{Flores:2020lji}
L.~J. Flores, N.~Nath, and E.~Peinado, ``{Non-standard neutrino interactions in
  U(1)' model after COHERENT data},''
  \href{http://dx.doi.org/10.1007/JHEP06(2020)045}{{\em JHEP} {\bfseries 06}
  (2020) 045}, \href{http://arxiv.org/abs/2002.12342}{{\ttfamily
  arXiv:2002.12342 [hep-ph]}}.

\bibitem{Cadeddu:2020nbr}
M.~Cadeddu, N.~Cargioli, F.~Dordei, C.~Giunti, Y.~F. Li, E.~Picciau, and Y.~Y.
  Zhang, ``{Constraints on light vector mediators through coherent elastic
  neutrino nucleus scattering data from COHERENT},''
  \href{http://dx.doi.org/10.1007/JHEP01(2021)116}{{\em JHEP} {\bfseries 01}
  (2021) 116}, \href{http://arxiv.org/abs/2008.05022}{{\ttfamily
  arXiv:2008.05022 [hep-ph]}}.

\bibitem{Amaral:2021rzw}
D.~W.~P. Amaral, D.~G. Cerdeno, A.~Cheek, and P.~Foldenauer, ``{Confirming
  $U(1)_{L_\mu -L_{\tau }}$ as a solution for $(g-2)_\mu $ with neutrinos},''
  \href{http://dx.doi.org/10.1140/epjc/s10052-021-09670-z}{{\em Eur. Phys. J.
  C} {\bfseries 81} (2021) 861},
  \href{http://arxiv.org/abs/2104.03297}{{\ttfamily arXiv:2104.03297
  [hep-ph]}}.

\bibitem{Flores:2021kzl}
L.~J. Flores, N.~Nath, and E.~Peinado, ``{CE\ensuremath{\nu}NS as a probe of
  flavored generalized neutrino interactions},''
  \href{http://dx.doi.org/10.1103/PhysRevD.105.055010}{{\em Phys. Rev. D}
  {\bfseries 105} no.~5, (2022) 055010},
  \href{http://arxiv.org/abs/2112.05103}{{\ttfamily arXiv:2112.05103
  [hep-ph]}}.

\bibitem{AtzoriCorona:2022moj}
M.~Atzori~Corona, M.~Cadeddu, N.~Cargioli, F.~Dordei, C.~Giunti, Y.~F. Li,
  E.~Picciau, C.~A. Ternes, and Y.~Y. Zhang, ``{Probing light mediators and (g
  \ensuremath{-} 2)$_{μ}$ through detection of coherent elastic neutrino
  nucleus scattering at COHERENT},''
  \href{http://dx.doi.org/10.1007/JHEP05(2022)109}{{\em JHEP} {\bfseries 05}
  (2022) 109}, \href{http://arxiv.org/abs/2202.11002}{{\ttfamily
  arXiv:2202.11002 [hep-ph]}}.

\bibitem{Schechter:1981hw}
J.~Schechter and J.~W.~F. Valle, ``{Majorana Neutrinos and Magnetic Fields},''
  \href{http://dx.doi.org/10.1103/PhysRevD.25.283}{{\em Phys.Rev.D} {\bfseries
  24} (1981) 1883--1889}.

\bibitem{Pal:1981rm}
P.~B. Pal and L.~Wolfenstein, ``{Radiative Decays of Massive Neutrinos},''
  \href{http://dx.doi.org/10.1103/PhysRevD.25.766}{{\em Phys. Rev. D}
  {\bfseries 25} (1982) 766}.

\bibitem{Kayser:1982br}
B.~Kayser, ``{Majorana Neutrinos and their Electromagnetic Properties},''
  \href{http://dx.doi.org/10.1103/PhysRevD.26.1662}{{\em Phys. Rev. D}
  {\bfseries 26} (1982) 1662}.

\bibitem{Nieves:1981zt}
J.~F. Nieves, ``{Electromagnetic Properties of Majorana Neutrinos},''
  \href{http://dx.doi.org/10.1103/PhysRevD.26.3152}{{\em Phys. Rev. D}
  {\bfseries 26} (1982) 3152}.

\bibitem{Shrock:1982sc}
R.~E. Shrock, ``{Electromagnetic Properties and Decays of Dirac and Majorana
  Neutrinos in a General Class of Gauge Theories},''
  \href{http://dx.doi.org/10.1016/0550-3213(82)90273-5}{{\em Nucl. Phys. B}
  {\bfseries 206} (1982) 359--379}.

\bibitem{Kosmas:2015sqa}
T.~S. Kosmas, O.~G. Miranda, D.~K. Papoulias, M.~Tortola, and J.~W.~F. Valle,
  ``{Probing neutrino magnetic moments at the Spallation Neutron Source
  facility},'' \href{http://dx.doi.org/10.1103/PhysRevD.92.013011}{{\em Phys.
  Rev.} {\bfseries D92} (2015) 013011},
  \href{http://arxiv.org/abs/1505.03202}{{\ttfamily arXiv:1505.03202
  [hep-ph]}}.

\bibitem{Canas:2015yoa}
B.~C. Canas, O.~G. Miranda, A.~Parada, M.~Tortola, and J.~W.~F. Valle,
  ``{Updating neutrino magnetic moment constraints},''
  \href{http://dx.doi.org/10.1016/j.physletb.2015.12.011}{{\em Phys. Lett. B}
  {\bfseries 753} (2016) 191--198},
  \href{http://arxiv.org/abs/1510.01684}{{\ttfamily arXiv:1510.01684
  [hep-ph]}}. [Addendum: Phys.Lett.B 757, 568--568 (2016)].

\bibitem{Miranda:2019wdy}
O.~G. Miranda, D.~K. Papoulias, M.~T\'ortola, and J.~W.~F. Valle, ``{Probing
  neutrino transition magnetic moments with coherent elastic neutrino-nucleus
  scattering},'' \href{http://dx.doi.org/10.1007/JHEP07(2019)103}{{\em JHEP}
  {\bfseries 07} (2019) 103}, \href{http://arxiv.org/abs/1905.03750}{{\ttfamily
  arXiv:1905.03750 [hep-ph]}}.

\bibitem{Miranda:2020kwy}
O.~Miranda {\em et~al.}, ``{XENON1T signal from transition neutrino magnetic
  moments},'' \href{http://dx.doi.org/10.1016/j.physletb.2020.135685}{{\em
  Phys.Lett.B} {\bfseries 808} (2020) 135685},
  \href{http://arxiv.org/abs/2007.01765}{{\ttfamily arXiv:2007.01765
  [hep-ph]}}.

\bibitem{Kosmas:2017zbh}
T.~S. Kosmas, D.~K. Papoulias, M.~Tortola, and J.~W.~F. Valle, ``{Probing light
  sterile neutrino signatures at reactor and Spallation Neutron Source neutrino
  experiments},'' \href{http://dx.doi.org/10.1103/PhysRevD.96.063013}{{\em
  Phys. Rev. D} {\bfseries 96} no.~6, (2017) 063013},
  \href{http://arxiv.org/abs/1703.00054}{{\ttfamily arXiv:1703.00054
  [hep-ph]}}.

\bibitem{Brdar:2018qqj}
V.~Brdar, W.~Rodejohann, and X.-J. Xu, ``{Producing a new Fermion in Coherent
  Elastic Neutrino-Nucleus Scattering: from Neutrino Mass to Dark Matter},''
  \href{http://dx.doi.org/10.1007/JHEP12(2018)024}{{\em JHEP} {\bfseries 12}
  (2018) 024}, \href{http://arxiv.org/abs/1810.03626}{{\ttfamily
  arXiv:1810.03626 [hep-ph]}}.

\bibitem{Blanco:2019vyp}
C.~Blanco, D.~Hooper, and P.~Machado, ``{Constraining Sterile Neutrino
  Interpretations of the LSND and MiniBooNE Anomalies with Coherent Neutrino
  Scattering Experiments},''
  \href{http://dx.doi.org/10.1103/PhysRevD.101.075051}{{\em Phys. Rev. D}
  {\bfseries 101} (2020) 075051},
  \href{http://arxiv.org/abs/1901.08094}{{\ttfamily arXiv:1901.08094
  [hep-ph]}}.

\bibitem{Miranda:2020syh}
O.~G. Miranda, D.~K. Papoulias, O.~Sanders, M.~T\'ortola, and J.~W.~F. Valle,
  ``{Future CEvNS experiments as probes of lepton unitarity and light-sterile
  neutrinos},'' \href{http://dx.doi.org/10.1103/PhysRevD.102.113014}{{\em Phys.
  Rev. D} {\bfseries 102} (2020) 113014},
  \href{http://arxiv.org/abs/2008.02759}{{\ttfamily arXiv:2008.02759
  [hep-ph]}}.

\bibitem{Miranda:2021kre}
O.~G. Miranda, D.~K. Papoulias, O.~Sanders, M.~T\'ortola, and J.~W.~F. Valle,
  ``{Low-energy probes of sterile neutrino transition magnetic moments},''
  \href{http://dx.doi.org/10.1007/JHEP12(2021)191}{{\em JHEP} {\bfseries 12}
  (2021) 191}, \href{http://arxiv.org/abs/2109.09545}{{\ttfamily
  arXiv:2109.09545 [hep-ph]}}.

\bibitem{Bolton:2021pey}
P.~D. Bolton, F.~F. Deppisch, K.~Fridell, J.~Harz, C.~Hati, and S.~Kulkarni,
  ``{Probing active-sterile neutrino transition magnetic moments with photon
  emission from CE\ensuremath{\nu}NS},''
  \href{http://dx.doi.org/10.1103/PhysRevD.106.035036}{{\em Phys. Rev. D}
  {\bfseries 106} (2022) 035036},
  \href{http://arxiv.org/abs/2110.02233}{{\ttfamily arXiv:2110.02233
  [hep-ph]}}.

\bibitem{Chang:2020jwl}
W.-F. Chang and J.~Liao, ``{Constraints on light singlet fermion interactions
  from coherent elastic neutrino-nucleus scattering},''
  \href{http://dx.doi.org/10.1103/PhysRevD.102.075004}{{\em Phys. Rev. D}
  {\bfseries 102} no.~7, (2020) 075004},
  \href{http://arxiv.org/abs/2002.10275}{{\ttfamily arXiv:2002.10275
  [hep-ph]}}.

\bibitem{Chao:2021bvq}
W.~Chao, T.~Li, J.~Liao, and M.~Su, ``{Loop effects with a vector mediator in
  coherent neutrino-nucleus scattering},''
  \href{http://dx.doi.org/10.1103/PhysRevD.104.095017}{{\em Phys. Rev. D}
  {\bfseries 104} (2021) 095017},
  \href{http://arxiv.org/abs/2108.02341}{{\ttfamily arXiv:2108.02341
  [hep-ph]}}.

\bibitem{Chen:2021uuw}
Z.~Chen, T.~Li, and J.~Liao, ``{Constraints on general neutrino interactions
  with exotic fermion from neutrino-electron scattering experiments},''
  \href{http://dx.doi.org/10.1007/JHEP05(2021)131}{{\em JHEP} {\bfseries 05}
  (2021) 131}, \href{http://arxiv.org/abs/2102.09784}{{\ttfamily
  arXiv:2102.09784 [hep-ph]}}.

\bibitem{AristizabalSierra:2021fuc}
D.~Aristizabal~Sierra, O.~G. Miranda, D.~K. Papoulias, and G.~S. Garcia,
  ``{Neutrino magnetic and electric dipole moments: From measurements to
  parameter space},'' \href{http://dx.doi.org/10.1103/PhysRevD.105.035027}{{\em
  Phys. Rev. D} {\bfseries 105} (2022) 035027},
  \href{http://arxiv.org/abs/2112.12817}{{\ttfamily arXiv:2112.12817
  [hep-ph]}}.

\bibitem{Calabrese:2022mnp}
R.~Calabrese, J.~Gunn, G.~Miele, S.~Morisi, S.~Roy, and P.~Santorelli,
  ``{Constraining scalar leptoquarks using COHERENT data},''
  \href{http://dx.doi.org/10.1103/PhysRevD.107.055039}{{\em Phys. Rev. D}
  {\bfseries 107} no.~5, (2023) 055039},
  \href{http://arxiv.org/abs/2212.11210}{{\ttfamily arXiv:2212.11210
  [hep-ph]}}.

\bibitem{Candela:2023rvt}
P.~M. Candela, V.~De~Romeri, and D.~K. Papoulias, ``{COHERENT production of a
  Dark Fermion},'' \href{http://arxiv.org/abs/2305.03341}{{\ttfamily
  arXiv:2305.03341 [hep-ph]}}.

\bibitem{Felkl:2023nan}
T.~Felkl, T.~Li, J.~Liao, and M.~A. Schmidt, ``{Probing general $U(1)'$ models
  with non-universal lepton charges at FASER/FASER2, COHERENT and long-baseline
  oscillation experiments},'' \href{http://arxiv.org/abs/2306.09569}{{\ttfamily
  arXiv:2306.09569 [hep-ph]}}.

\bibitem{Breso-Pla:2023tnz}
V.~Bres\'o-Pla, A.~Falkowski, M.~Gonz\'alez-Alonso, and K.~Mons\'alvez-Pozo,
  ``{EFT analysis of New Physics at COHERENT},''
  \href{http://dx.doi.org/10.1007/JHEP05(2023)074}{{\em JHEP} {\bfseries 05}
  (2023) 074}, \href{http://arxiv.org/abs/2301.07036}{{\ttfamily
  arXiv:2301.07036 [hep-ph]}}.

\bibitem{Barranco:2007tz}
J.~Barranco, O.~G. Miranda, and T.~I. Rashba, ``{Low energy neutrino
  experiments sensitivity to physics beyond the Standard Model},''
  \href{http://dx.doi.org/10.1103/PhysRevD.76.073008}{{\em Phys. Rev. D}
  {\bfseries 76} (2007) 073008},
  \href{http://arxiv.org/abs/hep-ph/0702175}{{\ttfamily arXiv:hep-ph/0702175}}.

\bibitem{Crivellin:2021bkd}
A.~Crivellin, M.~Hoferichter, M.~Kirk, C.~A. Manzari, and L.~Schnell,
  ``{First-generation new physics in simplified models: from low-energy parity
  violation to the LHC},''
  \href{http://dx.doi.org/10.1007/JHEP10(2021)221}{{\em JHEP} {\bfseries 10}
  (2021) 221}, \href{http://arxiv.org/abs/2107.13569}{{\ttfamily
  arXiv:2107.13569 [hep-ph]}}.

\bibitem{Asaadi:2022ojm}
J.~Asaadi {\em et~al.}, ``{Physics Opportunities in the ORNL Spallation Neutron
  Source Second Target Station Era},'' in {\em {Snowmass 2021}}.
\newblock 9, 2022.
\newblock \href{http://arxiv.org/abs/2209.02883}{{\ttfamily arXiv:2209.02883
  [hep-ex]}}.

\bibitem{Akimov:2022oyb}
D.~Akimov {\em et~al.}, ``{The COHERENT Experimental Program},'' in {\em
  {Snowmass 2021}}.
\newblock 4, 2022.
\newblock \href{http://arxiv.org/abs/2204.04575}{{\ttfamily arXiv:2204.04575
  [hep-ex]}}.

\bibitem{Bouchiat:1974kt}
M.~A. Bouchiat and C.~C. Bouchiat, ``{Weak Neutral Currents in Atomic
  Physics},'' \href{http://dx.doi.org/10.1016/0370-2693(74)90656-X}{{\em Phys.
  Lett. B} {\bfseries 48} (1974) 111--114}.

\bibitem{Safronova:2017xyt}
M.~S. Safronova, D.~Budker, D.~DeMille, D.~F.~J. Kimball, A.~Derevianko, and
  C.~W. Clark, ``{Search for New Physics with Atoms and Molecules},''
  \href{http://dx.doi.org/10.1103/RevModPhys.90.025008}{{\em Rev. Mod. Phys.}
  {\bfseries 90} no.~2, (2018) 025008},
  \href{http://arxiv.org/abs/1710.01833}{{\ttfamily arXiv:1710.01833
  [physics.atom-ph]}}.

\bibitem{Wood:1997zq}
C.~S. Wood, S.~C. Bennett, D.~Cho, B.~P. Masterson, J.~L. Roberts, C.~E.
  Tanner, and C.~E. Wieman, ``{Measurement of parity nonconservation and an
  anapole moment in cesium},''
  \href{http://dx.doi.org/10.1126/science.275.5307.1759}{{\em Science}
  {\bfseries 275} (1997) 1759--1763}.

\bibitem{Guena:2004sq}
J.~Guena, M.~Lintz, and M.~A. Bouchiat, ``{Measurement of the parity violating
  6S-7S transition amplitude in cesium achieved within 2 x 10(-13) atomic-unit
  accuracy by stimulated-emission detection},''
  \href{http://dx.doi.org/10.1103/PhysRevA.71.042108}{{\em Phys. Rev. A}
  {\bfseries 71} (2005) 042108},
  \href{http://arxiv.org/abs/physics/0412017}{{\ttfamily
  arXiv:physics/0412017}}.

\bibitem{Langacker:1990jf}
P.~Langacker, ``{Parity violation in muonic atoms and cesium},''
  \href{http://dx.doi.org/10.1016/0370-2693(91)90688-M}{{\em Phys. Lett. B}
  {\bfseries 256} (1991) 277--283}.

\bibitem{Davidson:1993qk}
S.~Davidson, D.~C. Bailey, and B.~A. Campbell, ``{Model independent constraints
  on leptoquarks from rare processes},''
  \href{http://dx.doi.org/10.1007/BF01552629}{{\em Z. Phys. C} {\bfseries 61}
  (1994) 613--644}, \href{http://arxiv.org/abs/hep-ph/9309310}{{\ttfamily
  arXiv:hep-ph/9309310}}.

\bibitem{Barger:2000gv}
V.~D. Barger and K.-m. Cheung, ``{Atomic parity violation, leptoquarks, and
  contact interactions},''
  \href{http://dx.doi.org/10.1016/S0370-2693(00)00401-9}{{\em Phys. Lett. B}
  {\bfseries 480} (2000) 149--154},
  \href{http://arxiv.org/abs/hep-ph/0002259}{{\ttfamily arXiv:hep-ph/0002259}}.

\bibitem{NuTeV:2001whx}
{\bfseries NuTeV} Collaboration, G.~P. Zeller {\em et~al.}, ``{A Precise
  Determination of Electroweak Parameters in Neutrino Nucleon Scattering},''
  \href{http://dx.doi.org/10.1103/PhysRevLett.88.091802}{{\em Phys. Rev. Lett.}
  {\bfseries 88} (2002) 091802},
  \href{http://arxiv.org/abs/hep-ex/0110059}{{\ttfamily arXiv:hep-ex/0110059}}.
  [Erratum: Phys.Rev.Lett. 90, 239902 (2003)].

\bibitem{Escrihuela:2011cf}
F.~J. Escrihuela, M.~Tortola, J.~W.~F. Valle, and O.~G. Miranda, ``{Global
  constraints on muon-neutrino non-standard interactions},''
  \href{http://dx.doi.org/10.1103/PhysRevD.83.093002}{{\em Phys. Rev. D}
  {\bfseries 83} (2011) 093002},
  \href{http://arxiv.org/abs/1103.1366}{{\ttfamily arXiv:1103.1366 [hep-ph]}}.

\bibitem{H1:2001ezk}
{\bfseries H1} Collaboration, C.~Adloff {\em et~al.}, ``{A Search for
  leptoquark bosons in e- p collisions at HERA},''
  \href{http://dx.doi.org/10.1016/S0370-2693(01)01262-X}{{\em Phys. Lett. B}
  {\bfseries 523} (2001) 234--242},
  \href{http://arxiv.org/abs/hep-ex/0107038}{{\ttfamily arXiv:hep-ex/0107038}}.

\bibitem{ZEUS:2003uzd}
{\bfseries ZEUS} Collaboration, S.~Chekanov {\em et~al.}, ``{A Search for
  resonance decays to lepton + jet at HERA and limits on leptoquarks},''
  \href{http://dx.doi.org/10.1103/PhysRevD.68.052004}{{\em Phys. Rev. D}
  {\bfseries 68} (2003) 052004},
  \href{http://arxiv.org/abs/hep-ex/0304008}{{\ttfamily arXiv:hep-ex/0304008}}.

\bibitem{ZEUS:2012pwm}
{\bfseries ZEUS} Collaboration, H.~Abramowicz {\em et~al.}, ``{Search for
  first-generation leptoquarks at HERA},''
  \href{http://dx.doi.org/10.1103/PhysRevD.86.012005}{{\em Phys. Rev. D}
  {\bfseries 86} (2012) 012005},
  \href{http://arxiv.org/abs/1205.5179}{{\ttfamily arXiv:1205.5179 [hep-ex]}}.

\bibitem{L3:1991sow}
{\bfseries L3} Collaboration, B.~Adeva {\em et~al.}, ``{Search for leptoquarks
  in Z0 decays},'' \href{http://dx.doi.org/10.1016/0370-2693(91)91346-W}{{\em
  Phys. Lett. B} {\bfseries 261} (1991) 169--176}.

\bibitem{OPAL:1991xaz}
{\bfseries OPAL} Collaboration, G.~Alexander {\em et~al.}, ``{A Search for
  scalar leptoquarks in Z0 decays},''
  \href{http://dx.doi.org/10.1016/0370-2693(91)91717-A}{{\em Phys. Lett. B}
  {\bfseries 263} (1991) 123--134}.

\bibitem{OPAL:1998gjo}
{\bfseries OPAL} Collaboration, G.~Abbiendi {\em et~al.}, ``{Tests of the
  standard model and constraints on new physics from measurements of fermion
  pair production at 183-GeV at LEP},''
  \href{http://dx.doi.org/10.1007/s100529801027}{{\em Eur. Phys. J. C}
  {\bfseries 6} (1999) 1--18},
  \href{http://arxiv.org/abs/hep-ex/9808023}{{\ttfamily arXiv:hep-ex/9808023}}.

\bibitem{OPAL:1998znn}
{\bfseries OPAL} Collaboration, K.~Ackerstaff {\em et~al.}, ``{Search for
  stable and longlived massive charged particles in e+ e- collisions at
  s**(1/2) = 130-GeV - 183-GeV},''
  \href{http://dx.doi.org/10.1016/S0370-2693(98)00518-8}{{\em Phys. Lett. B}
  {\bfseries 433} (1998) 195--208},
  \href{http://arxiv.org/abs/hep-ex/9803026}{{\ttfamily arXiv:hep-ex/9803026}}.

\bibitem{L3:2000bql}
{\bfseries L3} Collaboration, M.~Acciarri {\em et~al.}, ``{Search for
  manifestations of new physics in fermion pair production at LEP},''
  \href{http://dx.doi.org/10.1016/S0370-2693(00)00887-X}{{\em Phys. Lett. B}
  {\bfseries 489} (2000) 81--92},
  \href{http://arxiv.org/abs/hep-ex/0005028}{{\ttfamily arXiv:hep-ex/0005028}}.

\bibitem{UA2:1991ovi}
{\bfseries UA2} Collaboration, J.~Alitti {\em et~al.}, ``{A Search for scalar
  leptoquarks at the CERN anti-p p collider},''
  \href{http://dx.doi.org/10.1016/0370-2693(92)92024-B}{{\em Phys. Lett. B}
  {\bfseries 274} (1992) 507--512}.

\bibitem{CDF:1993dmp}
{\bfseries CDF} Collaboration, F.~Abe {\em et~al.}, ``{A Search for first
  generation leptoquarks in $\bar{p}p$ collisions at $\sqrt{s} = 1.8$ TeV},''
  \href{http://dx.doi.org/10.1103/PhysRevD.48.R3939}{{\em Phys. Rev. D}
  {\bfseries 48} (1993) R3939--R3944}.

\bibitem{Norman:1994tp}
{\bfseries D0} Collaboration, D.~Norman, ``{Search for first and second
  generation leptoquarks at D0},'' in {\em {1994 Meeting of the American
  Physical Society, Division of Particles and Fields (DPF 94)}},
  pp.~1165--1168.
\newblock 8, 1994.
\newblock \href{http://arxiv.org/abs/hep-ex/9409008}{{\ttfamily
  arXiv:hep-ex/9409008}}.

\bibitem{D0:1997eun}
{\bfseries D0} Collaboration, B.~Abbott {\em et~al.}, ``{Search for scalar
  leptoquark pairs decaying to electrons and jets in $\bar{p}p$ collisions},''
  \href{http://dx.doi.org/10.1103/PhysRevLett.79.4321}{{\em Phys. Rev. Lett.}
  {\bfseries 79} (1997) 4321--4326},
  \href{http://arxiv.org/abs/hep-ex/9707033}{{\ttfamily arXiv:hep-ex/9707033}}.

\bibitem{Barfuss:2009zz}
A.-F. Barfuss, ``{Searches for leptoquarks at Tevatron},''
  \href{http://dx.doi.org/10.1088/1742-6596/171/1/012084}{{\em J. Phys. Conf.
  Ser.} {\bfseries 171} (2009) 012084}.

\bibitem{ATLAS:2020dsk}
{\bfseries ATLAS} Collaboration, G.~Aad {\em et~al.}, ``{Search for pairs of
  scalar leptoquarks decaying into quarks and electrons or muons in $ \sqrt{s}
  $ = 13 TeV $pp$ collisions with the ATLAS detector},''
  \href{http://dx.doi.org/10.1007/JHEP10(2020)112}{{\em JHEP} {\bfseries 10}
  (2020) 112}, \href{http://arxiv.org/abs/2006.05872}{{\ttfamily
  arXiv:2006.05872 [hep-ex]}}.

\bibitem{CMS:2018lab}
{\bfseries CMS} Collaboration, A.~M. Sirunyan {\em et~al.}, ``{Search for pair
  production of second-generation leptoquarks at $\sqrt{s}=$ 13 TeV},''
  \href{http://dx.doi.org/10.1103/PhysRevD.99.032014}{{\em Phys. Rev. D}
  {\bfseries 99} no.~3, (2019) 032014},
  \href{http://arxiv.org/abs/1808.05082}{{\ttfamily arXiv:1808.05082
  [hep-ex]}}.

\bibitem{CMS:2018ncu}
{\bfseries CMS} Collaboration, A.~M. Sirunyan {\em et~al.}, ``{Search for pair
  production of first-generation scalar leptoquarks at $\sqrt{s} =$ 13 TeV},''
  \href{http://dx.doi.org/10.1103/PhysRevD.99.052002}{{\em Phys. Rev. D}
  {\bfseries 99} no.~5, (2019) 052002},
  \href{http://arxiv.org/abs/1811.01197}{{\ttfamily arXiv:1811.01197
  [hep-ex]}}.

\bibitem{ATLAS:2020yat}
{\bfseries ATLAS} Collaboration, G.~Aad {\em et~al.}, ``{Search for new
  non-resonant phenomena in high-mass dilepton final states with the ATLAS
  detector},'' \href{http://dx.doi.org/10.1007/JHEP11(2020)005}{{\em JHEP}
  {\bfseries 11} (2020) 005}, \href{http://arxiv.org/abs/2006.12946}{{\ttfamily
  arXiv:2006.12946 [hep-ex]}}. [Erratum: JHEP 04, 142 (2021)].

\bibitem{CMS:2021ctt}
{\bfseries CMS} Collaboration, A.~M. Sirunyan {\em et~al.}, ``{Search for
  resonant and nonresonant new phenomena in high-mass dilepton final states at
  $ \sqrt{s} $ = 13 TeV},''
  \href{http://dx.doi.org/10.1007/JHEP07(2021)208}{{\em JHEP} {\bfseries 07}
  (2021) 208}, \href{http://arxiv.org/abs/2103.02708}{{\ttfamily
  arXiv:2103.02708 [hep-ex]}}.

\bibitem{CMS:2021far}
{\bfseries CMS} Collaboration, A.~Tumasyan {\em et~al.}, ``{Search for new
  particles in events with energetic jets and large missing transverse momentum
  in proton-proton collisions at $ \sqrt{s} $ = 13 TeV},''
  \href{http://dx.doi.org/10.1007/JHEP11(2021)153}{{\em JHEP} {\bfseries 11}
  (2021) 153}, \href{http://arxiv.org/abs/2107.13021}{{\ttfamily
  arXiv:2107.13021 [hep-ex]}}.

\bibitem{Kirk:2023fin}
M.~Kirk, S.~Okawa, and K.~Wu, ``{A $\nu$ window onto leptoquarks?},''
  \href{http://arxiv.org/abs/2307.11152}{{\ttfamily arXiv:2307.11152
  [hep-ph]}}.

\bibitem{Tomalak:2020zfh}
O.~Tomalak, P.~Machado, V.~Pandey, and R.~Plestid, ``{Flavor-dependent
  radiative corrections in coherent elastic neutrino-nucleus scattering},''
  \href{http://dx.doi.org/10.1007/JHEP02(2021)097}{{\em JHEP} {\bfseries 02}
  (2021) 097}, \href{http://arxiv.org/abs/2011.05960}{{\ttfamily
  arXiv:2011.05960 [hep-ph]}}.

\bibitem{ParticleDataGroup:2020ssz}
{\bfseries Particle Data Group} Collaboration, P.~A. Zyla {\em et~al.},
  ``{Review of Particle Physics},''
  \href{http://dx.doi.org/10.1093/ptep/ptaa104}{{\em PTEP} {\bfseries 2020}
  no.~8, (2020) 083C01}.

\bibitem{COHERENT:2020ybo}
{\bfseries COHERENT} Collaboration, D.~Akimov {\em et~al.}, ``{COHERENT
  Collaboration data release from the first detection of coherent elastic
  neutrino-nucleus scattering on argon},''
  \href{http://arxiv.org/abs/2006.12659}{{\ttfamily arXiv:2006.12659
  [nucl-ex]}}.

\bibitem{Picciau:2022xzi}
E.~Picciau, {\em {Low-energy signatures in DarkSide-50 experiment and neutrino
  scattering processes}}.
\newblock PhD thesis, Cagliari U., 2022.

\bibitem{Dzuba:2012kx}
V.~A. Dzuba, J.~C. Berengut, V.~V. Flambaum, and B.~Roberts, ``{Revisiting
  parity non-conservation in cesium},''
  \href{http://dx.doi.org/10.1103/PhysRevLett.109.203003}{{\em Phys. Rev.
  Lett.} {\bfseries 109} (2012) 203003},
  \href{http://arxiv.org/abs/1207.5864}{{\ttfamily arXiv:1207.5864 [hep-ph]}}.

\bibitem{Cadeddu:2018izq}
M.~Cadeddu and F.~Dordei, ``{Reinterpreting the weak mixing angle from atomic
  parity violation in view of the Cs neutron rms radius measurement from
  COHERENT},'' \href{http://dx.doi.org/10.1103/PhysRevD.99.033010}{{\em Phys.
  Rev. D} {\bfseries 99} no.~3, (2019) 033010},
  \href{http://arxiv.org/abs/1808.10202}{{\ttfamily arXiv:1808.10202
  [hep-ph]}}.

\bibitem{Gresham:2012wc}
M.~I. Gresham, I.-W. Kim, S.~Tulin, and K.~M. Zurek, ``{Confronting Top AFB
  with Parity Violation Constraints},''
  \href{http://dx.doi.org/10.1103/PhysRevD.86.034029}{{\em Phys. Rev. D}
  {\bfseries 86} (2012) 034029},
  \href{http://arxiv.org/abs/1203.1320}{{\ttfamily arXiv:1203.1320 [hep-ph]}}.

\bibitem{Dorsner:2014axa}
I.~Dorsner, S.~Fajfer, and A.~Greljo, ``{Cornering Scalar Leptoquarks at
  LHC},'' \href{http://dx.doi.org/10.1007/JHEP10(2014)154}{{\em JHEP}
  {\bfseries 10} (2014) 154}, \href{http://arxiv.org/abs/1406.4831}{{\ttfamily
  arXiv:1406.4831 [hep-ph]}}.

\bibitem{Erler:2013xha}
J.~Erler and S.~Su, ``{The Weak Neutral Current},''
  \href{http://dx.doi.org/10.1016/j.ppnp.2013.03.004}{{\em Prog. Part. Nucl.
  Phys.} {\bfseries 71} (2013) 119--149},
  \href{http://arxiv.org/abs/1303.5522}{{\ttfamily arXiv:1303.5522 [hep-ph]}}.

\bibitem{Workman:2022ynf}
{\bfseries Particle Data Group} Collaboration, R.~L. Workman and Others,
  ``{Review of Particle Physics},''
  \href{http://dx.doi.org/10.1093/ptep/ptac097}{{\em PTEP} {\bfseries 2022}
  (2022) 083C01}.

\bibitem{Blondel:1989ev}
A.~Blondel {\em et~al.}, ``{Electroweak Parameters From a High Statistics
  Neutrino Nucleon Scattering Experiment},''
  \href{http://dx.doi.org/10.1007/BF01549665}{{\em Z. Phys. C} {\bfseries 45}
  (1990) 361--379}.

\bibitem{CHARM:1987pwr}
{\bfseries CHARM} Collaboration, J.~V. Allaby {\em et~al.}, ``{A Precise
  Determination of the Electroweak Mixing Angle from Semileptonic Neutrino
  Scattering},'' \href{http://dx.doi.org/10.1007/BF01630598}{{\em Z. Phys. C}
  {\bfseries 36} (1987) 611}.

\bibitem{Schmaltz:2018nls}
M.~Schmaltz and Y.-M. Zhong, ``{The leptoquark Hunter\textquoteright{}s guide:
  large coupling},'' \href{http://dx.doi.org/10.1007/JHEP01(2019)132}{{\em
  JHEP} {\bfseries 01} (2019) 132},
  \href{http://arxiv.org/abs/1810.10017}{{\ttfamily arXiv:1810.10017
  [hep-ph]}}.

\bibitem{Dreiner:2023bvs}
H.~K. Dreiner, Y.~S. Koay, D.~K\"ohler, V.~M. Lozano, J.~Montejo~Berlingen,
  S.~Nangia, and N.~Strobbe, ``{The ABC of RPV: Classification of R-Parity
  Violating Signatures at the LHC for Small Couplings},''
  \href{http://arxiv.org/abs/2306.07317}{{\ttfamily arXiv:2306.07317
  [hep-ph]}}.

\bibitem{Diaz:2017lit}
B.~Diaz, M.~Schmaltz, and Y.-M. Zhong, ``{The leptoquark
  Hunter\textquoteright{}s guide: Pair production},''
  \href{http://dx.doi.org/10.1007/JHEP10(2017)097}{{\em JHEP} {\bfseries 10}
  (2017) 097}, \href{http://arxiv.org/abs/1706.05033}{{\ttfamily
  arXiv:1706.05033 [hep-ph]}}.

\bibitem{Alwall:2014hca}
J.~Alwall, R.~Frederix, S.~Frixione, V.~Hirschi, F.~Maltoni, O.~Mattelaer,
  H.~S. Shao, T.~Stelzer, P.~Torrielli, and M.~Zaro, ``{The automated
  computation of tree-level and next-to-leading order differential cross
  sections, and their matching to parton shower simulations},''
  \href{http://dx.doi.org/10.1007/JHEP07(2014)079}{{\em JHEP} {\bfseries 07}
  (2014) 079}, \href{http://arxiv.org/abs/1405.0301}{{\ttfamily arXiv:1405.0301
  [hep-ph]}}.

\bibitem{Frederix:2018nkq}
R.~Frederix, S.~Frixione, V.~Hirschi, D.~Pagani, H.~S. Shao, and M.~Zaro,
  ``{The automation of next-to-leading order electroweak calculations},''
  \href{http://dx.doi.org/10.1007/JHEP11(2021)085}{{\em JHEP} {\bfseries 07}
  (2018) 185}, \href{http://arxiv.org/abs/1804.10017}{{\ttfamily
  arXiv:1804.10017 [hep-ph]}}. [Erratum: JHEP 11, 085 (2021)].

\bibitem{Borschensky:2020hot}
C.~Borschensky, B.~Fuks, A.~Kulesza, and D.~Schwartl\"ander, ``{Scalar
  leptoquark pair production at hadron colliders},''
  \href{http://dx.doi.org/10.1103/PhysRevD.101.115017}{{\em Phys. Rev. D}
  {\bfseries 101} no.~11, (2020) 115017},
  \href{http://arxiv.org/abs/2002.08971}{{\ttfamily arXiv:2002.08971
  [hep-ph]}}.

\bibitem{Borschensky:2021hbo}
C.~Borschensky, B.~Fuks, A.~Kulesza, and D.~Schwartl\"ander, ``{Scalar
  leptoquark pair production at the LHC: precision predictions in the era of
  flavour anomalies},'' \href{http://dx.doi.org/10.1007/JHEP02(2022)157}{{\em
  JHEP} {\bfseries 02} (2022) 157},
  \href{http://arxiv.org/abs/2108.11404}{{\ttfamily arXiv:2108.11404
  [hep-ph]}}.

\bibitem{urllq}
\url{https://www.uni-muenster.de/Physik.TP/research/kulesza/leptoquarks.html}.

\bibitem{COHERENT:2021yvp}
{\bfseries COHERENT} Collaboration, D.~Akimov {\em et~al.}, ``{Simulating the
  neutrino flux from the Spallation Neutron Source for the COHERENT
  experiment},'' \href{http://dx.doi.org/10.1103/PhysRevD.106.032003}{{\em
  Phys. Rev. D} {\bfseries 106} no.~3, (2022) 032003},
  \href{http://arxiv.org/abs/2109.11049}{{\ttfamily arXiv:2109.11049
  [hep-ex]}}.

\bibitem{Abele:2022iml}
H.~Abele {\em et~al.}, ``{Particle Physics at the European Spallation
  Source},'' \href{http://arxiv.org/abs/2211.10396}{{\ttfamily arXiv:2211.10396
  [physics.ins-det]}}.

\bibitem{Chatterjee:2022mmu}
S.~S. Chatterjee, S.~Lavignac, O.~G. Miranda, and G.~Sanchez~Garcia,
  ``{Constraining nonstandard interactions with coherent elastic
  neutrino-nucleus scattering at the European Spallation Source},''
  \href{http://dx.doi.org/10.1103/PhysRevD.107.055019}{{\em Phys. Rev. D}
  {\bfseries 107} no.~5, (2023) 055019},
  \href{http://arxiv.org/abs/2208.11771}{{\ttfamily arXiv:2208.11771
  [hep-ph]}}.

\bibitem{Dreiner:2021ext}
H.~K. Dreiner, V.~M. Lozano, S.~Nangia, and T.~Opferkuch, ``{Lepton PDFs and
  multipurpose single-lepton searches at the LHC},''
  \href{http://dx.doi.org/10.1103/PhysRevD.107.035011}{{\em Phys. Rev. D}
  {\bfseries 107} no.~3, (2023) 035011},
  \href{http://arxiv.org/abs/2112.12755}{{\ttfamily arXiv:2112.12755
  [hep-ph]}}.

\bibitem{Husek:2021isa}
T.~Husek, K.~Monsalvez-Pozo, and J.~Portoles, ``{Constraints on leptoquarks
  from lepton-flavour-violating tau-lepton processes},''
  \href{http://dx.doi.org/10.1007/JHEP04(2022)165}{{\em JHEP} {\bfseries 04}
  (2022) 165}, \href{http://arxiv.org/abs/2111.06872}{{\ttfamily
  arXiv:2111.06872 [hep-ph]}}.

\end{thebibliography}\endgroup

\end{document}